\documentclass[aps,pre,reprint,twocolumn,superscriptaddress,showpacs]{revtex4-1}
\usepackage{amssymb,amsmath}
\usepackage[table]{xcolor}
\usepackage{graphicx}
\usepackage{mathtools}
\usepackage[export]{adjustbox}
\usepackage{overpic}
\usepackage[colorlinks=true,
 linkcolor=black,
 urlcolor=blue,
 citecolor=blue]{hyperref}
\usepackage{nicefrac}
\usepackage{multirow}
\usepackage{lipsum} 
\usepackage[normalem]{ulem}
\usepackage{pbox}
\newcolumntype{C}[1]{>{\centering\let\newline\\\arraybackslash\hspace{0pt}}m{#1}}

\usepackage{enumitem}

\usepackage{framed,color}
\definecolor{shadecolor}{rgb}{0.85,0.80,0.80}

\definecolor{myorange}{RGB}{253, 184, 99}
\definecolor{mypurple}{RGB}{178, 171, 210}

\newcommand{\comments}[1]{}
\usepackage{multirow}

\definecolor{ao}{rgb}{0.0, 0.4, 0.1}

\def\d{\mathrm{d}}
\def\star{*}
\def\xx{x}
\def\E{\mathcal{E}}

\def\M{\mathcal{M}}

\def\nhalf{{\nicefrac{-1}{2}}}

\renewcommand{\vec}[1]{\mathbf{#1}}
\newcommand{\be}{\begin{equation}}
\newcommand{\ee}{\end{equation}}
\newcommand{\bd}{\begin{displaymath}}
\newcommand{\ed}{\end{displaymath}}
\newcommand{\BE}{\begin{eqnarray}}
\newcommand{\EE}{\end{eqnarray}}

\newcommand{\bx}{\ensuremath{\mathbf{x}}}

\newcommand{\avg}[1]{\left\langle{#1}\right\rangle}

\newcommand{\cM}{{\cal M}}
\newcommand{\cL}{{\cal L}}

\newcommand{\cMbar}{{\cal M}_{\rm avg}}

\newcommand{\cF}{{\cal F}}
\newcommand{\cFbar}{{\cal F}_{\rm avg}}

\begin{document}
\title{Model reduction methods for classical stochastic systems with fast-switching environments: reduced master equations, stochastic differential equations, and applications}
\author{Peter G. Hufton}
\affiliation{Theoretical Physics, School of Physics and Astronomy, The University of Manchester, Manchester M13 9PL, United Kingdom}
\author{Yen Ting Lin}
\affiliation{Theoretical Physics, School of Physics and Astronomy, The University of Manchester, Manchester M13 9PL, United Kingdom}
\affiliation{Center for Nonlinear Studies and Theoretical and Biophysics Group, Theoretical Division, Los Alamos National Laboratory, Los Alamos, New Mexico 87545, USA}
\author{Tobias Galla}
\affiliation{Theoretical Physics, School of Physics and Astronomy, The University of Manchester, Manchester M13 9PL, United Kingdom}

\date{\today}

\begin{abstract}
We study classical stochastic systems with discrete states, coupled to switching external environments. For fast environmental processes we derive reduced dynamics for the system itself, focusing on corrections to the adiabatic limit of infinite time scale separation. In some cases, this leads to master equations with negative transition `rates' or bursting events. We devise a simulation algorithm in discrete time to unravel these master equations into sample paths, and provide an interpretation of bursting events. Focusing on stochastic population dynamics coupled to external environments, we discuss a series of approximation schemes combining expansions in the inverse switching rate of the environment, and a Kramers--Moyal expansion in the inverse size of the population. This places the different approximations in relation to existing work on piecewise-deterministic and piecewise-diffusive Markov processes. We apply the model reduction methods to different examples including systems in biology and a model of crack propagation.
\end{abstract}
\maketitle

\section{Introduction}

Physical and biological systems can never be fully isolated from their environment. This includes the dynamics of microbes in time-varying external conditions (e.g., antibiotic treatment) \cite{kussell2005bacterial, kussell2005phenotypic, leibler2010individual, thomas2014phenotypic}, or protein production in gene regulatory networks, influenced by the stochastic binding and unbinding of promoters \cite{kepler2001stochasticity,thattai2004stochastic, swain2002intrinsic,assaf2013extrinsic,duncan2015noise}. Other examples can be found in models of evolutionary dynamics \cite{assaf2, ashcroft2014fixation,wienand,west2018}, the spread of diseases \cite{black2010stochastic}, and in ecology and population dynamics \cite{escudero2008persistence, assaf2008population,luo2007stochastic, zhu2009competitive}. Many of models of these phenomena contain two types of randomness: one intrinsic to the system itself, and another generated by the noise in the environmental dynamics.  Applications of systems coupled to stochastic external environments go as far as reliability analysis and crack propagation in materials, where environmental states correspond to different strains due to external loading \cite{chiquet2008method,chiquet2008modelling,chiquet2009piecewise, lorton2013methodology, zhang2008piecewise,lorton2013computation}. The study of open quantum systems defines an entire area of research \cite{Breuer2002, Breuer2016,Vega2017}.

These examples share a common structure: there is the system proper and the environment, and a coupling between them; this interaction can act either in one way or in both directions. In such situations it is often not possible (or desirable) to track and analyse in detail the dynamics of the system {\em and} that of the environment. Instead the focus is on deriving reduced dynamics for the system itself, which in some way account for the influence of the environment on the system. 
Work on open quantum systems for example focuses on understanding the dynamics of reduced density matrices after integrating out the environment \cite{Breuer2002,Breuer2016,Vega2017}.  

Existing work on open classical systems includes those described by stochastic differential equations (SDEs) coupled to continuous environments \cite{Bressloff2016,assaf1,assaf2,assaf3}, and deterministic models with discrete external noise \cite{Bressloff2014,Bressloff2017,Bressloff2017a}. A specific case of Brownian particles, subject to random external gating is considered in Ref. \cite{Bressloffgated}. In chemical or biological systems the quasi-steady-state approximation or related adiabatic reduction techniques can be used to eliminate fast reactions \cite{QSSA1,QSSA2}.

In this paper we consider open stochastic systems with discrete states. While some of our theory is applicable more generally, we mostly focus on populations of interacting `individuals'. We will often use the words `system' and `population' synonymously. Examples we have in mind are chemical reaction system with discrete molecules, or populations in biological systems, composed of members of different species. For a fixed environment, such a system is described by a (classical) master equation defined by the transition rates between its discrete states. These transitions are typically events in which particles are produced or removed from the population, or in which a particle of one type is converted into another type. In biological populations they can represent birth or death events. We are interested in cases in which such a population is coupled to an external environment, which also takes discrete states. The environmental states in turn affect the transition rates within the population. 

Our aim is to study the reduced dynamics of such systems after the environmental dynamics are integrated out. In particular we focus on the limit in which the environmental dynamics are fast compared to those of the population, but where the separation of time scales is not infinite. We show how reduced master equations can be derived systematically; interestingly negative transition `rates' can emerge in these reduced dynamics. This is similar to what is observed in the theory of open quantum systems \cite{Breuer2016,breuer2009stochastic,piiloprl}, but there are also key differences. We provide an approximation at the level of sample paths in discrete-time, and comment on different numerical schemes to address master equations with negative rates. 
In addition, we describe in more detail how expansions in the inverse time scale of the environmental dynamics can be combined with expansions in the inverse system size of the population. These are effectively weak-noise expansions for the extrinsic and intrinsic stochasticity of the problem. Finally we apply the formalism to a number of examples ranging from gene regulatory networks to crack propagation in materials.

The remainder of the paper is organised as follows. In Sec.~\ref{sec:def} we introduce the type of model we address, a classical stochastic system with discrete states coupled to an external environment, also with discrete states. In Sec.~\ref{sec:analysis} we present the detailed mathematics used for the analysis and derive an effective master equation in the limit of fast time scales of the environmental switching; specifically, our analysis includes next-order corrections to the adiabatic limit of infinitely fast environments. We illustrate this using a set of simple examples. In Sec.~\ref{sec:twospecies} we use this general result to show how master equations with negative transition `rates' arise, and comment on their interpretation and on a numerical scheme to sample its statistics at ensemble level. Sec.~\ref{sec:paths} describes the effective dynamics on the level of sample paths, and provides more insight into reduced master equations with negative `rates'. In Sec.~\ref{sec:km} we combine expansions in the inverse size of the population with that in the time scale of the switching dynamics, and provide a systematic classification of the different resulting model reduction schemes. We discuss a set of applications in Sec.~\ref{sec:appl}, before we summarise and present our conclusions in Sec.~\ref{sec:concl}.

\section{General definitions}\label{sec:def}
\subsection{Model}
We focus on a classical system with discrete states, labelled $\ell$, which is coupled to an environment also taking discrete states, which we label $\sigma$. The system and the environment evolve in continuous time.
The dynamics of the system itself depend on the current state of the environment.
The environment in turn switches between its states, with transition rates which can depend on the state $\ell$ of the system.
The combined dynamics of system and environment are then governed by the master equation
\begin{align}
\frac{\d}{\d t}p(\ell,\sigma,t)={}&
\M_\sigma p(\ell,\sigma,t)\nonumber\\
&+\lambda\sum_{\sigma'} A_{\sigma'\to\sigma}{(\ell)} p(\ell,\sigma',t),
\label{eq:mastermaster}
\end{align}
where $p(\ell,\sigma,t)$ is the joint probability of finding the system in state $\ell$ and the environment in state $\sigma$ at time $t$.
The object $\M_\sigma$ is an operator, and determines how the state of the system can change when the environment is in state $\sigma$.
More specifically, the effect of the operator can be written in the form
\begin{equation}
\M_\sigma p(\ell,\sigma,t)\equiv\sum_{\ell'} {R}^{\left(\sigma\right)}_{\ell'\to\ell} p(\ell',\sigma,t).
\end{equation}
The matrix element ${R}_{\ell'\rightarrow\ell}^{(\sigma)}$ describes the rate at which the system transitions from state $\ell'$ to state $\ell$ when the environment is in state $\sigma$. For a chemical reaction system, the types of allowed transitions are specified by the stoichiometric coefficients; together with associated reaction rates these determine the transition matrix. In the context of population dynamics the matrix ${R}_{\ell'\rightarrow\ell}^{(\sigma)}$ is defined by the underlying birth and death processes (e.g., see Refs.~\cite{ewens,TraulsenHauert}).

The second term in Eq.~\eqref{eq:mastermaster}, proportional to $\lambda$, characterises the environmental switching. The rate with which the environment transitions from state $\sigma$ to state $\sigma'$ is $\lambda A_{\sigma\to\sigma'}(\ell)$. In the most general setup, these can depend on the state $\ell$ of the system. We write $\lambda\mathbf{A}(\ell)$ for the corresponding transition matrix. The pre-factor $\lambda>0$ has been introduced to parametrise the time scale of the environment, relative to the internal dynamics of the population. Large values of $\lambda\gg 1$ indicate a fast environmental process.
To fix the diagonal elements of both transition matrices, we use the convention $R_{\ell\to\ell}^{(\sigma)}=-\sum_{\ell'\neq\ell }R_{\ell\to\ell'}^{(\sigma)}$, and $A_{\sigma\to\sigma}(\ell)=-\sum_{\sigma'\neq\sigma}A_{\sigma\to\sigma'}(\ell)$. 
\subsection{Simplification in the adiabatic limit}
We first consider the so-called `adiabatic' limit of infinitely fast environmental switching, $\lambda\to \infty$. In this limit we find from Eq.~\eqref{eq:mastermaster}
\begin{equation}\label{eq:astat}
\sum_{\sigma'} A_{\sigma'\to\sigma}(\ell)p(\ell, \sigma', t)=0,
\end{equation}
for all $\ell$.
We introduce the notation $\Pi(\ell,t)=\sum_\sigma p(\ell,\sigma,t)$ for the marginal of the probability distribution after integrating out the environment.
We also write the joint distribution in terms of this marginal and a conditional probability: $p(\ell,\sigma,t)=\rho(\sigma|\ell,t)\Pi(\ell,t)$.
Substituting this into Eq.~\eqref{eq:astat} we find 
\begin{equation}
\sum_{\sigma'} A_{\sigma'\to\sigma}(\ell)\rho^\star(\sigma'|\ell)=0,
\label{eq:adiabatic_environment}
\end{equation}
for all $\ell$, for the stationary distribution of the environment conditioned on the state of the system. We label this stationary distribution by an asterisk. In the adiabatic limit we then have 
\begin{equation}\label{eq:statad}
p(\ell,\sigma,t)=\rho^\star(\sigma|\ell)\Pi(\ell,t). 
\end{equation}
We will use this relation as a starting point for further analysis; in this context we also obtain the reduced dynamics for $\Pi(\ell,t)$ in the adiabatic limit.

\section{Analysis for fast but finite environments}\label{sec:analysis}
\subsection{General formalism}
\label{sec:genform}
Our next aim is to derive reduced dynamics in the limit of fast environmental switching, but keeping the time-scale separation finite (i.e., $\lambda$ large, but finite). Specifically, the objective is to derive a closed equation for the time-evolution of the distribution of states $\Pi(\ell,t)$. This is done, in essence, by performing an expansion of the joint master equation for system and environment in powers of the time-scale separation $\lambda^{-1}$.
We then retain the leading and sub-leading terms, and integrate out the environment. The algebraic steps are similar to those in Ref.~\cite{Bressloff2014}, in which the authors work in the context of piecewise-deterministic Markov processes. We carry out the calculation starting from a system with discrete states $\ell$. As we will see below, this leads to interesting features of the reduced dynamics, not necessarily seen for continuous states.

To separate leading-order terms from sub-leading contributions we start with the decomposition
\begin{equation}\label{eq:ansatz}
p(\ell,\sigma,t)=\rho^\star(\sigma|\ell)\Pi(\ell,t)+\frac{1}{\lambda}w_\sigma(\ell,t).
\end{equation}
The sub-leading order term $w_\sigma(\ell,t)$ describes deviations from the adiabatic limit [Eq.~\eqref{eq:statad}], due to a finite time scale of the environment. Because of normalisation, this ansatz requires $\sum_\sigma w_\sigma(\ell,t)=0$, for all $\ell$.
We proceed by inserting Eq.~\eqref{eq:ansatz} into Eq.~\eqref{eq:mastermaster}, and obtain
\begin{align}
\rho^\star(&\sigma|\ell) \frac{\d}{\d t} \Pi(\ell,t)+\frac{1}{\lambda} \frac{\d}{\d t} w_\sigma(\ell,t) \nonumber \\
={}&\cM_\sigma \left[\rho^\star(\sigma|\ell)\Pi(\ell,t) \right] +\sum_{\sigma'} A_{\sigma'\to\sigma}(\ell) w_{\sigma'}(\ell,t) \nonumber\\
& + \frac{1}{\lambda}\cM_\sigma w_\sigma(\ell,t),\label{eq:bhelp1}
\end{align}
where one further term has been eliminated using Eq.~\eqref{eq:adiabatic_environment}.
Next, we sum over the environmental states $\sigma$ for each $\ell$. We find
\begin{equation}
\frac{\d}{\d t}\Pi(\ell,t)
=\sum_\sigma\cM_\sigma \left[\rho^\star(\sigma|\ell)\Pi(\ell,t) \right]+ \frac{1}{\lambda}\sum_\sigma \cM_\sigma w_\sigma(\ell,t). 
\label{eq:bhelp3}
\end{equation}
Once the $w_\sigma(\ell,t)$ are expressed in terms of $\Pi(\ell,t)$, this equation describes the time-evolution of $\Pi(\ell,t)$, valid to sub-leading order in $\lambda^{-1}$.

To find the $w_\sigma(\ell,t)$ we collect the terms of order $(1/\lambda)^0$ in Eq.~\eqref{eq:bhelp1},
\begin{align}
\sum_{\sigma'}A_{\sigma'\to\sigma}(\ell) w_{\sigma'}(\ell,t) =&{}
\rho^\star(\sigma|\ell)\sum_{\sigma'}\cM_{\sigma'} \left[ \rho^\star(\sigma'|\ell)\Pi(\ell,t) \right] \nonumber \\
&-\cM_\sigma \left[\rho^\star(\sigma|\ell)\Pi(\ell,t) \right], \label{eq:bhelp6}
\end{align}
where we have used Eq.~\eqref{eq:bhelp3} to further simplify the result. Effectively, we have disregarded terms of order $\lambda^{-1}$ in Eq.~\eqref{eq:bhelp1}.
This procedure indicates that the $w_\sigma(\ell,t)$ are to be obtained as the solution of Eq.~\eqref{eq:bhelp6}, subject to $\sum_\sigma w_\sigma(\ell,t)=0$ for all $\ell$ and $t$. The truncation of higher order terms of leads to an error in Eq.~\eqref{eq:bhelp6} of order $\lambda^{-1}$. We note that in specific cases master equations for the system can be obtained in closed form without truncation (examples can be found in Refs.~\cite{Sancho1984,Hernandez-Garcia1989}). These usually rely on specific properties of the model at hand, such as linearity. Eqs.~(\ref{eq:bhelp3}) and (\ref{eq:bhelp6}), while constituting an approximation to sub-leading order in $\lambda^{-1}$, hold more generally; we have not made significant restrictions on the dynamics of the system (i.e., on the operators $\cM_\sigma$).

\subsection{Switching dynamics independent of state of the system with two environmental states}
\label{sec:two_environments}
We now make a simplifying assumption, and consider the case in which the environmental switching dynamics are independent of the state of the population; that is to say, the transition rate matrix $A_{\sigma\to\sigma'}$ does not depend on $\ell$.
In this case, the stationary distribution of the environment in the adiabatic limit is independent of the state of population, i.e., $\rho^\star(\sigma|\ell)=\rho^\star_\sigma$. The more general case is discussed further in Appendix \ref{sec:gen} and below in Sec.~\ref{sec:appl}.

In this simplified case the dynamics in the adiabatic limit are given by
\be\label{eq:adiab}
\frac{\d}{\d t}\Pi(\ell,t) = \cMbar \Pi(\ell,t), 
\ee
where $\cMbar=\sum_{\sigma'}\rho^\star_{\sigma'}\cM_{\sigma'}$ is an effective, average operator.
Equation~\eqref{eq:adiab} is obtained from Eq.~\eqref{eq:bhelp3} by sending $\lambda\to\infty$, and using $\rho^\star(\sigma|\ell)=\rho^\star_\sigma$.

Equation~\eqref{eq:bhelp6}, on the other hand, reduces to
\begin{equation}
\sum_{\sigma'}A_{\sigma'\to\sigma} w_{\sigma'}(\ell,t) = \rho^\star_\sigma \left[ \cMbar -\cM_\sigma \right]\Pi(\ell,t).\label{eq:bhelp7}
\end{equation}
This relation indicates a balance of the form $\sum_{\sigma'}A_{\sigma'\to\sigma} w_{\sigma'}(\ell,t) + \left[\cM_\sigma -\cMbar \right]\rho^\star_\sigma \Pi(\ell,t)=0$.
To understand this in more detail, we recall that $w_\sigma(\ell,t)$ describes the next-order deviation of the solution of Eq.~(\ref{eq:mastermaster}) from the adiabatic limit when the environmental switching is finite [Eq.~\eqref{eq:ansatz}]. When the environment is in state $\sigma$, the first term in the above balance relation is the influx of probability into state $\ell$ induced by these deviations and due to the environmental switching.
Secondly, for finite environmental switching, the dynamics of the population are governed not by $\cMbar$, but by $\cM_\sigma$ when the environment is in state $\sigma$.
The second term in the above relation reflects this; self-consistency of the ansatz requires that these contributions balance.

While the above procedure applies to an arbitrary number of discrete environmental states, it is useful to look at the case of two states, which we label $\sigma=0$ and $\sigma=1$.
We then have $w_0(\ell,t)=-w_1(\ell,t)$ for all $\ell$ and $t$. To shorten the notation, we write $k_0$ and $k_1$ for the switching rates $A_{1\to 0}$ and $A_{0\to 1}$ respectively. In the adiabatic limit, the probabilities of finding the environment in each of its two states are then given by
\begin{align}
\rho^\star_0=\frac{k_0}{k_0+k_1}, && \rho^\star_1=\frac{k_1}{k_0+k_1}.\label{eq:rhostat}
\end{align}
From Eq.~\eqref{eq:bhelp7} one obtains
\begin{equation}
w_\sigma(\ell,t) = \frac{k_\sigma}{(k_0+k_1)^2}\left[\cM_\sigma-\cMbar \right]\Pi(\ell,t).
\end{equation}
Substituting in Eq.~\eqref{eq:bhelp3} and simplifying, we arrive at
\begin{equation}\label{eq:2states}
\frac{\d}{\d t}\Pi(\ell,t) = \cMbar \Pi(\ell,t) + \frac{1}{2} \frac{\theta^2}{\lambda}(\cM_0-\cM_1)^2\Pi(\ell,t) ,
\end{equation}
where we have introduced the constant
\begin{equation}\label{eq:varsigma}
\theta^2=\frac{2k_0k_1}{(k_0+k_1)^3}.
\end{equation}
For systems with two environmental states and with population-independent environmental switching, Eq.~\eqref{eq:2states} is a general result approximating the dynamics in the limit of fast switching. It captures the time-evolution of $\Pi(\ell,t)$ up to and including sub-leading terms in $\lambda^{-1}$. We will refer to this equation (and its analogue for more complicated setups) as a reduced master equation. An expression similar to Eq.~(\ref{eq:2states}) was derived in Ref.~\cite{Bressloff2014} for systems with continuous states. We note that $(\cM_0-\cM_1)^2=(\cM_0^2-1)+(\cM_1^2-1)-(\cM_0\cM_1-1)-(\cM_1\cM_0-1)$, indicating that Eq.~(\ref{eq:2states}) preserves total probability, i.e., $\frac{\d}{\d t}\sum_\ell \Pi(\ell,t)=0$.

We will next illustrate this result in the context of two simple, but instructive examples.

\subsection{Basic example}\label{sec:example1}
We consider a population of discrete individuals who all belong to a single species.
The state of the population is specified by the number $n$ of individuals.
Discrete events involve the removal (death) of existing individuals ($n\to n-1$), or the production (birth) of new individuals ($n\to n+1$).
In this first example we assume that the per capita death rate $\delta$ does not depend on the state of the environment.
The birth rate, however, does: it is $\Omega\beta_0$ in environmental state $\sigma=0$, and $\Omega\beta_1$ in environmental state $\sigma=1$.
The scale parameter $\Omega>0$ sets the typical number of particles; see also the next paragraph. This simple setup is widely used as an elementary model of protein production controlled by the state of a gene~\cite{kepler2001stochasticity, zeiser2010autocatalytic, grima2012steady, duncan2015noise, LinHufton2018}.

For this model, the operator $\M_\sigma$ can be written as
\begin{equation}
\cM_\sigma=\Omega\beta_\sigma ({\cal E}^{-1}-1) + \delta ({\cal E}-1)n, \label{eq:cm0cm1}
\end{equation}
where we have introduced the raising operator $\E$, defined by its action on a function of $n$: $\E f(n)=f(n+1)$.
Operators act on everything to their right.
We find
\begin{equation}
\cMbar= \Omega\beta_{\rm avg} ({\cal E}^{-1}-1) + \delta({\cal E}-1)n,
\end{equation}
where $\beta_{\rm avg} = (k_1 \beta_0 + k_0 \beta_1)/(k_0+k_1)$. This operator describes the dynamics in the limit of infinitely fast switching ($\lambda\to\infty$). The resulting birth rate, $\Omega\beta_{\rm avg}$, is the weighted average of the birth rates in the two environments. The total rate with which deaths occur in the population is $\delta n$. These rates balance when $n=(\beta_{\rm avg}/\delta)\Omega$. It is in this sense that $\Omega$ sets the typical scale for the population size.

Inserting the expression for $\cM_\sigma$ into Eq.~\eqref{eq:2states} and reorganising terms we find
\begin{align}
\frac{\d}{\d t}\Pi(n) ={}&\delta({\cal E}-1)n\Pi(n) \nonumber \\
&+\left[\Omega\beta_{\rm avg}- \frac{\Omega^2\theta^2}{\lambda} (\beta_0-\beta_1)^2\right] ({\cal E}^{-1}-1) \Pi(n) \nonumber \\
&+\frac{1}{2}\frac{\Omega^2\theta^2}{\lambda} (\beta_0-\beta_1)^2 \left[{\cal E}^{-2}-1\right]\Pi(n), \label{eq:aw10}
\end{align}
where we have suppressed the explicit dependence of $\Pi(n)$ on time.
This equation captures terms up to (and including) order $\lambda^{-1}$; higher-order terms have been discarded.

Each term in the reduced master equation can be seen as describing a particular reaction (or type of event) in the population.
The first term on the right-hand side (RHS) of Eq.~\eqref{eq:aw10} describes death events which occur with per capita rate $\delta$. These events occur in either of the two environments, and appear in the reduced dynamics unaltered. The second term indicates birth events occurring with a rate $\beta_{\rm eff}\equiv \Omega\beta_{\rm avg}- (\Omega^2\theta^2/\lambda) (\beta_0-\beta_1)^2$. The reduced dynamics are derived for $\lambda\gg 1$, and we will always assume that $\lambda$ is large enough so that effective rates such as $\beta_{\rm eff}$ are non-negative. The third term on the right-hand side of Eq.~\eqref{eq:aw10} describes events in which {\em two} individuals are created at the same time. This occurs with rate $\frac{1}{2}(\Omega^2\theta^2/\lambda) (\beta_0-\beta_1)^2$; we note that this rate is proportional to $\lambda^{-1}$. Such events are not part of the original dynamics in either environment (neither $\cM_0$ nor $\cM_1$ contain events of this type). They come about due to the fast switching with large, but finite time scale separation, and indicate `bursting' behaviour. This is discussed in more detail in Sections~\ref{sec:twospecies} and \ref{sec:paths}.  We stress that this type of bursting is different from that discussed for example in Refs.~\cite{friedman,Shahrezaei2008,lin2016gene,Lin2016bursting}; there, bursting in protein production is due to short-lived mRNA as a source of protein.

For further illustration, we briefly consider a second, albeit similar, example. We assume now that the birth rate is equal in the two environments ($\beta_0=\beta_1\equiv\beta$), but that there are different death rates, $\delta_0$ and $\delta_1$. We find
\begin{align}
\frac{\d}{\d t}\Pi(n) ={}& \Omega\beta ({\cal E}^{-1}-1)\Pi(n) \nonumber \\
&+ ({\cal E}-1) [ \delta_{\rm avg} - \frac{\theta^2}{\lambda} (\delta_0-\delta_1)^2(2n-1) ] n\Pi(n) \nonumber \\
&+\frac{1}{2}\frac{\theta^2}{\lambda} (\delta_0-\delta_1)^2 \left[{\cal E}^{2}-1\right]n(n-1)\Pi(n)
\label{eq:masterexample2}\end{align}
for the reduced dynamics.
Again we note bursting behaviour, the last term in Eq.~\eqref{eq:masterexample2} describes `double death' events, which are not present in the original dynamics.
The factor $n(n-1)$ ensures that such events can only occur when there are at least two individuals in the population.

\section{Several species and reduced master equations with negative transition rates}
\label{sec:twospecies}
\subsection{Model}\label{sec:twospecies:model}
We next consider an example in which there are two types of particles, labelled $A$ and $B$. This is still a relatively simple setup, but it will help reveal a number of interesting features which can emerge in the reduced dynamics.

Particles of either type are removed with constant per capita rates $\gamma$ and $\delta$, respectively, and are created with rates $\Omega\alpha_\sigma$ and $\Omega\beta_\sigma$.
These production rates depend on the state of the environment, as indicated by the subscript.
The population takes states $\ell=(n_A,n_B)$, where $n_A$ is the number of particles of type $A$, and $n_B$ the number of particles of type $B$.
We then have operators
\begin{align}
 \cM_\sigma ={}&\Omega\alpha_\sigma(\E_{A}^{-1}-1)+\gamma (\E_{A}-1)n_A \nonumber \\
 &+\Omega\beta_\sigma(\E_{B}^{-1}-1)+\delta (\E_{B}-1)n_B, \label{eq:simple}
\end{align}
where $\E_A f(n_A,n_B)=f(n_A+1,n_B)$, and similarly for $\E_B$.
The switching between environmental states is the same as in the previous section.
Using Eq.~\eqref{eq:2states} we find, to sub-leading order in $\lambda^{-1}$,
\begin{align}
\frac{\d}{\d t} \Pi={}&\gamma(\E_{A}-1) n_A \Pi+\delta(\E_{B} -1) n_B \Pi \nonumber \\
&+\Omega\alpha_{\rm eff} (\E_{A}^{-1}-1) \Pi+\Omega\beta_{\rm eff}(\E_{B}^{-1}-1) \Pi \nonumber \\
&+ \frac{\Omega^2\theta^2}{2\lambda}(\Delta\alpha)^2 (\E_{A}^{-2}-1) \Pi \nonumber \\
&+ \frac{\Omega^2\theta^2}{2\lambda}(\Delta\beta)^2 (\E_{B}^{-2}-1) \Pi \nonumber \\
&+\frac{\Omega^2\theta^2}{\lambda}\Delta\alpha\Delta\beta(\E_{A}^{-1}\E_{B}^{-1}-1) \Pi,
\label{eq:cme_ab_approx}
\end{align}
where $\Delta \alpha\equiv\alpha_0-\alpha_1$ and $\Delta\beta\equiv\beta_0-\beta_1$, and where
\begin{align}\begin{split}
\alpha_{\rm eff}={}& \alpha_\text{avg} - \frac{\Omega\theta^2}{\lambda}(\Delta\alpha)^2 - \frac{\Omega\theta^2}{\lambda}\Delta\alpha\Delta\beta,\ \\
\beta_{\rm eff}={}& \beta_\text{avg} - \frac{\Omega\theta^2}{\lambda}(\Delta\beta)^2 - \frac{\Omega\theta^2}{\lambda} \Delta\alpha\Delta\beta.
\label{eq:effrates}
\end{split}\end{align}
The quantity $\alpha_{\rm avg}$ is defined as above, and similar for $\beta_{\rm avg}$. We have suppressed the explicit dependence of $\Pi$ on $n_A, n_B$ and $t$ to keep the notation compact.

Again, we can interpret the reduced master equation as a set of reactions.
The first two terms on the RHS of Eq.~\eqref{eq:cme_ab_approx} describe particle removal, present already in the original model, and independent of the state of the environment.
The terms in the second line are single-birth reactions, as appeared originally in the model, but now with effective birth rates in the reduced dynamics as indicated in Eq.~\eqref{eq:effrates}. Similar to the example in the previous Section, the effective rates $\alpha_{\rm eff}$ and $\beta_{\rm eff}$ are non-negative, provided the switching is fast enough. Given that the reduced dynamics are derived in the limit $\lambda\gg 1$, we always assume that the time-scale separation $\lambda$ is large enough so that $\alpha_{\rm eff}, \beta_{\rm eff}\geq 0$.

\begin{figure}[t!]
\includegraphics[width=0.95\columnwidth,valign=t]{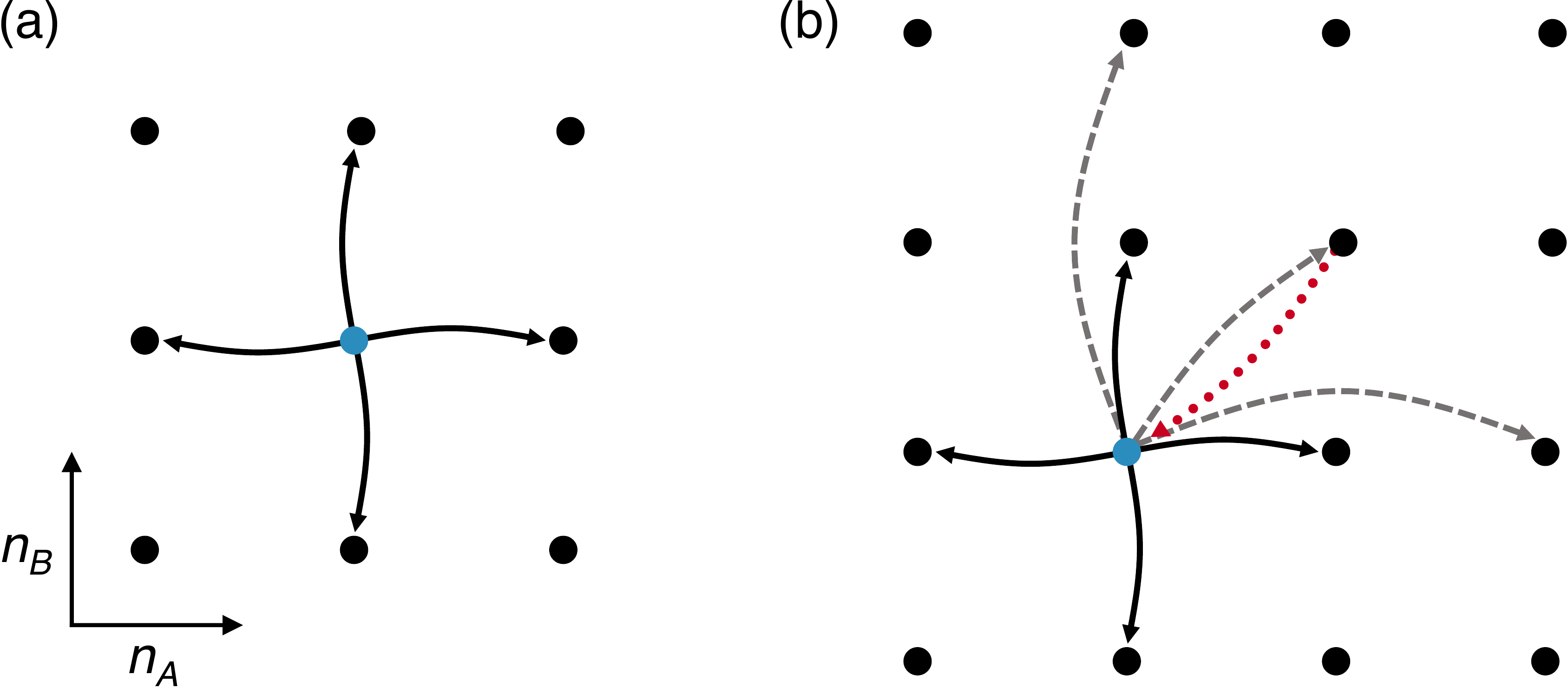}
\caption{Illustration of the possible reactions for (a) the model described by equations (\ref{eq:simple}), and (b) the approximation to the model described by Eq.~(\ref{eq:cme_ab_approx}). In the original model the next event can take the population from $(n_A,n_B)$ to four possible destinations: $(n_A\pm 1,n_B), (n_A,n_B\pm 1)$. The bursting reactions in the reduced model lead to further states which can be reached, indicated by grey dashed arrows; these are $(n_A+2,n_B), (n_A, n_B+2), (n_A+1, n_B+1)$. For certain choices of parameters the transition to $(n_A+1, n_B+1)$ can have a negative `rate'. In this case the flow of probability is from $(n_A+1, n_B+1)$ to $(n_A, n_B)$ as indicated by the red dotted arrow; see Sec.~\ref{sec:renormalisation} for details.
}
\label{fig:scheme}
\end{figure}

The remaining terms in Eq.~\eqref{eq:cme_ab_approx} represent reactions which are not present in the original model;
they arise from the effects of integrating out the environment.
These terms represent `bursting' reactions; they describe events in which two particles of type $A$ are produced simultaneously, or two particles of type $B$, or one of either type.
This is illustrated in Fig.~\ref{fig:scheme}.
Panel (a) is a schematic showing the four states that the population can reach from a given state in the next event in the original model. Panel (b) shows that the reduced dynamics allow three additional destinations (indicated by grey dashed arrows). 
The rates of the first two bursting reactions in Eq.~\eqref{eq:cme_ab_approx} are proportional to $(\Delta\alpha)^2$ and $(\Delta\beta)^2$, and are always positive [lines three and four on the right-hand side of  Eq.~\eqref{eq:cme_ab_approx}].
The rate of the third bursting reaction [last term on RHS of  Eq.~\eqref{eq:cme_ab_approx}] is positive only if $\Delta\alpha$ and $\Delta\beta$ have the same sign.
If $\Delta\alpha\Delta\beta<0$, this reaction will have a negative (pseudo-) rate, no matter how large the time scale separation $\lambda$.
In this case, it is not immediately clear how to proceed with the interpretation of Eq.~\eqref{eq:cme_ab_approx}. We will return to this below in Sec.~\ref{sec:renormalisation}, after we first briefly consider the case $\Delta\alpha\Delta\beta>0$. 

\begin{figure*}[t!]
\includegraphics[width=0.82\textwidth]{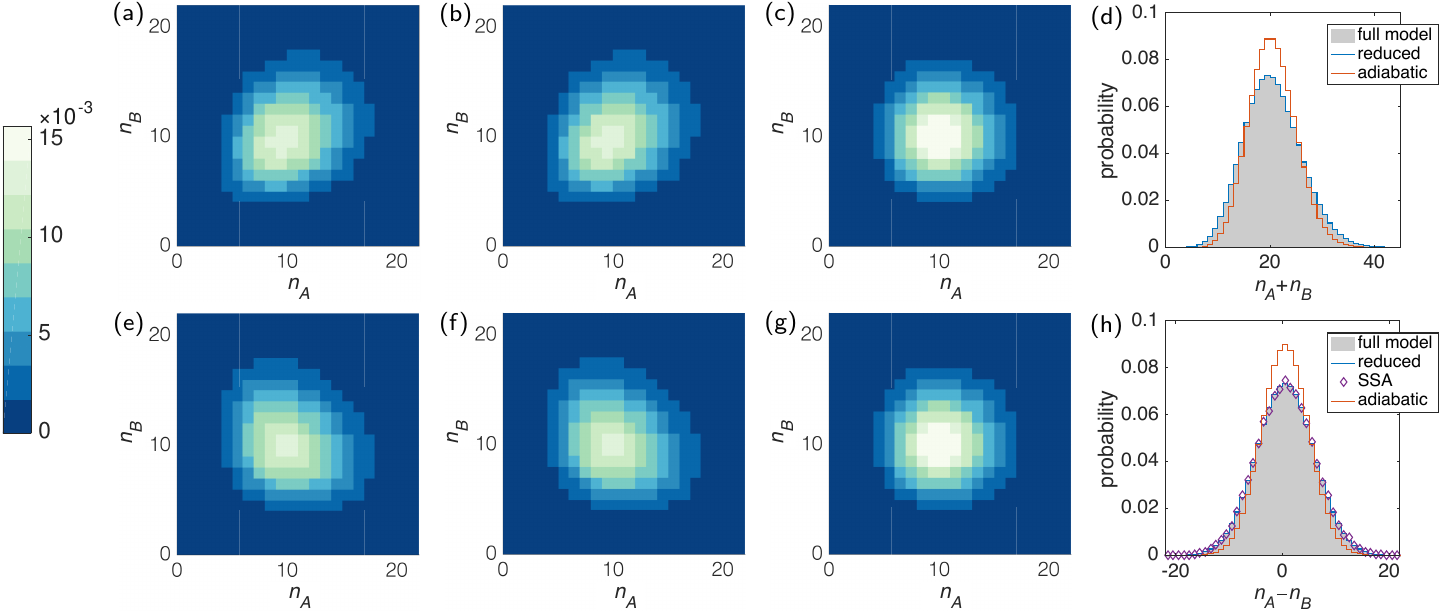}
\caption{Stationary distribution of the model defined in Sec.~\ref{sec:twospecies:model} [see Eqs.~\eqref{eq:simple}]. The upper panels are for $\Delta\alpha\Delta\beta>0$, the lower row for $\Delta\alpha\Delta\beta<0$. The distributions are obtained by numerical integration of: (a,e) the full master equation with explicit environment, (b,f) the reduced master Eq.~\eqref{eq:cme_ab_approx}, and (c,g) the adiabatic approximation; (d) shows the marginal distribution of $n_A+n_B$; panel (h) shows $n_A-n_B$. Markers labelled `SSA' in panel (h) are from the stochastic simulation algorithm (SSA) described in Sec.~\ref{sec:sim_paths}. Parameters are $\alpha_0=0, \alpha_1=1, \beta_0=0, \beta_1=1$ in the upper row, and $\alpha_0=0, \alpha_1=1, \beta_0=1, \beta_1=0$ in the lower row. Remaining parameters are $\Omega=20, \lambda=20, k_0=k_1=1$.}
\label{fig:distrib}
\end{figure*}
\subsection{Positive correlation between the species}\label{sec:pos}
In the case $\Delta\alpha\Delta\beta>0$, the correlations between $n_A$ and $n_B$ are positive. There is one state of the environment which favours both species, i.e., they each have a higher birth rate in this environmental state than in the other. All rates in Eq.~\eqref{eq:cme_ab_approx} are positive (provided $\lambda$ is sufficiently large, so that $\alpha_{\rm eff}, \beta_{\rm eff}\geq 0$), and mathematically there is then a clear and unique way of interpreting this equation as a continuous-time Markov process. The events described by the various terms are then as above: single deaths, single births and bursting reactions in which two particles are produced. The notion of sample paths is well-defined; they can be generated using the standard Gillespie algorithm \cite{gillespie1976general,gillespie1977exact}.  

Some support for the validity of the reduced master equation is given in Fig.~\ref{fig:distrib}, panels (a)--(c).
In panel (a) we show the stationary distribution obtained from numerically integrating the full master equation Eq.~\eqref{eq:mastermaster}, i.e., from the full dynamics of population and environment.
Panel (b) shows the corresponding distribution from numerical integration of the reduced master equation \eqref{eq:cme_ab_approx}.
In panel (c) we have taken the adiabatic limit $\lambda\rightarrow\infty$. In each case the numerical integration is carried out using a Runge--Kutta scheme (RK4). The reduced dynamics capture the correlations between $n_A, n_B$ in the original model; this correlation is no longer seen in the adiabatic approximation. Panel (d) shows the marginal distribution for the quantity $n_A+n_B$ to allow better comparison.

\subsection{Anti-correlations and negative transition rates}
\label{sec:renormalisation}
For cases in which $\Delta\alpha$ and $\Delta\beta$ have opposite signs, the interpretation of Eq.~\eqref{eq:cme_ab_approx} presents an interesting feature.
In this situation the (pseudo-) rate of the last reaction $(\Omega^2 \theta^2/\lambda)\Delta\alpha\Delta\beta$ is negative, irrespective of the value of $\lambda$. The interpretation of this term is then not clear {\it a priori}, and Eq.~\eqref{eq:cme_ab_approx} is not a master equation in the usual sense. We will nevertheless refer to it as the reduced master equation, quotation marks or a prefix `pseudo-' are implied. Similarly, we will continue to speak of rates, even if these are negative.

In order to better understand a master equation with negative rates, we focus on a pair of states, which we label $\ell$ and $\ell'$, and on a single reaction of type $\ell\to\ell'$ occurring with a rate $R_{\ell\to \ell'}$.
In the specific example above one would have $\ell=(n_A,n_B)$ and $\ell'=(n_A+1,n_B+1)$.
The corresponding terms in the master equation are then
\begin{subequations}\begin{align}
\frac{\d}{\d t} \Pi(\ell,t) ={}& -R_{\ell\to \ell'} \Pi(\ell,t), \\
\frac{\d}{\d t} \Pi(\ell',t) ={}& R_{\ell\to \ell'}\Pi(\ell,t).
\end{align}\label{eq:masterplus}\end{subequations}
In conventional cases the rate is positive, $R_{\ell\to \ell'}>0$. The master equation then describes a non-negative probability flow $R_{\ell\to \ell'}\,\Pi(\ell)$ from $\ell$ to $\ell'$ (we suppress the time dependence of $\Pi(\ell)$ for convenience).

For $R_{\ell\to \ell'}<0$, the flow of probability per unit time in Eqs.~\eqref{eq:masterplus} is $|R_{\ell\to \ell'}|\,\Pi(\ell)\geq 0$ from $\ell'$ to $\ell$. This is not a standard Markovian situation: the flow is directed from $\ell'$ to $\ell$, but proportional to the probability already present at $\ell$. Furthermore, the magnitude of this flow does not depend on $\Pi(\ell')$. In making this argument, we have assumed $\Pi(\ell)\geq 0$. This assumption is not always justified in master equations with negative rates. However the above argument holds more generally: a negative value of $R_{\ell\to \ell'}\,\Pi(\ell)$ indicates a positive probability flux $|R_{\ell\to \ell'}\,\Pi(\ell)|$ from $\ell'$ to $\ell$. 
\begin{figure}[t!]
\includegraphics[width=0.85\columnwidth,valign=t]{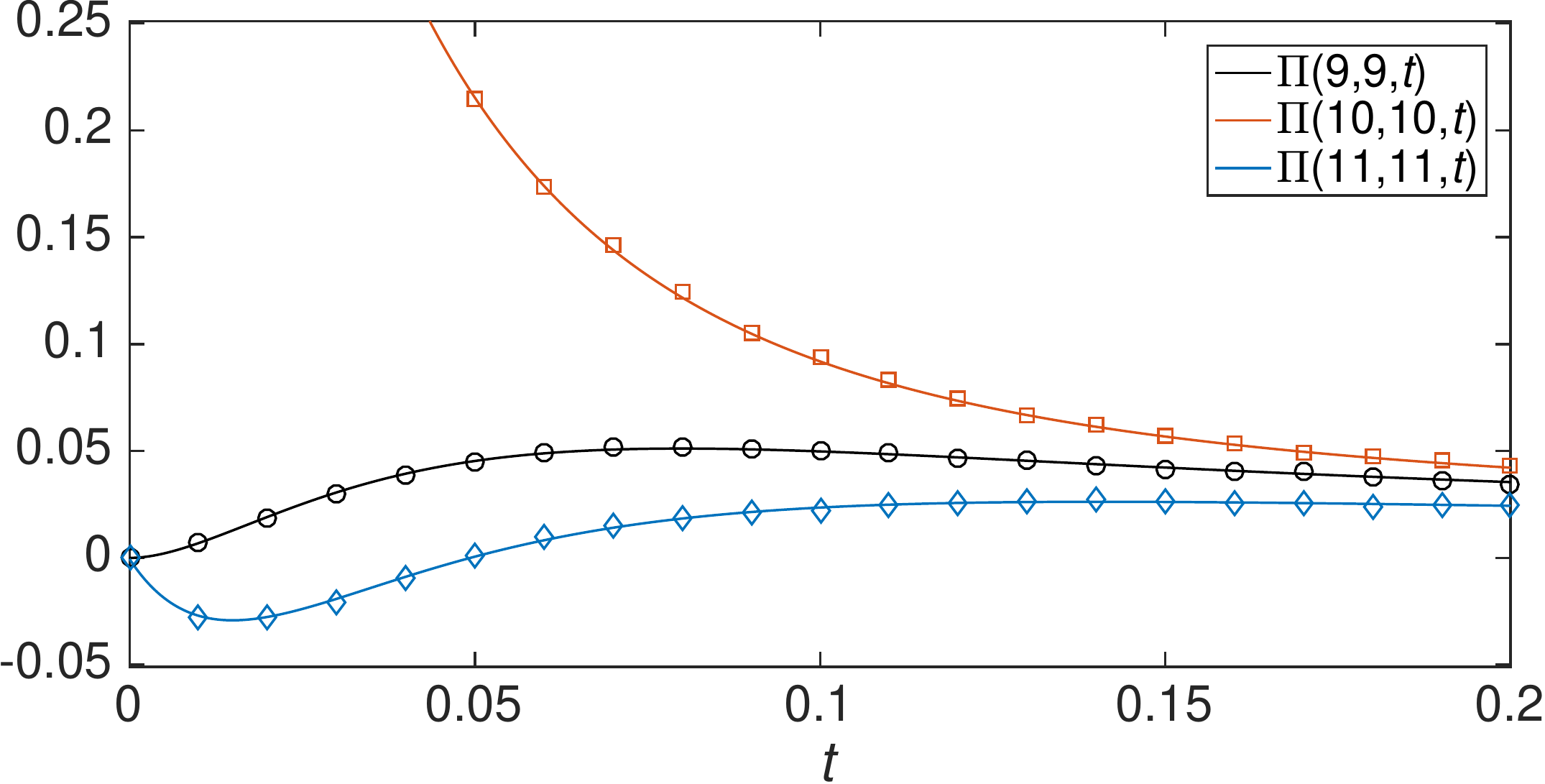}
\caption{Time evolution of several entries $\Pi(n_A,n_B,t)$ for the example defined in Sec.~\ref{sec:twospecies:model}. The solid lines show results from integrating the reduced master equation \eqref{eq:cme_ab_approx}, starting from a delta-distribution $\Pi(n_A,n_B,t=0)=\delta_{n_A,10}\delta_{n_B,10}$. 
Markers are from the numerical simulation scheme described in Sec.~\ref{sec:sim_ensemble}.
Model parameters are $\alpha_0=0, \alpha_1=1, \beta_0=1, \beta_1=0$, $\Omega=20, \lambda=20, k_0=k_1=1$.}
\label{fig:negprob}
\end{figure}
Similar structures with negative rates are found in open quantum systems, and an approach renormalising master equations of this type has been proposed for example in Refs.~\cite{piiloprl,breuer2009stochastic}. We illustrate this using Eqs.~(\ref{eq:masterplus}), assuming again $R_{\ell\to\ell'}<0$. For $\Pi(\ell')>0$ one defines the renormalised transition rate
\begin{align}
T_{\ell'\to \ell}(t)\equiv\frac{\Pi(\ell,t)}{\Pi(\ell',t)}\left|R_{\ell\to \ell'}\right|.
\label{eq:renormalised_rates}
\end{align}
The master equation \eqref{eq:masterplus} can be then written as
\begin{subequations}\begin{align}
\frac{\d}{\d t} \Pi(\ell,t) ={}& T_{\ell'\to \ell}(t)\Pi(\ell',t), \\
\frac{\d}{\d t} \Pi(\ell',t) ={}& -T_{\ell'\to \ell}(t)\Pi(\ell',t).\end{align}\label{eq:master_reversed}
\end{subequations}
Equations~\eqref{eq:master_reversed}, then, resemble a more traditional master equation, and $T_{\ell'\to \ell}$ is the rate for transitions from $\ell'$ to $\ell$. However, this rate depends on the probability distribution $\Pi$, in particular $T_{\ell'\to \ell}$ is a function of $\Pi(\ell)$. This indicates non-Markovian properties \cite{piiloprl,breuer2009stochastic,Breuer2016}.

\subsection{Lack of positivity in initial transients}
\label{sec:lack_of_pos}
\begin{figure}[t!]
\includegraphics[width=0.75\columnwidth,valign=t]{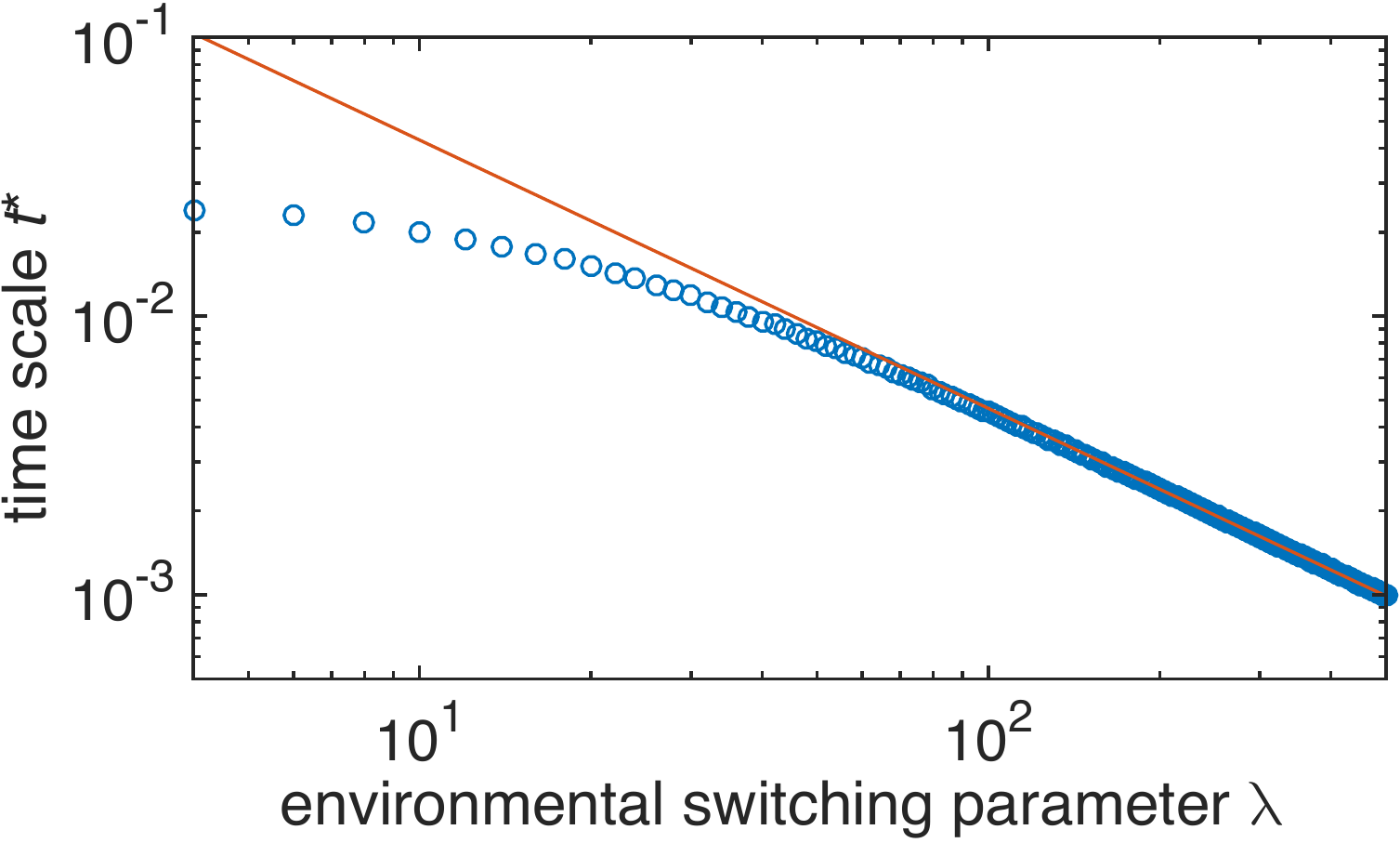}
\caption{Time scale $t^\star$ over which negative probabilities are accumulated.
Specifically, $t^\star$ is the time at which the sum of all negative entries in $\Pi$ has maximal magnitude; data is from numerical integration of Eq.~\eqref{eq:cme_ab_approx}. The solid line is a guide and corresponds to $t^\star\propto 1/\lambda$.
Parameters and initial condition are the same as Fig.~\ref{fig:negprob}.}
\label{fig:tstar}
\end{figure}

Numerically integrating the reduced master equation \eqref{eq:cme_ab_approx}, we find transient regimes of negative (pseudo-) probabilities. For example, if the initial condition is chosen as a delta-peak concentrated on one state $\ell=(n_A,n_B)$, the numerical solution for $\Pi(n_A+1,n_B+1)$ is negative for a limited time as shown in Fig.~\ref{fig:negprob}. We analyse this further in Fig.~\ref{fig:tstar}, where we show the duration $t^\star$ of the initial transient in which negative probabilities are accumulated.
The data suggests that this time window is limited to a duration of order $\lambda^{-1}$.

While we have not attempted to establish formal conditions under which Eq.~\eqref{eq:cme_ab_approx} preserves positivity, we note that negative transients have been observed before in reduced dynamics for open classical and quantum systems \cite{Suarez1992, Pechukas1994, Gnutzmann1996, Benatti2003}. Indeed, it is not surprising that Eq.~\eqref{eq:cme_ab_approx} should become unphysical on short time scales. The typical time between switches of the environmental state is of order $\lambda^{-1}$, and the reduced dynamics were derived by integrating out the fast environmental dynamics.
We cannot expect Eq.~\eqref{eq:cme_ab_approx} to resolve the physics of the problem on time scales shorter than order $\lambda^{-1}$, as then the detailed mechanics of the environment become important. 

We have verified that the appearance of transient negative solutions can be cured by first integrating the full master equation describing the population and the environment for a short period of time, and then subsequently changing to the reduced master equation (\ref{eq:cme_ab_approx}). Alternatively, the reduced dynamics can be started from `slipped' initial conditions \cite{Suarez1992,Gnutzmann1996}.

Focusing on long times, we find that the stationary distribution obtained from numerical integration of Eq.~\eqref{eq:cme_ab_approx} for $\Delta\alpha\Delta\beta<0$ captures the negative correlation of $n_A$ and $n_B$ in the original dynamics. This can be seen in Fig.~\ref{fig:distrib}(e) and (f). Working in the adiabatic limit, however, produces significant deviations [panels (g) and (h)].

\section{Numerical approaches to a master equation with negative rates}
\subsection{Distribution-level simulation}
\label{sec:sim_ensemble}
 The time-dependent solution $\Pi(\ell,t)$ can be obtained by direct numerical integration of the reduced master equation, for example using a Runge--Kutta scheme. However for large state spaces this approach can become slow. The technique described in this Section can, in some cases, provide a faster alternative. 

We consider a large number $M$ of discrete units of probability, $1/M$. At each point in time the state of the simulation is defined by the `occupation numbers' $N_\ell$ for all states $\ell$; some of the $N_\ell$ may be negative. One has $\sum_\ell N_{\ell}=M$. The algorithms proceeds along the following steps:
\begin{enumerate}[nosep]
\item[1.] For given occupation numbers $N_\ell$ at time $t$, make a list of all possible reactions, labelled by index $\gamma$. Each reaction has a site of origin, $\ell_{\gamma}$, a destination site, $\ell'_{\gamma}$, and rate $r_\gamma=R_{\ell_{\gamma}\to\ell'_{\gamma,}} N_{\ell_{\gamma}}$. Some of the $r_\gamma$ may be negative.
\item[2.] Draw a random number $\tau$ from an exponential distribution with parameter $\sum_\gamma |r_\gamma|$.
\item[3.] Pick a reaction from the list created in 1. The probability to pick $\gamma$ is $|r_\gamma|/\sum_{\gamma'} |r_{\gamma'}|$.
\item[4.] If $r_\gamma>0$ increase $N_{\ell'_{\gamma}}$ by one and reduce $N_{\ell_{\gamma}}$ by one. If $r_\gamma<0$ reduce $N_{\ell'_{\gamma}}$ by one and increase $N_{\ell_{\gamma}}$ by one.
\item[5.] Increment time by $\tau$, and go to 1.
\end{enumerate}
The process in step $4$ allows occupation numbers to go negative. The typical time step of this scheme is given by $1/\sum_\gamma |r_\gamma|$, and reaction $\gamma$ is triggered with probability $|r_\gamma|/(\sum_{\gamma'} |r_{\gamma'}|)$. Thus $|r_\gamma|$ reactions of type $\gamma$ are triggered per unit time. The sign convention in step 4 ensures correct sampling of the reduced master equation.

We tested this procedure for the example given by Eq.~\eqref{eq:cme_ab_approx}. Results are shown in
Fig.~\ref{fig:negprob}; there is near perfect agreement between the Monte Carlo procedure and direct numerical integration of the reduced master equation.

We stress that this algorithm does not generate sample paths for the reduced master equation. This motivates us to ask whether the notation of a sample path is valid for a master equation involving negative transition rates.

\subsection{`Path-level' simulation}
\label{sec:sim_paths}

As discussed in Sec.~\ref{sec:pos} the reduced master equation (\ref{eq:cme_ab_approx}) defines a Markovian process for $\Delta\alpha\Delta\beta>0$. All rates in the reduced master equation are non-negative, and sample paths can be simulated using the conventional Gillespie method \cite{gillespie1976general,gillespie1977exact}. The solution of Eq.~(\ref{eq:cme_ab_approx}) can be recovered from the statistics of a large ensemble of such realisations.

In Sec.~\ref {sec:lack_of_pos} we have seen that reduced master equations with negative rates can---for certain initial conditions---lead to negative transient solutions. These can be delivered by the ensemble-level algorithm in Sec.~\ref{sec:sim_ensemble}. A simulation generating sample paths cannot capture these negative (pseudo-) probabilities.

However, this does not preclude a meaningful notion of sample paths in situations where the reduced master equation is started from an initial distribution which avoids subsequent negative transients. For example, one could focus on the stationary state of Eq.~(\ref{eq:cme_ab_approx}). 

A stochastic simulation algorithm was discussed in Ref.~\cite{piiloprl} for non-Markovian jumps in quantum systems. This method simulates processes defined by quantum master equations with temporarily negative decay rates. Realisations are generated by combining non-Markovian quantum jumps with the deterministic evolution of quantum states between jumps \cite{breuer2009stochastic}. The central idea is to represent the solution of the master equation by an ensemble of sample paths, which are generated {\em in parallel}. In contrast with standard methods \cite{gillespie1976general,gillespie1977exact} these paths are correlated with each other. 

In order to test the principles of this approach we have adapted it to the case of the classical master equation
\begin{equation}
\frac{\d}{\d t}\Pi(\ell,t)=\sum_{\ell'} R_{\ell'\to\ell} \Pi(\ell',t),
\label{eq:master_general}
\end{equation}
where some of the rates $R_{\ell'\to\ell}$ may be negative. The algorithm uses Eqs.~\eqref{eq:renormalised_rates} and~\eqref{eq:master_reversed} to convert reactions with negative rates into reactions in the opposite direction, and with positive renormalised rates. In order to do this we need the entries of the probability distribution, $\Pi(\ell)$ and $\Pi(\ell')$, see Eq.~\eqref{eq:renormalised_rates}. These in turn are estimated from the ensemble of sample paths. In this way, the trajectories are correlated with each other, because the evolution of a single sample path depends on the ensemble \cite{piiloprl,breuer2009stochastic}. 

We index each trajectory individually, so that we can follow the time evolution of each sample path. At each point in time the ensemble is specified by the state of each of the sample paths. We write $N_\ell$ for the number of sample paths in state $\ell$. To keep the notation compact we suppress the time dependence of $N_\ell$. One has $\sum_\ell N_\ell=M$ at all times, where $M$ is the size of the ensemble.

Before we detail the algorithm we describe the construction of a matrix $\mathbf{S}$ with elements $S_{\ell\to\ell'}$ which give the rate of a reaction $\ell\to\ell'$ to occur {\em in the ensemble}. 
The matrix is needed frequently in the algorithm, and is constructed as follows:
(i) start with $S_{\ell\to\ell'}=0$ for all $\ell, \ell'$;
(ii) for all reactions $\ell \to \ell'$ with positive rate $R_{\ell\to\ell'}$ increase $S_{\ell\to\ell'}$ by $R_{\ell\to\ell'}$;
(iii) for reactions with negative rate $R_{\ell\to\ell'}$ and $N_{\ell'}>0$ construct $T_{\ell'\to\ell}$ as in Eq.~\eqref{eq:renormalised_rates}, where $N_{\ell}/N_{\ell'}$ is used as a proxy for $\Pi(\ell)/\Pi(\ell')$. If $N_{\ell'}=0$ set $T_{\ell'\to\ell}=0$. Increase $S_{\ell'\to\ell}$ by $T_{\ell'\to\ell}$;
(iv) finally, for all pairs $\ell, \ell'$ multiply $S_{\ell\to\ell'}$ by $N_\ell$.
For a given master equation (i.e., a given matrix $\mathbf{R}$) the matrix $\mathbf{S}$ is a function of the current state of the ensemble, i.e., of the $\{N_\ell\}$. All entries $S_{\ell\to\ell'}$ ($\ell\neq\ell')$ are non-negative. The diagonal elements are zero. The element $S_{\ell\to\ell'}$ indicates the rate for a reaction $\ell\to\ell'$ to occur, given the current state of the ensemble. One has $S_{\ell\to\ell'}=0$ if no sample path in the ensemble is in state $\ell$. We also note that the total rate for a reaction of any type to happen, $\sum_{\ell\neq\ell'} S_{\ell\to\ell'}$, scales linearly with $M$. This guarantees that each time step in the procedure below is of order $M^{-1}$, or in other words, that order $M$ reactions occur per unit time.

The algorithm proceeds as follows:
\begin{enumerate}[nosep]
\item[1.] Given the current state of the ensemble compute the matrix $\mathbf{S}$ as described above.
\item[2.] Draw a random time increment $\tau$ from an exponential distribution with parameter $s=\sum_{\ell,\ell'} S_{\ell\to\ell'}$.
\item[3.] Randomly select an origin $\ell$ and a destination $\ell'$ with a probability weighted by $S_{\ell\to\ell'}$ (i.e., the probability that $\ell$ is picked as an origin and $\ell'$ as a destination is $S_{\ell\to\ell'}/s$).
\item[4.] Randomly (with equal probabilities) pick one of the sample paths currently in state $\ell$ and change its state to $\ell'$.
\item[5.] Increment time by $\tau$ and go to step 1.
\end{enumerate}

We note that this algorithm does not allow for any state $\ell$ to ever have a negative occupancy $N_\ell$. Furthermore, if all $R_{\ell\to\ell'}$ are non-negative the simulation reduces to the standard Gillespie algorithm \cite{gillespie1976general,gillespie1977exact}. In this case the sample paths remain uncorrelated from each other.
\begin{figure*}[t!]
\includegraphics[width=1.0\textwidth,valign=t]{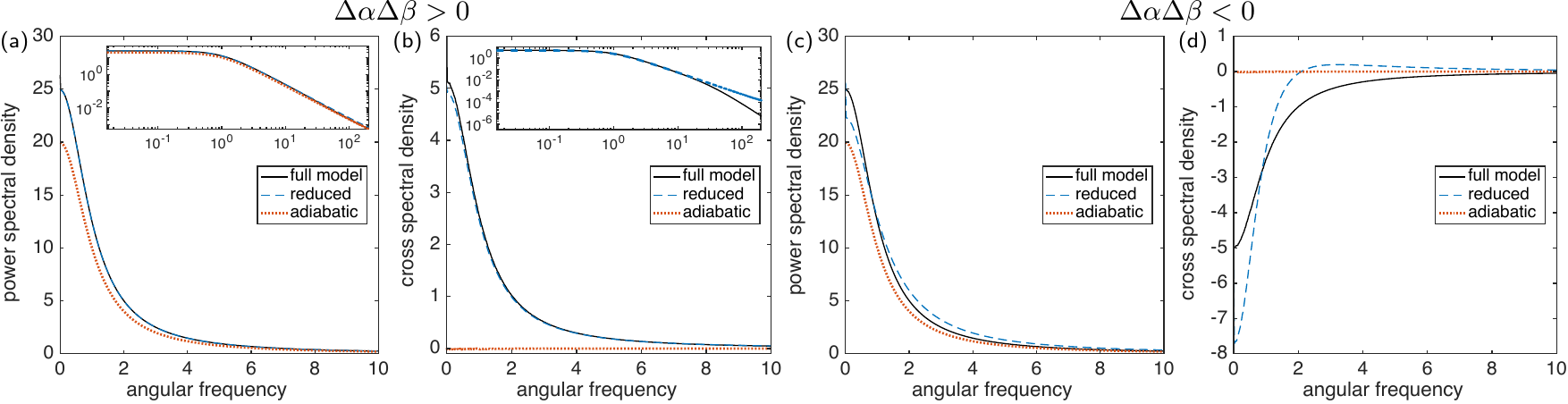}
\caption{
Spectra of fluctuations for the model defined in Sec.~\ref{sec:twospecies:model}. Panels (a) and (b) are for $\Delta\alpha\Delta\beta>0$; (c) and (d) for $\Delta\alpha\Delta\beta<0$. We show the power spectral density $S_{AA}(\omega)$ in (a) and (c), and the cross spectral density $S_{AB}(\omega)$ in (b) and (d); the insets show the same quantities on a logarithmic scale. Parameters: $\alpha_0=0, \alpha_1=1$, $\lambda k_0 = \lambda k_1 = 20, \Omega=20$ in all panels; $\beta_0=0, \beta_1=1$ in (a,b); $\beta_0=1, \beta_1=0$ in (c,d). }
\label{fig:corspec}
\end{figure*}

To test the algorithm we use the example in Eq.~\eqref{eq:cme_ab_approx}. The algorithm captures the stationary distribution accurately, as illustrated by the markers in Fig.~\ref{fig:distrib} (h). Next, we test whether the simulation reproduces dynamical properties of the sample paths of the full model.

To this end, we define the power spectral density $S_{AA}(\omega)=\avg{|\hat n_A(\omega)|^2}$, where $\hat n_A(\omega)$ is the Fourier transform of the random process $n_A(t)$. Similarly, we also look at the cross power spectral density $S_{AB}(\omega)=\avg{n_A^\dagger(\omega)n_B(\omega)}$ (the superscript ${}^\dagger$ denotes complex conjugation). These are the Fourier transforms of the autocorrelation and cross-correlation functions respectively. Figure~\ref{fig:corspec} shows these quantities, measured in the regime when $\Pi(n_A,n_B)$ has reached the stationary state, and averaged over a large ensemble of trajectories.

Panels (a) and (b) serve as a benchmark, and show the case $\Delta\alpha\Delta\beta>0$ when all rates in the reduced master equation are positive. Thus the above simulation scheme reduces to the standard Gillespie method. As seen in the figure the power and cross spectra $S_{AA}(\omega)$ and $S_{AB}(\omega)$ obtained from simulating paths of the reduced master equation agree well with those from simulations of the full model, at least at sufficiently low frequencies $\omega$. At larger frequencies deviations are seen, this is particularly visible for the cross spectrum; see the inset of panel (b). These deviations between reduced and the full model are not surprising; the reduced model does not resolve the mechanics of the environment on short time scales. Spectra obtained from sample paths of the master equation in the adiabatic limit show significant deviations from those of the full model; we note in particular that the cross spectrum $S_{AB}(\omega)$ vanishes [dotted red line in Fig.~\ref{fig:corspec} (b)]; see also Appendix~\ref{app:xspec} for further analysis. 

Results for the case with negative rates in the reduced master equation are shown in panels (c) and (d) of Fig.~\ref{fig:corspec}.
We find marked differences between the spectra generated from the reduced master equation with the above algorithm and those of sample paths of the full model. This is particularly noticeable in the cross spectrum in panel (d). Further details can also be found in Appendix \ref{app:saa}. 

We conclude that the trajectories generated by the above simulation algorithm do not represent sample paths of the full model when the reduced master equation contains negative rates. Our findings invite the question whether algorithms of this type \cite{piiloprl,breuer2009stochastic} provide a faithful representation of the full dynamics of open quantum systems and their environment. It would be interesting to compare the structure of the reduced dynamics in the classical and quantum cases, and to relate our observations to the quantum regression theorem \cite{guarnieri}.

\section{An intuition to our expansion on the level of sample paths}
\label{sec:paths}
In Sec.~\ref{sec:genform} we derived the general formalism for approximating the dynamics of a system coupled to a fast-switching environment. We found this resulted in an effective reduced master equation, which can describe `bursting' events not present in the dynamics of the original model. From a physical perspective, however, it is not obvious how such bursting events can arise as a consequence of the coupling to a fast environment when such events do not occur in any (fixed) state of the environment.
In this section we look at this problem from viewpoint of single trajectories of the full model in discrete time, in order to provide intuition to this result. Taking this view also allows us to develop a method to use the reduced dynamics to approximate sample paths of the full model in discrete time; using a time step larger than $\lambda^{-1}$ we avoid the issues highlighted in the previous section.

\subsection{Effective time-averaged reaction rates}
We focus again on the two-species example given in Sec.~\ref{sec:twospecies}.
An interpretation of the terms in Eq.~\eqref{eq:cme_ab_approx} can be obtained by looking at one sample path of the full model (population and environment) for a time interval $I\equiv[t_0,t_0+\Delta t]$.
We focus on the birth reactions.
If the production rate $\Omega \alpha$ of particles of type $A$ were constant in time, the number of birth events in the interval would be a Poissonian random variable with parameter $\Omega \alpha\Delta t$, and similarly for particles of type $B$ (see also Ref.~\cite{tauleaping}).
In the present model, the production rates are not constant in time as they depend on the state of the environment.
For a given trajectory of the environment we introduce the quantity
\begin{equation}\label{eq:ovlalpha}
\overline \alpha = \frac{1}{\Delta t}\int_{t_0}^{t_0+\Delta t} \d t' ~ \alpha_{\sigma(t')},
\end{equation}
and a similar definition for $\overline{\beta}$; the quantities $\Omega\overline\alpha$ and $\Omega\overline\beta$ are time-averaged production rates in the time interval $I$.

The number of production events of particles of type $A$ in $I$ can then be expected to be Poissonian with parameter $\Omega \overline{\alpha}\Delta t$, and similarly for $B$. We note that $\overline\alpha$ and $\overline\beta$ are random variables when $\Delta t$ is finite, as they depend on the random path of the environment, $\sigma(t'), ~ t'\in I$.
The quantities $\overline{\alpha}$ and $\overline{\beta}$ will in general be correlated, as they derive from the same realisation of the environment.
The main principle of the calculation that follows is to approximate $\overline{\alpha}$ and $\overline{\beta}$ as correlated Gaussian random variables, while capturing their first and second moments.
This Gaussian approximation is justified provided that there is a large number of switches of the environment during the time interval $I$, i.e., when $\lambda\Delta t\gg 1$.

\subsection{Averaging out the environmental process}
Correlations of the environmental process decay on time scales proportional to $\lambda^{-1}$. This means that the environment is in its stationary distribution, except for a short period of order $\lambda^{-1}$ at the beginning. For $\lambda\Delta t\gg 1$ this period constitutes a negligibly small fraction of the time interval, and the distribution of $\sigma(t')$ can hence be assumed to be the stationary one at all times $t'$ during the interval. Writing $\avg{\dots}$ for averages over the environmental process we have $\avg{\overline\alpha}=\alpha_\text{avg}$ and $\avg{\overline{\beta}}=\beta_\text{avg}$. 

Moving on to the second moments we find
\begin{align}\begin{split}
\avg{\overline \alpha^2} ={}&(\Delta t)^{-2}\int_{I} \int_{I} \d t\,\d t' \avg{\alpha_{\sigma(t)}\alpha_{\sigma(t')}}, \\
={}&(\Delta t)^{-2}\sum_{\sigma\sigma'} \alpha_\sigma \alpha_{\sigma'}\ \\
& \times \int_I \int_I \d t ~\d t' \rho[\sigma,\min(t,t')] ~ \rho\left(\sigma',\left|t-t'\right|\big|\sigma\right),\label{eq:aux1}
\end{split}\end{align}
where $\rho[\sigma,\min(t,t')]$ is the probability distribution of $\sigma$ at the earlier of the two times $t$ and $t'$. It is given by the stationary distribution of the environment, $\rho[\sigma,\min(t,t')]=\rho_\sigma^\star$, with $\rho_\sigma^\star$ as in Eq.~\eqref{eq:rhostat}. The notation $\rho(\sigma',\tau|\sigma)$ in Eq.~(\ref{eq:aux1}) indicates the probability of finding the environment in state $\sigma'$ if $\tau$ units of time earlier it was in state $\sigma$ ($\tau>0$). These can be obtained straightforwardly from the asymmetric telegraph process for the environment, $ \rho(0,\tau|0)=\rho_0^\star\left[1+\frac{k_1}{k_0}e^{-\lambda(k_0+k_1) \tau} \right]$, and $\rho(0,\tau|1)=\rho_0^\star\left[1-e^{-\lambda(k_0+k_1) \tau}\right]$. Using this in Eq.~(\ref{eq:aux1}) we find
\begin{align}
\avg{\overline\alpha^2} ={}& \alpha_\text{avg}^2+\bigg[ \frac{2}{\lambda (k_0+k_1) \Delta t}
+\frac{2}{\lambda^2(k_0+k_1)^2 \Delta t^2} \nonumber \\
&\times \left(e^{-\lambda (k_0+k_1) \Delta t}-1\right)\bigg]\frac{k_0k_1}{(k_0+k_1)^2}(\alpha_0-\alpha_1)^2.
\end{align}
For $\lambda\Delta t \gg1$ the first term in the square bracket dominates relative to the second, so we can approximate
\begin{subequations}\begin{equation}
\avg{\overline \alpha^2}-\alpha_\text{avg}^2 \approx \frac{\theta^2}{\lambda\Delta t}(\Delta\alpha)^2,
\end{equation}
with $\theta^2=2k_0k_1/(k_0+k_1)^3$ as before [see Eq.~\eqref{eq:varsigma}]. Following similar steps one finds
\begin{align}
\avg{\overline \beta^2}-\beta_\text{avg}^2 \approx \frac{\theta^2}{\lambda \Delta t}(\Delta\beta)^2,\ \\
\avg{\overline \alpha\overline\beta}-\alpha_\text{avg}\beta_\text{avg} \approx \frac{\theta^2}{\lambda \Delta t}\Delta\alpha\Delta\beta.
\end{align}\label{eq:aa_etc}\end{subequations}
We therefore approximate the joint probability distribution of $\bar\alpha$ and $\bar\beta$ in the fast switching limit as a bivariate normal distribution with these parameters. 
\subsection{Resulting event statistics}
The probability that exactly $m_A$ production events for species $A$ occur during the time interval $\Delta t$, and $m_B$ for species $B$, is given by \begin{equation}
{\rm P}(m_A,m_B)=\left<e^{-\Delta t\Omega\left(\overline \alpha+\overline\beta\right)}\frac{(\Delta t \Omega \overline\alpha)^{m_A}}{m_A!}\frac{(\Delta t \Omega\overline\beta)^{m_B}}{m_B!}\right>_{\overline\alpha,\overline\beta},
\end{equation}
resulting from Poissonian statistics for given $\overline\alpha, \overline\beta$, subsequently averaged over the Gaussian distribution for $\overline\alpha$ and $\overline\beta$ (this average is indicated as $\avg{\dots}_{\overline\alpha,\overline\beta}$).
Expanding in powers of $\Delta t$, and carrying out the Gaussian average we find
\begin{align}\begin{split}
{\rm P}(m_A\!=\!1,m_B\!=\!0)={}&\Delta t \Omega\left[\alpha_\text{avg}-\tfrac{\Omega\theta^2}{\lambda}(\Delta\alpha)^2-\tfrac{\Omega\theta^2}{\lambda}\Delta\alpha\Delta\beta\right]\\
&-\Delta t^2\Omega^2(\alpha_\text{avg}^2+\beta_\text{avg}^2),\\
{\rm P}(m_A\!=\!0,m_B\!=\!1)={}&\Delta t \Omega \left[\beta_\text{avg}-\tfrac{\Omega\theta^2}{\lambda}(\Delta\beta)^2-\tfrac{\Omega\theta^2}{\lambda}\Delta\alpha\Delta\beta\right]\\
&-\Delta t^2\Omega^2(\alpha_\text{avg}^2+\beta_\text{avg}^2),\\
{\rm P}(m_A\!=\!2,m_B\!=\!0)={}&\tfrac{1}{2}\Delta t \tfrac{\Omega^2\theta^2}{\lambda} (\Delta\alpha)^2+\tfrac{1}{2}\Delta t^2 \Omega^2 \alpha_\text{avg}^2,\\
{\rm P}(m_A\!=\!0,m_B\!=\!2)={}&\tfrac{1}{2}\Delta t \tfrac{\Omega\theta^2}{\lambda} (\Delta\beta)^2+\tfrac{1}{2}\Delta t^2 \Omega^2 \beta_\text{avg}^2,\\
{\rm P}(m_A\!=\!1,m_B\!=\!1)={}&\tfrac{\Omega^2\theta^2}{\lambda} \Delta t \Delta\alpha\Delta\beta+\Delta t^2 \Omega^2 \alpha_\text{avg}\beta_\text{avg},
\label{eq:mm1}
\end{split}\end{align}
where we have ignored higher-order terms (those which go like $\Delta t^3$ or $\Delta t^2/\lambda$). Larger numbers of production events ($m_A+m_B\geq 3$) do not contribute at this order.

It is tempting to consider the limit of infinitesimally small $\Delta t$, and to use the first-order terms in $\Delta t$ in Eq.~(\ref{eq:mm1}) to construct reaction rates. If one does so, one recovers the rates exactly as they appear in the reduced master equation \eqref{eq:cme_ab_approx}; for example one would infer a rate of $\frac{1}{2}(\Omega^2\theta^2/\lambda) (\Delta\alpha)^2$ for events in which two particles of type $A$ are produced and none of type $B$ ($m_A=2, m_B=0$). The rate of an event in which one $A$ and one $B$ are produced simultaneously would be $(\Omega^2\theta^2/\lambda) \Delta\alpha\Delta\beta$, which is negative if $\Delta\alpha\Delta\beta<0$. 

\begin{figure}[t!]
\includegraphics[width=1\columnwidth,valign=t]{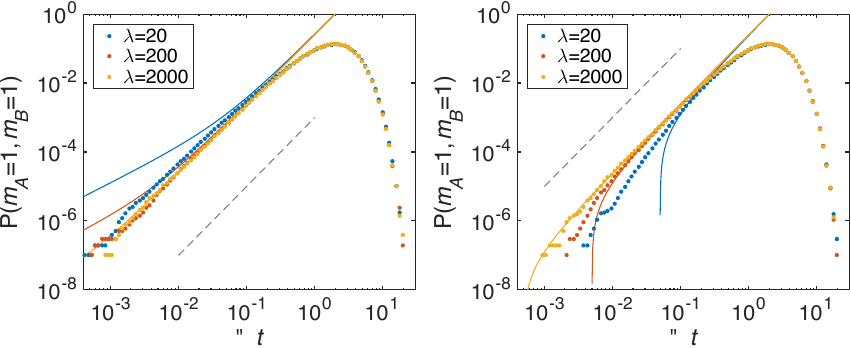} 
\caption{Probability of seeing $m_A\!=\!1, m_B\!=\!1$ in a given time interval of duration $\Delta t$.
Circles show the results of simulation of the full model (population and environment); full lines show Eq.~\eqref{eq:mm1}. The dashed line shows a slope of $2$ for comparison. Data is shown for different values of $\lambda$, all other model parameters are as in the earlier figures. Left: $\Delta\alpha\Delta\beta>0$, right: $\Delta\alpha\Delta\beta<0$.}
\label{fig:birth_rates}
\end{figure}

However, taking the limit $\Delta t\to 0$ at fixed $\lambda$ is not compatible with the assumption that a large number of environmental switching events occur in a given time-step, i.e., $\lambda\Delta t\gg1$. To illustrate this further we carried out simulations of the full model of population and environment, and measured how many birth events of either particle type occur in a given time interval $\Delta t$. Specifically we focus on the probability $P(m_A=1,m_B=1)$ of seeing exactly one birth event of type $A$ and one birth event of type $B$ during such a time interval; note that in the full model these births occur in two separate events. The lines in Fig.~\ref{fig:birth_rates} show the predictions of Eqs.~\eqref{eq:mm1}, results from simulations of the full model are shown as markers. We first notice that simulations deviate from the results of Eqs.~\eqref{eq:mm1} at large values of $\Delta t$. This is to be expected as Eqs.~(\ref{eq:mm1}) are derived neglecting higher-order terms in $\Delta t$. Simulations and the above expressions agree to good accuracy at intermediate values of the time step; we write $\Delta t_*$ for the lower end of this range, and $\Delta t^*$ for the upper end. As seen in Fig.~\ref{fig:birth_rates}, the lower threshold $\Delta t_*$ decreases as the switching of the environment becomes faster (i.e., $\lambda$ is increased). The reduction of the threshold is in-line with the requirement $\lambda\Delta t \gg 1$ for the theoretical analysis above. As seen in the figure the results of Eqs.~\eqref{eq:mm1} are largely determined by the term of order $\Delta t^2$ when they match simulations of the full model (the slope of the simulation data in the log-log plot of Fig.~\ref{fig:birth_rates} is then approximately two). This term, $\Delta t^2 \Omega^2 \alpha_\text{avg}\beta_\text{avg}$, is positive, irrespective of the sign of $\Delta\alpha\Delta\beta$. At low values of $\Delta t\lesssim \Delta t_*$, we observe systematic deviations between simulations of the full model and the expressions in Eqs.~\eqref{eq:mm1}. For the case $\Delta\alpha\Delta\beta<0$ it is obvious that this must occur: at small $\Delta t$, Eqs.~\eqref{eq:mm1} predict ${\rm P}(m_A=1,m_B=1)\approx(\Omega^2\theta^2/\lambda)\Delta t \Delta\alpha\Delta\beta<0$, whereas $P(m_A=1,m_B=1)$ is non-negative in simulations by definition. Deviations at small time steps are also seen in the left-hand panel of Fig.~\ref{fig:birth_rates}, the expression in Eqs.~\eqref{eq:mm1} shows a cross-over to linear scaling in $\Delta t$, whereas simulation results scale approximately as $\Delta t^2$.

\subsection{Simulation procedure for discrete-time sample paths}\label{sec:tauleap}
The analysis of the previous section is based on a discretisation of time into intervals of length $\Delta t$. In the limit of fast switching of the environment it then assumes that the time-averaged birth rates $\Omega\overline\alpha$ and $\Omega\overline\beta$ are Gaussian random variables with statistics given in Eqs. (\ref{eq:aa_etc}). We will now use this interpretation to define an algorithm with which to approximate sample paths of the full model in discrete time. We note that $\overline\alpha$ and $\overline\beta$ can take negative values in this Gaussian approximation. This issue arises irrespective of the sign of $\Delta\alpha\Delta\beta$ and is separate from the problem of negative rates in the reduced master equation. The probability for $\overline\alpha$ and/or $\overline{\beta}$ to be negative is exponentially suppressed in $\lambda\Delta t$, as the mean of the Gaussian distribution, $(\alpha_{\rm avg}, \beta_{\rm avg})$, does not depend on $\lambda$ or $\Delta t$, and the covariance matrix is of order $(\lambda\Delta t)^{-1}$ [Eq.~(\ref{eq:aa_etc})].  As the switching of the environment becomes faster the distributions of $\overline\alpha$ and $\overline\beta$ become increasingly peaked around their mean. For the purposes of the numerical scheme we truncate the distribution at zero.

The algorithm uses ideas from the $\tau$-leaping variant of the Gillespie algorithm \cite{tauleaping}, and proceeds as follows:

\begin{enumerate}[nosep]
\item[1.] Assume the simulation has reached time $t$ and that the current particle numbers are $n_A$ and $n_B$. Draw correlated Gaussian random numbers $\overline\alpha$ and $\overline\beta$, from a distribution with $\avg{\overline\alpha}=\alpha_\text{avg}$, and $\avg{\overline\beta}=\beta_\text{avg}$, and with second moments as in Eqs.~\eqref{eq:aa_etc}. If $\overline\alpha<0$ set $\overline\alpha=0$ and similar for $\overline\beta$.

\item[2.] Using the $\overline\alpha$ and $\overline\beta$ just generated, draw independent integer random numbers $m_A$ and $m_B$ from Poissonian distributions with parameters $\Omega\overline \alpha\Delta t$ and $\Omega\overline\beta\Delta t$, respectively. 
\item[3.] For the death processes draw Poissonian random variables $m_A'$ and $m_B'$ from Poissonian distributions with parameters $\gamma n_A \Delta t$ and $\delta n_B\Delta t$ respectively.
\item[4.] Update the particle numbers to $n_A+m_A-m_A'$ and $n_B+m_B-m_B'$, respectively (if this results in $n_A<0$ set $n_A=0$, and similar for $n_B$).
\item[5.] Increment time by $\Delta t$ and go to 1.
\end{enumerate}

We have introduced a cutoff procedure in step 4 of the algorithm, in order to prevent particle numbers from going negative. This is necessary due to the discrete-time nature of the process, and well-known in the context of $\tau$-leaping \cite{tauleaping}. In particular this is not related to the appearance of negative rates in the reduced master equation, and applies in the case $\Delta\alpha\Delta\beta>0$ as well. 
\begin{figure}[t!]
\includegraphics[width=0.85\columnwidth,valign=t]{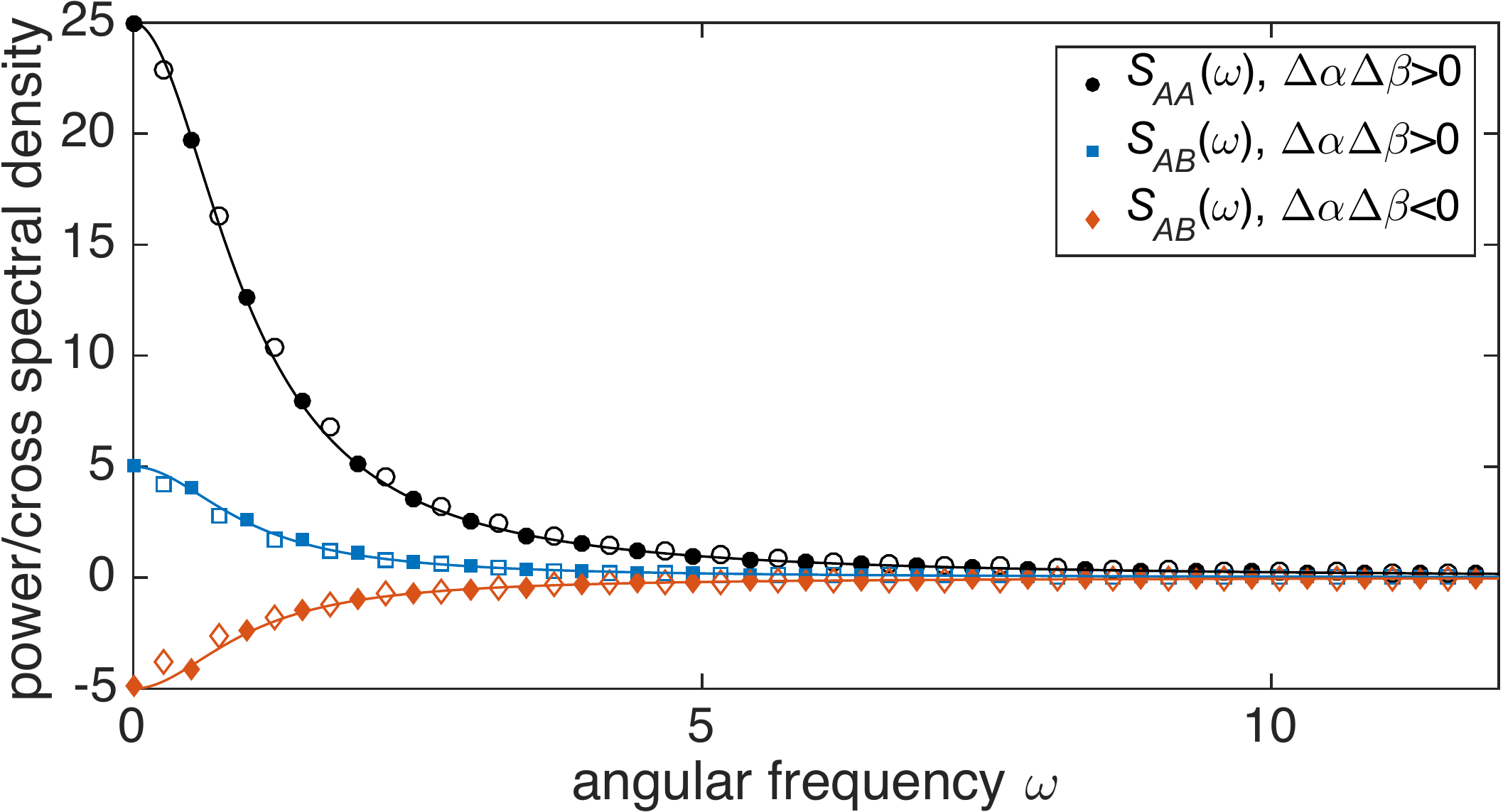} 
\caption{Spectra of fluctuations from direct simulations of the full model (filled symbols), and using the discrete-time algorithm in Sec.~\ref{sec:tauleap} (open symbols). The model is the same as in previous figures ($k_0=k_1=1, \Omega=20, \lambda=20, \alpha_0=0, \alpha_1=1$, $\Delta t=0.1$). Solid lines show the power spectrum/cross spectrum obtained from the linear-noise approximation of the reduced dynamics, Eq.~\eqref{eq:spectoy}.}
\label{fig:tauleap}
\end{figure}

We have carried out simulations using this algorithm for both cases $\Delta\alpha\Delta\beta>0$ and $\Delta\alpha\Delta\beta<0$.
As shown in Fig.~\ref{fig:tauleap} the resulting spectra of fluctuations are in agreement with those of the full model, at least to reasonable approximation.
In particular the cross spectrum $S_{AB}(\omega)$ comes out negative in the anti-correlated case. We attribute remaining discrepancies to the discretisation of time and the assumptions of Gaussian effective birth rates.

 It is important to stress that agreement with the full model requires a careful choice of the time step $\Delta t$.
On the one hand, one needs $\Delta t \gtrsim 1/\lambda$, otherwise it is not justified to replace $\overline\alpha$ and $\overline\beta$ by Gaussian random variables.
On the other hand, the so-called `leap condition' for $\tau$-leaping must be fulfilled \cite{tauleaping}, that is, the time step $\Delta t$ must not be long enough for the population to change significantly in one step. More precisely the changes in particle numbers must remain of order $\Omega^0$ in each step.

\section{Expansion in system size}\label{sec:km}
In the previous Sections we started from a microscopic process in a population of discrete individuals, subject to a randomly switching environment. We then carried out an expansion in the limit of fast environmental switching. We discussed different levels of coarse graining: the switching of the environment was either kept in its original form (full model), treated as fast but not infinitely so (reduced master equation), or the adiabatic limit of infinitely fast switching was taken (master equation with effective, average rates). So far we have considered discrete populations; its intrinsic stochastic dynamics, due to production and removal events, were not approximated.

Another approximation method for Markov jump processes with small jump sizes involves carrying out an asymptotic expansion in powers of the inverse population size. In the context of large populations, and without the complication of environmental switching, this typically is achieved by performing either the Kramers--Moyal expansion or van Kampen's system-size expansion \cite{gardiner1985handbook,van1992stochastic}. These techniques are commonly used in a number of applications of population dynamics; they have recently been extended to the case of jump processes in switching environments \cite{mao2006stochastic,potoyan2015dichotomous,hufton2016intrinsic}.
Following such an expansion, the state of the population is continuous and, for a fixed environmental state, described by a stochastic or ordinary differential equation. Alternative approaches, based on the WKB method, have been pursued for example in Refs.~\cite{kamenev2008, assaf2008population,assaf1,assaf2,assaf2013extrinsic,assaf3}.

The purpose of this Section is to combine Kramers--Moyal-type expansions with an expansion in the time scale separation between environment and population. This leads to different levels of description depending on how the environmental switching and the discreteness and intrinsic stochasticity of the population are treated.
Studying these different levels of approximation is also useful to put our results of the previous Sections into the context with existing work \cite{mao2006stochastic,friedman,potoyan2015dichotomous,hufton2016intrinsic, davis1984piecewise,zeiser2008simulation,ge2015stochastic,jia2017simplification,jia2017emergent,herbach2017inferring,LinHufton2018,Lin2018, Newby2010,Bressloff2014,Bressloff2017,Bressloff2017a,segel1989quasi,haseltine2002approximate,rao2003stochastic,goutsias2005quasiequilibrium,qian2009stochastic,pahlajani2011stochastic,kim2014,duncan2015noise}. We will first give a general overview, and then consider a specific example.

\subsection{Overview}
\begin{figure}[t!]
{\centering
\begin{overpic}[width=0.49\textwidth,valign=t]{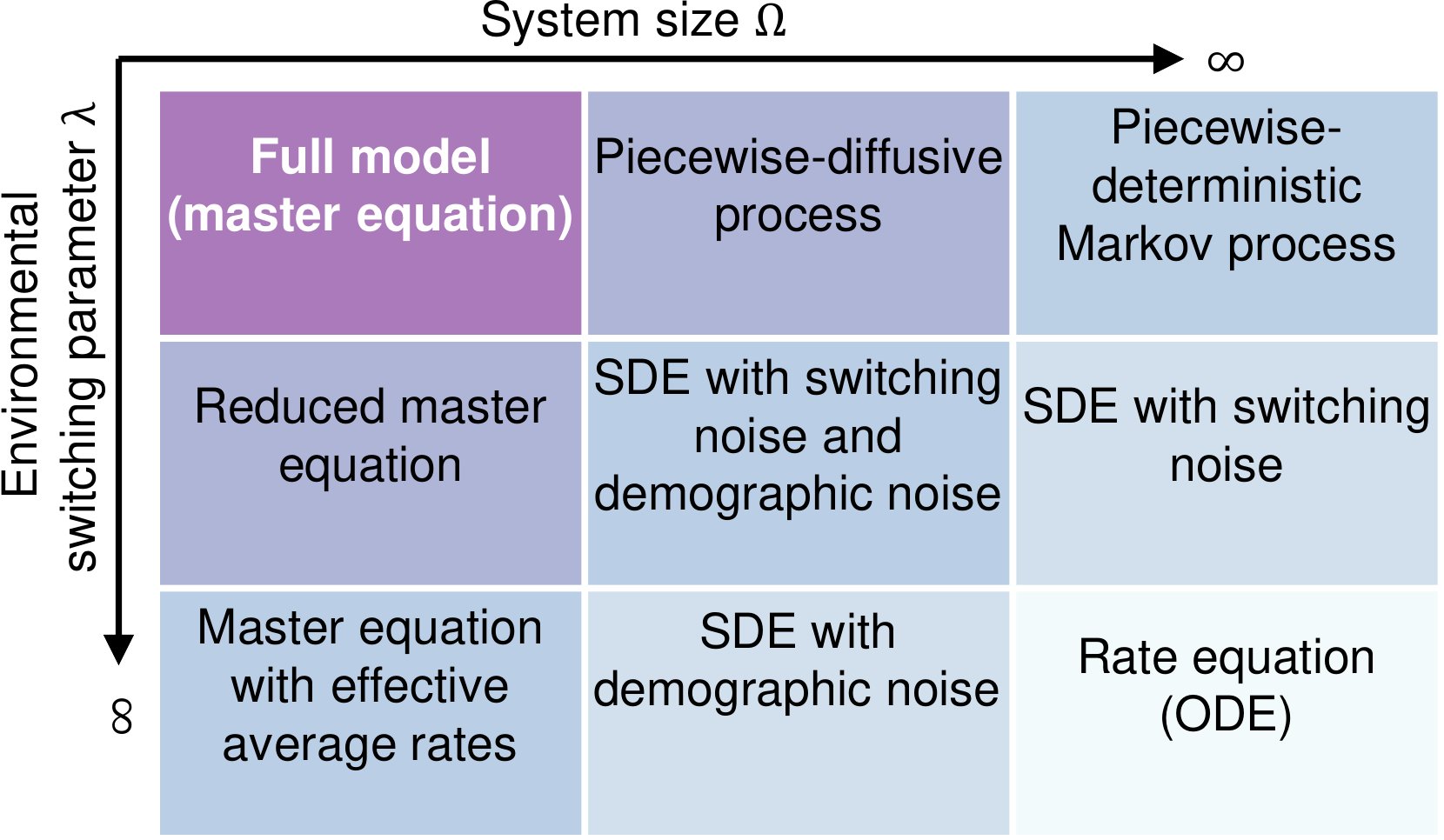}
 \put(50.8,36.5){\textsf{\scriptsize\textcolor{black}{\cite{mao2006stochastic,potoyan2015dichotomous,hufton2016intrinsic}}}}  
\put(78,36.5){\textsf{\scriptsize\textcolor{black}{\cite{davis1984piecewise,zeiser2008simulation,friedman,ge2015stochastic,jia2017simplification,jia2017emergent,herbach2017inferring,LinHufton2018,Lin2018}}}} 
\put(77.5,19.75){\textsf{\scriptsize\textcolor{black}{ \cite{Newby2010,Bressloff2014,Bressloff2017,Bressloff2017a} }}} 
\put(20.2,2){\textsf{\scriptsize\textcolor{black}{\cite{segel1989quasi,haseltine2002approximate,rao2003stochastic,goutsias2005quasiequilibrium,qian2009stochastic,pahlajani2011stochastic,kim2014}}}} 
\put(45.8,2){\textsf{\scriptsize\textcolor{black}{\cite{duncan2015noise,risken1984fokker,gardiner1985handbook,van1992stochastic}}}}

\end{overpic}}
\caption{Schematic overview of the different model-reduction schemes for populations coupled to external environments with discrete states. Each column and row corresponds to a successive layer of approximation.}
\label{fig:table}
\end{figure}
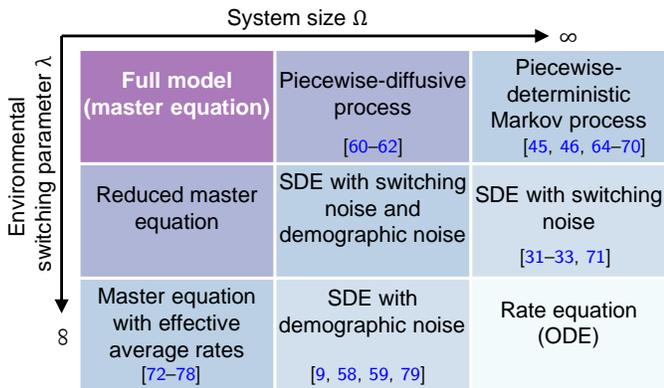

A schematic overview is given in Fig.~\ref{fig:table}.
Broadly speaking the overall picture involves expansions in the inverse switching time scale ($\lambda^{-1}$) and/or the inverse typical size of the population ($\Omega^{-1}$).
The parameters $\lambda$ and $\Omega$ correspond to the vertical and horizontal directions in Fig.~\ref{fig:table}.
In the upper row we perform no expansion in $\lambda^{-1}$ (i.e, we keep all terms), in the middle row we assume $\lambda \gg 1$ but finite (keeping leading and sub-leading terms), and in the lower row the adiabatic limit has been taken ($\lambda\to\infty$), i.e., the noise due to the environmental switching is discarded altogether.
The left-hand column describes models with a discrete population (arbitrary $\Omega$), in the middle column we assume $\Omega\gg 1$, but finite, and in the right-hand column the limit $\Omega\to\infty$ has been taken, i.e., all intrinsic noise in the population is disregarded. We now discuss the relation between the different levels of approximation in more detail.

\subsubsection{Expansion in environmental time scale} In the previous Sections we have focused on the left-hand column of Fig.~\ref{fig:table}.
The upper left box is the full microscopic model, involving a discrete population of typical size $\Omega$, and an environmental process associated with a switching time scale set by $\lambda$.
This full model is defined by the master equation \eqref{eq:mastermaster}.
Expanding to sub-leading order in $\lambda^{-1}$, but keeping $\Omega$ fixed and general, one obtains the reduced master equation \eqref{eq:bhelp3}.
Here, we restrict the discussion to processes in which the environmental switching is independent of the state of the population; the more general case is discussed briefly in Appendix \ref{sec:gen}.
In the case of only two environmental states the reduced master equation is given by Eq.~\eqref{eq:2states}.
It describes the process with bursting as discussed in Sec.~\ref{sec:example1}.

The limit $\lambda\to\infty$ is the adiabatic limit; restricting the master equation to leading-order terms in $\lambda^{-1}$ produces a process described by the master equation 
\begin{equation}\label{eq:mead}
\frac{\d}{\d t}\Pi(\ell,t)=\cMbar\Pi(\ell,t).
\end{equation}
For the case of two environmental states this can be obtained from Eq.~\eqref{eq:2states}, but the general form is applicable for multiple environmental states as well. Eq.~(\ref{eq:mead}) describes a process with the same types of reactions as the original dynamics, but with rates that are weighted averages over the stationary distribution of the environmental states.
This is the lower left-hand box in Fig.~\ref{fig:table}. This is conceptually similar to the quasi-steady-state approximation\cite{QSSA1,segel1989quasi,kim2014,haseltine2002approximate,rao2003stochastic,goutsias2005quasiequilibrium},
in which the fast-reacting species  are regarded as constant at values obtained from an appropriate weighted average. Another approach to approximating environmental noise in the fast-switching limit involves assuming a large number of environmental states, so that the environment may be approximated as continuous \cite{Thomas2012,Thomas2012a}.

\subsubsection{Expansion in powers of inverse system size}
In a different approach one can first approximate the intrinsic noise for large system size ($\Omega\gg 1$), starting from the full model (environment and population), without  any expansion in the environmental switching time scale. This is done by carrying out a Kramers--Moyal expansion on the dynamics of the population, while simultaneously maintaining the discrete environmental states \cite{hufton2016intrinsic}.
This corresponds to travelling horizontally along the first row of Fig.~\ref{fig:table}. 

If sub-leading order terms in powers of the inverse system size are retained, one obtains piecewise-diffusive dynamics \cite{hufton2016intrinsic,potoyan2015dichotomous,mao2006stochastic,hufton2017bacteria}, corresponding to the middle box in the first row of Fig.~\ref{fig:table}.
Between switches of the environmental state, the population is then described by a stochastic differential equation. The process is described by 
\begin{equation}\label{eq:masterfp}
\frac{\partial}{\partial t}p_\sigma(\bx,t)={\cal F}_\sigma p_\sigma(\bx,t) + \lambda\sum_{\sigma'} A_{\sigma'\to\sigma} p_\sigma(\bx,t),
\end{equation}
where $p_\sigma(\vec{\bx},t)$ is a probability density over continuous states $\bx$, obtained from discrete states $\ell$ in the limit of large $\Omega$ (see Sec.~\ref{sec:kmexample} for a specific example). The ${\cal F}_\sigma$ are Fokker--Planck operators obtained from a Kramers--Moyal expansion on $\cM_\sigma$.

\subsubsection{Combined expansion}
Starting from the piecewise-diffusive process [upper row, middle in Fig.~\ref{fig:table}, Eq.~(\ref{eq:masterfp})] one can follow the same steps as in Sec.~\ref{sec:genform} and consider the limit of fast but not infinitely fast environmental switching.
In Fig.~\ref{fig:table} this means working down the central column. We simultaneously consider the limit of large $\Omega$ and the limit of large $\lambda$.
In taking these limits we assume that the ratio $\Omega/\lambda$ remains finite, this will become more clear in the example discussed below (Sec.~\ref{sec:kmexample}).
For the case of two environmental states, the result can be read off from Eq.~\eqref{eq:2states} simply replacing $\cM_\sigma$\ by ${\cal F}_\sigma$, i.e.,
\begin{equation}\label{eq:2statesfpe}
\frac{\partial}{\partial t}\Pi(\bx,t)= \cFbar\Pi(\bx,t)+\frac{1}{2}\frac{\theta^2}{\lambda} (\cF_0-\cF_1)^2\Pi(\bx,t).
\end{equation}
An interpretation of Eq.~\eqref{eq:2statesfpe} in terms of a stochastic differential equation can be obtained by expanding the term $(\cF_0-\cF_1)^2$ further and keeping only terms to order $1/\Omega$. This stochastic differential equation contains two different sources of Gaussian noise, one representing demographic noise and the other the stochasticity of the environmental switching. 

Finally, we could also take the adiabatic limit $\lambda\to\infty$; this leads to $\frac{\partial}{\partial t}\Pi(\bx,t)= \cFbar\Pi(\bx,t)$. In this limit the noise due to the environmental process has been eliminated entirely, and the resulting SDE contains only Gaussian noise coming from the intrinsic fluctuations in the population.

\subsubsection{Piecewise-deterministic process}
Finally, we can also take the limit of an infinite population $\Omega\to\infty$ first, keeping $\lambda$ general. Thus, we neglect intrinsic fluctuations altogether.
This is achieved by retaining only the leading-order term in the Kramers--Moyal expansion of the population.
In each fixed environment the dynamics of the population are then described by an ordinary differential equation.
This constitutes what is known as a piecewise-deterministic Markov process (PDMP) \cite{davis1984piecewise,zeiser2008simulation}.
In Fig.~\ref{fig:table} this is the right-hand box in the upper row. Mathematically, the PDMP is described by
\begin{equation}\label{eq:master_lou}
\frac{\partial}{\partial t}p_\sigma(\bx,t)={\cal L}_\sigma p_\sigma(\bx,t) + \lambda\sum_{\sigma'} A_{\sigma'\to\sigma} p_\sigma(\bx,t),
\end{equation}
with Liouville operators $\cL_\sigma$; they are first-order differential operators which describe the deterministic drift of the system in a given environmental state.

We can now use the PDMP as a starting point, and work down the right-hand column of Fig.~\ref{fig:table}, following the same steps as in Sec.~\ref{sec:genform}, replacing $\cM_\sigma$\ by ${\cal L}_\sigma$. For two environmental states and keeping terms of order $\lambda^{-1}$, the result is analogous to Eq.~\eqref{eq:2states}. One finds
\begin{equation}\label{eq:2statesliouville}
\frac{\partial}{\partial t}\Pi(\bx,t)= \cL_\text{avg}\Pi(\bx,t)+\frac{1}{2}\frac{\theta^2}{\lambda} (\cL_0-\cL_1)^2\Pi(\bx,t).
\end{equation}
This is a Fokker--Planck equation and corresponds to an SDE in which Gaussian noise reflects the effects of the fast-switching environment.
This result was previously reported in Ref.~\cite{Bressloff2014}.

A further approximation to the dynamics would again involve taking the adiabatic limit: this is equivalent to ignoring the final term in Eq.~\eqref{eq:2statesliouville}.
The resulting Liouville equation corresponds to an ODE description of the system. Its dynamics is then governed by a rate equation, where the reaction rates are weighted averages over the different environmental states.
In such an approximation all stochasticity, both intrinsic and environmental, has been eliminated.
This is the lower entry in the right-hand column of Fig.~\ref{fig:table}.

\subsection{Example}\label{sec:kmexample}
We now focus on one of the single-species models in Sec.~\ref{sec:example1}. The purpose of this basic example is purely illustrative; specific applications will be discussed in Sec.~\ref{sec:appl}.
Particles are produced at constant rate $\beta$, and they are removed with per capita rates $\delta_\sigma$ in environments $\sigma\in\{0,1\}$. We have
\begin{equation}
\cM_\sigma= \beta\Omega ({\cal E}^{-1}-1) + \delta_\sigma ({\cal E}-1)n,
\end{equation}
where $n$ is the number of particles in the population.
Keeping the system-size parameter $\Omega$ fixed, and taking the limit of large but finite $\lambda$, one obtains Eq.~\eqref{eq:masterexample2}.
This corresponds to the middle box in the left-hand column of Fig.~\ref{fig:table}.
Taking $\lambda\to\infty$ one has
\begin{equation}\label{eq:bottomleft}
\frac{\d}{\d t}\Pi(n,t) = \Omega\beta ({\cal E}^{-1}-1)\Pi(n,t)+ ({\cal E}-1) \delta_{\rm avg}n\Pi(n,t),
\end{equation}
where $\delta_{\rm avg} = (k_1 \delta_0 + k_0 \delta_1)/(k_0+k_1)$; this is the master equation with effective average rates (lower box on the left in Fig.~\ref{fig:table}).

Next, writing $x=n/\Omega$, and starting again from the full model of population and environment, we carry out a Kramers--Moyal expansion first (keeping terms up to sub-leading order in $1/\Omega$). One has the Fokker--Planck operators
\begin{align}\begin{split}
\cF_0={}&\beta\left(-\partial_x+\frac{1}{2\Omega}\partial_x^2\right)+\delta_0\left(\partial_x+\frac{1}{2\Omega}\partial_x^2\right)x, \\
\cF_1={}&\beta\left(-\partial_x+\frac{1}{2\Omega}\partial_x^2\right)+\delta_1\left(\partial_x+\frac{1}{2\Omega}\partial_x^2\right)x.
\end{split}\end{align}
These operators together with Eq.~\eqref{eq:masterfp} describe a piecewise-diffusive process (upper row, central column in Fig.~\ref{fig:table}); in a given environmental state the dynamics are described by an Ito SDE
\begin{equation}
\dot x = \beta-\delta_{\sigma(t)}x+\sqrt{\frac{\beta+\delta_{\sigma(t)}x}{\Omega}}\eta(t),
\label{eq:PDiffExample}
\end{equation}
where $\eta(t)$ is Gaussian white noise of unit variance.

Further approximating the piecewise-diffusive process in the limit of fast environmental switching, we can insert the explicit form of $\cF_\sigma$ into Eq.~\eqref{eq:2statesfpe} to give
\BE
\frac{\partial}{\partial t}\Pi(x,t) &=& -\partial_x \left\{\left[\beta-\delta_{\rm avg}x+\frac{1}{2}g_{\rm e}\partial_x g_{\rm e}\right]\Pi(x,t) \right\} \nonumber \\
&&+\frac{1}{2}\partial_x^2\left\{\left[g^2_{\rm i }+g^2_{\rm e }\right]\Pi(x,t)\right\},
\label{eq:fpecentre}
\EE
where $\Delta=\delta_0-\delta_1$, and
\begin{align}
g_{\rm i}(x)^2={}&\frac{1}{\Omega}\left(\beta+\delta_{\rm avg}x\right), \\
g_{\rm e}(x)^2={}&\frac{\theta^2}{\lambda}\Delta^2 x^2.
\end{align}
The subscript `i' indicates intrinsic stochasticity (demographic noise), and `e' labels the contribution to the noise from environmental switching. We note that $g_{\rm i}(x)^2\propto \Omega^{-1}$, and $g(x)_{\rm e}^2\propto \lambda^{-1}$. It is interesting to note that the same Fokker--Planck equation is obtained by a direct Kramers--Moyal expansion of Eq.~\eqref{eq:masterexample2}. Details can be found in Appendix~\ref{sec:appkm}. The contribution $g_{\rm e}\partial_x g_{\rm e}/2$ to the drift term in Eq.~(\ref{eq:fpecentre}) is of order $\lambda^{-1}$, and it can safely be neglected to the order we are working at (see also Ref.~\cite{Bressloff2014}). Equation~\eqref{eq:fpecentre} then describes an Ito SDE of the form
\begin{equation}\label{eq:sdecentre}
\dot x = \beta-\delta_{\rm avg}x+g_{\rm i}(x)\eta_{\rm i}(t) + g_{\rm e}(x)\eta_{\rm e}(t),
\end{equation}
in which $\eta_{\rm i}(t)$ and $\eta_{\rm e}(t)$ are independent Gaussian processes of unit variance, and with no correlations in time. The SDE \eqref{eq:sdecentre}, corresponds to the central box in Fig.~\ref{fig:table}.

Equation~\eqref{eq:fpecentre} can be used as a starting point for further approximations. In the case of infinitely fast switching, $\lambda\to\infty$, the term $g_{\rm e}(x)$ can be neglected, and one finds
\begin{align}\label{eq:fpelow}
\frac{\partial}{\partial t} \Pi(x,t) ={}& -\partial_x \left[\left(\beta-\delta_{\rm avg}x\right)\Pi(x,t) \right] \nonumber \\
&+\frac{1}{2\Omega}\partial_x^2\left[\left(\beta+\delta_{\rm avg}x\right)\Pi(x,t)\right].
\end{align}
We note that this relation can also be obtained by direct Kramers--Moyal expansion of Eq.~\eqref{eq:bottomleft}. Only the Gaussian noise from the intrinsic stochasticity then remains in the SDE~\eqref{eq:sdecentre}. This is the lower box in the central column of Fig.~\ref{fig:table}.

In the case of an infinite population $\Omega\to\infty$, Eq.~\eqref{eq:fpecentre} turns into
\begin{align}\begin{split}\label{eq:centreright}
\frac{\partial}{\partial t} \Pi(x,t) ={}& -\partial_x \left[\left(\beta-\delta_{\rm avg}x\right)\Pi(x,t) \right] \\
&+\frac{\theta^2}{2\lambda}\Delta^2\partial_x^2\left[ x^2\Pi(x,t)\right], 
\end{split}\end{align}
so that the noise term containing $g_{\rm i}(x)$ is no longer present in the SDE~\eqref{eq:sdecentre}. This is the centre box in the right-hand column of Fig.~\ref{fig:table}.
Equation~\eqref{eq:centreright} can also be found from Eq.~\eqref{eq:2statesliouville} upon using $\cL_\sigma\Pi(x)=-\partial_x (\beta-\delta_\sigma x)\Pi(x)$, see Appendix \ref{sec:appliouville}.

If all stochasticity is ignored altogether ($\lambda\to\infty$ and $\Omega\to\infty$) one has $g_{\rm i}=g_{\rm e}=0$. In our example one then finds the rate equation
\begin{equation}
\dot x = \beta-\delta_{\rm avg}x.
\end{equation}
This corresponds to the lower box in the right-hand column of Fig.~\ref{fig:table}.

\subsection{Linear-noise approximation}
In order to obtain analytical results, for example approximations to the stationary distribution and power spectral density of fluctuations, an additional step---the linear noise approximation (LNA)---can be taken in Eq.~\eqref{eq:sdecentre}. 
The LNA simplifies an SDE with multiplicative noise into one with additive noise, and is applicable when the noise is sufficiently small \cite{van1992stochastic}, i.e., in our case $\lambda\gg 1$ and $\Omega\gg 1$.

The stochastic differential equation~\eqref{eq:sdecentre} is of the form
\begin{equation}\label{eq:gensde}
\dot x = v_{\rm avg}(x)+g_{\rm i}(x) \eta_i(t) + g_{\rm e}(x)\eta_e(t),
\end{equation}
where $v_{\rm avg}(x)=\beta-\delta_{\rm avg}x$, and where $g_{\rm i}={\cal O}(\Omega^{-1/2})$ and $g_{\rm e}={\cal O}(\lambda^{-1/2})$. The LNA can then be carried out using the ansatz
\begin{equation}\label{eq:LNA_ansatz}
x(t)=x_{\rm avg}(t)+\Omega^{-1/2}\xi_{\rm i}(t)+\lambda^{-1/2}\xi_{\rm e}(t),
\end{equation}
where $x_{\rm avg}(t)$ is a deterministic function to be determined self-consistently; the quantities $\xi_{\rm i}(t)$ and $\xi_{\rm i}(t)$ are each stochastic processes describing the deviations due to intrinsic and extrinsic noise, respectively.
Inserting into Eq.~\eqref{eq:gensde}, expanding in powers of $\Omega^{-1/2}$ and $\lambda^{-1/2}$ one obtains $\dot x_{\rm avg}=v_{\rm avg}(x_{\rm avg})$ from the lowest-order terms, and
\begin{align}\begin{split}
\dot \xi_{\rm i}={}& v_{\rm avg}'\left(x_{\rm avg}\right)\xi_{\rm i} + \Omega^{1/2}g_{\rm i}\left(x_{\rm avg}\right)\eta_i(t), \\
\dot \xi_{\rm e}={}& v_{\rm avg}'\left(x_{\rm avg}\right)\xi_{\rm e} + \lambda^{1/2}g_{\rm e}\left(x_{\rm avg}\right)\eta_e(t),
\end{split}\end{align}
for the sub-leading order terms, where $v_{\rm avg}'=\d v_{\rm avg}/\d x$. We note that the arguments of both $g_{\rm i}$ and $g_{\rm e}$ are now given by $x_{\rm avg}$, so that the multiplicative noise in Eq.~\eqref{eq:gensde} has been turned into additive noise.
Introducing $\zeta(t)=\Omega^{-1/2}\xi_{\rm i}(t)+\lambda^{-1/2}\xi_{\rm e}(t)$ describing the total amount of deviation caused by both sources of noise, we can write this more compactly as $x(t)=x_{\rm avg}(t)+\zeta(t)$, with
\begin{equation}
\dot \zeta = v'(x_{\rm avg})\zeta+\left[g_{\rm i}(x_{\rm avg})^2 + g_{\rm e}(x_{\rm avg})^2\right]^{1/2}\eta(t),
\end{equation}
where the two Gaussian processes have been combined, so that the stochasticity is described by a single white noise Gaussian process $\eta(t)$.
In the above example we have $\dot x_{\rm avg} = \beta-\delta_{\rm avg}x_{\rm avg}$, and
\begin{equation}
\dot \zeta=-\delta_{\rm avg}\zeta+\left[\frac{\beta+\delta_{\rm avg}x_{\rm avg}}{\Omega}+\frac{\theta^2}{\lambda}\left(x_{\rm avg}\Delta\right)^2\right]^{1/2}\eta(t).
\end{equation}

\subsection{Analytical approximation for power spectra}\label{sec:spectraanalytic}
We now return to the model with two species defined in Eq.~\eqref{eq:simple}. Carrying out a Kramers--Moyal expansion of the reduced master equation Eq.~\eqref{eq:cme_ab_approx} we arrive at the following stochastic differential equations for $x_A=n_A/\Omega$ and $x_B=n_B/\Omega$
\begin{align}\begin{split}
\dot x_A={}&\alpha_{\rm avg}-\gamma x_A+\eta_A(t), \\
\dot x_B={}&\beta_{\rm avg}-\delta x_B+\eta_B(t). 
\end{split}\label{eq:kmtoy1}\end{align}
For compactness, we have absorbed the diffusion coefficients (describing both intrinsic and extrinsic noise) into the white noise terms $\eta_A$ and $\eta_B$, so that they have the following covariance matrix:
\BE
\avg{\eta_A(t)\eta_A(t')}&=&\left(\frac{\alpha_{\rm avg}+\gamma x_A}{\Omega}+\frac{\theta^2}{\lambda}(\Delta\alpha)^2\right)\delta(t-t'),\nonumber \\
\avg{\eta_B(t)\eta_B(t')}&=&\left(\frac{\beta_{\rm avg}+\delta x_B}{\Omega}+\frac{\theta^2}{\lambda}(\Delta\beta)^2\right)\delta(t-t'),\nonumber \\
\avg{\eta_A(t)\eta_B(t')}&=&\frac{\theta^2}{\lambda}\Delta\alpha\Delta\beta\delta(t-t'), 
 \label{eq:kmtoy2}
 \EE
see also Appendix \ref{sec:kmab}. 

To simplify matters we now restrict the discussion to the case $\gamma=\delta$ and $\alpha_{\rm avg}=\beta_{\rm avg}$ (the latter does not imply $\Delta\alpha=\Delta\beta$).
In the long run the deterministic trajectory converges to the fixed point given by $x_A^*=x_B^*=\alpha_{\rm avg}/\gamma$. Applying the LNA at this fixed point, we find
\begin{align}\begin{split}
\dot \zeta_A={}&-\gamma\zeta_A+\eta_A(t), \\
\dot \zeta_B={}&-\gamma\zeta_B+\eta_B(t),
\end{split}\end{align}
where
\begin{align}\begin{split}
\avg{\eta_A(t)\eta_A(t')}={}&\left(\frac{2\alpha_{\rm avg}}{\Omega}+\frac{\theta^2}{\lambda}(\Delta\alpha)^2\right)\delta(t-t'), \\
\avg{\eta_B(t)\eta_B(t')}={}&\left(\frac{2\alpha_{\rm avg}}{\Omega}+\frac{\theta^2}{\lambda}(\Delta\beta)^2\right)\delta(t-t'), \\
\avg{\eta_A(t)\eta_A(t')}={}&\frac{\theta^2}{\lambda}\Delta\alpha\Delta\beta\delta(t-t').
\end{split}\end{align}
In order to find the power spectral density of fluctuations, we perform a Fourier transform and obtain
\begin{align}\begin{split}
\avg{\zeta_A(\omega)\zeta_A^*(\omega')}=\delta(\omega+\omega') \Omega^{-2}S_{AA}(\omega), \\
\avg{\zeta_A(\omega)\zeta_B^*(\omega')}=\delta(\omega+\omega') \Omega^{-2}S_{AB}(\omega),
\end{split}\end{align}
with
\begin{align}\label{eq:spectoy}\begin{split}
S_{AA}(\omega)={}&\Omega^2\frac{\frac{2\alpha_{\rm avg}}{\Omega}+\frac{\theta^2}{\lambda}(\Delta\alpha)^2}{\gamma^2+\omega^2}, \\
S_{AB}(\omega)={}&\Omega^2\frac{\frac{\theta^2}{\lambda}\Delta\alpha\Delta\beta}{\gamma^2+\omega^2}.
\end{split}\end{align}
This result matches well with the results of Gillespie simulating the full model (for $\Omega=20$, $\lambda=20$). A comparison is shown in Fig.~\ref{fig:tauleap}. In the adiabatic limit ($\lambda\to\infty$) Eq.~(\ref{eq:spectoy}) reduces to
\begin{align}\begin{split}
S_{AA}(\omega)={}&\Omega\frac{2\alpha_{\rm avg}}{\gamma^2+\omega^2}, \\
S_{AB}(\omega)={}&0,
\end{split}\end{align}
confirming again the absence of correlations between $n_A$ and $n_B$ in the limit of infinitely fast environments.

\section{Further applications}\label{sec:appl}
In this Section we will apply the formalism we have developed to a series of specific examples.

\subsection{Model of protein production}\label{sec:prot}

\begin{figure*}
\includegraphics[width=1\textwidth,valign=t]{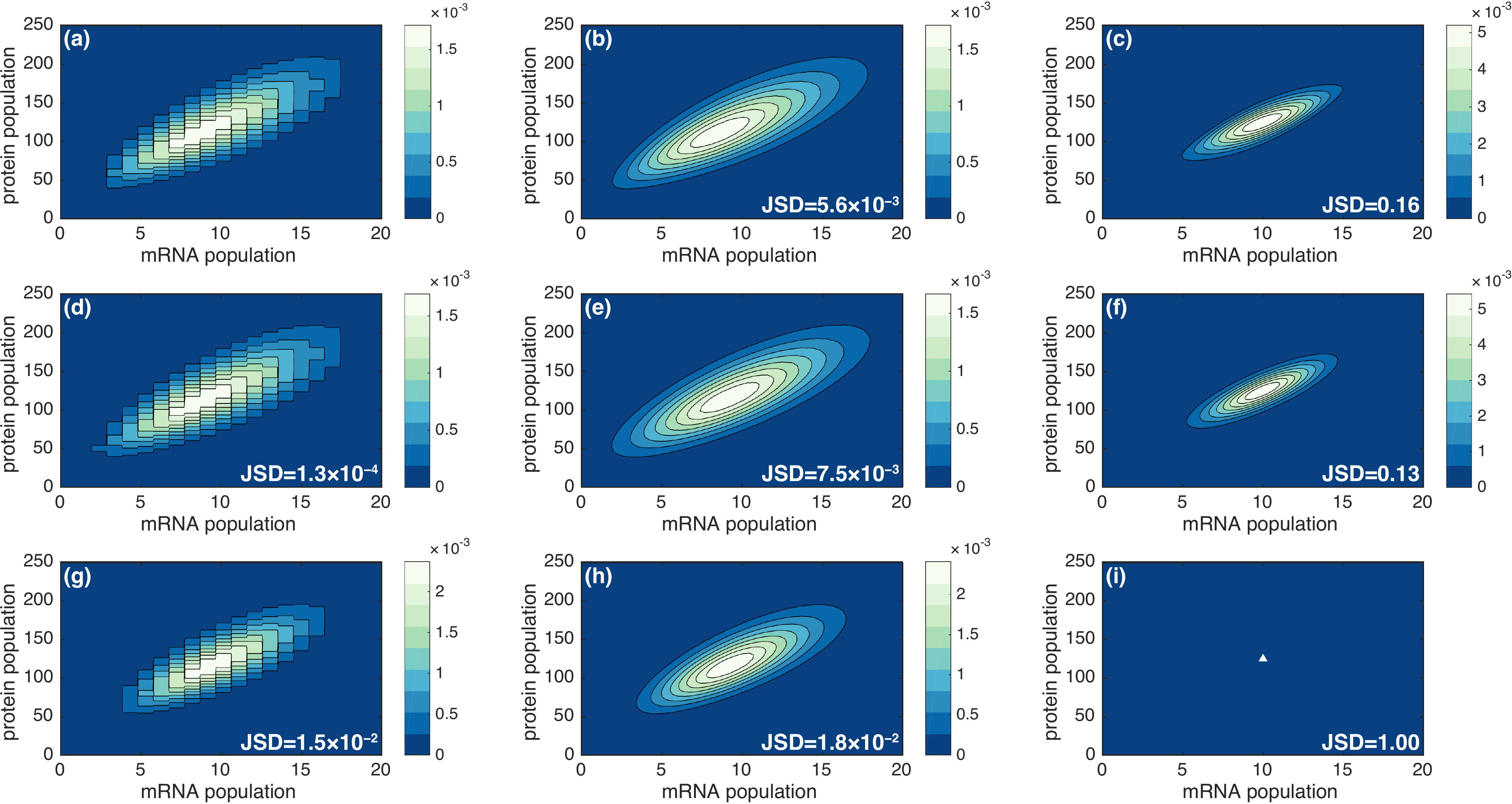} 
\caption{The stationary probability distribution of the populations of mRNA and protein molecules for the model in Sec. \ref{sec:prot}. Data is from Monte Carlo simulations of each different level of approximation in Fig.~\ref{fig:table}: (a) full model; (b) piecewise-diffusive process; (c) piecewise-deterministic Markov process; (d) reduced master equation with bursting; (e) SDE with switching noise and demographic noise; (f) SDE with switching noise; (g) master equation with average rates; (h) SDE with demographic noise; and (i) rate equation ($\blacktriangle$ represents a delta peak). Parameters: $\Omega=20, b_0=0, b_1=1, d=1, \beta=25, \delta=2,$ and $\lambda=10, k_0=k_1 = 1$.}
\label{fig:threestage_comparison}
\end{figure*}
\subsubsection{Motivation and model definitions}
The dynamics of gene expression are inherently noisy \cite{swain2002intrinsic,Cai2006}, and stochastic approaches are hence most appropriate to model such processes. They also frequently exhibit a separation of time scales, see e.g. Refs.~\cite{vilar,weinberger,friedman,kim2014}. Here, we consider a commonly-used model which describes two essential steps for gene expression, the transcription into mRNA  and the translation into protein \cite{thattai2001intrinsic,swain2002intrinsic,lipniacki2006transcriptional,bobrowski2007asymptotic,zeiser2008simulation,thomas2014phenotypic,sherman2014computational}. 
The model describes a single gene ${G}$, which can be in two different states, labelled `on' ($\sigma=1$) and `off' ($\sigma=0$).
The gene switches between these states with rates $k_0$ and $k_1$, respectively.
In each state, mRNA molecules are produced with a rate $\Omega b_\sigma$; they decay with rate $d$. The presence of mRNA also leads to the production of protein molecules; this occurs with rate $\beta$ (per mRNA molecule).
Protein molecules finally decay with rate $\delta$. The model can be summarised by the following reactions 
\begin{equation}\begin{gathered}
{G}_{\rm off} \xrightleftharpoons[\lambda k_0]{\mathmakebox[5mm]{\lambda k_1}}{} {G}_{\rm on}, \quad
\emptyset \xrightarrow{\mathmakebox[5mm]{\Omega b_\sigma}}{} {M}, \quad
{M} \xrightarrow{\mathmakebox[5mm]{d}}{} \emptyset, \\
{M} \xrightarrow{\mathmakebox[5mm]{\beta}}{} {M} + {P}, \quad
{P} \xrightarrow{\mathmakebox[5mm]{\delta}}{} \emptyset,\label{eq:proteinmodel}\end{gathered}\end{equation}
where $M$ and $P$ refer to mRNA and protein molecules, respectively.

\subsubsection{Comparison of different approximation schemes}
We proceed to consider the full model, and each of the eight levels of approximation in Fig.~\ref{fig:table}.
The reduced master equation for large $\lambda$ can be derived following the procedure outlined in Sec.~\ref{sec:analysis}. The details of this are very similar to the example in Sec.~\ref{sec:example1}; we do not report them in full.
The reduced master equation describes a set of effective reactions in which mRNA molecules are made in bursts of sizes one or two. We stress again that the origin of this type of bursting is different from the one discussed in Refs. \cite{friedman,Shahrezaei2008,lin2016gene,Lin2016bursting}. These effective reactions can then be simulated by the standard Gillespie method, because the reduced master equation for this model does not contain negative rates.

Similarly, for the adiabatic limit $\lambda\to\infty$, effective production rates are obtained by replacing the rates $b_\sigma$ in Eq.~\eqref{eq:proteinmodel} by their weighted average, $b_{\rm avg}$. This can then be used in the Gillespie simulation.

For large but finite $\Omega$, the piecewise-diffusive process for this model is given by
\begin{align}\begin{split}
\dot m={}&\left(b_{\sigma(t)}-d m\right) +\Omega^\nhalf\sqrt{b_{\sigma(t)}+dm}~\eta^{\rm m}(t) \\
\dot p={}&\left(\beta m -\delta p\right) +\Omega^\nhalf\sqrt{\beta m + \delta p}~\eta^{\rm p}(t),
\label{eq:three_stage_PDiffP}
\end{split}\end{align}
where $\Omega m$ and $\Omega p$ are the numbers of mRNA molecules and protein molecules, respectively, and where $\sigma(t)$ is the stochastic trajectory of the switching process for the gene; $\eta^\text{m}(t)$ and $\eta^\text{m}(t)$ are independent Gaussian white noise processes. For both $\Omega$ and $\lambda$ large but finite we find the following description in terms of stochastic differential equations (corresponding to the central box in Fig.~\ref{fig:table}):
\begin{align}\begin{split}
\dot m={}&\left(b_{\rm avg}-d m\right) +\left[g^{\rm m}_{\rm i}(m,p)^2+g^{\rm m}_{\rm e}(m,p)^2\right]^{1/2}\eta^{\rm m}(t), \\
\dot p={}&\left(\beta m -\delta p\right) +g^{\rm p}_{\rm i}(m,p)\eta^{\rm p}(t),
\label{eq:three_stage_central}
\end{split}\end{align}
where
\begin{align}\begin{split}
g^{\rm m}_{\rm i}(m,p)={}&\Omega^\nhalf\sqrt{b_\text{avg}+dm}, \\
g^{\rm m}_{\rm e}(m,p)={}&\lambda^\nhalf\sqrt{\frac{2k_1 k_0\left(b_0-b_1\right)^2}{\left(k_0 + k_1\right)^3}}, \\
g^{\rm p}_{\rm i}(m,p)={}&\Omega^\nhalf\sqrt{\beta m + \delta p}.
\end{split}\end{align}
From these it is straightforward to obtain the remaining approximations in Fig.~\ref{fig:table}, by either sending the amplitude of the environmental noise $g^{\rm m}_{\rm e}$ to zero, or that of the intrinsic noise ($g^{\rm p}_{\rm i}$, and $g^{\rm m}_{\rm i}$), or both.

Figure~\ref{fig:threestage_comparison} shows the stationary distributions obtained from Monte Carlo simulations of the full model and the eight different approximations.  The arrangement in the figure corresponds to that in Fig.~\ref{fig:table}. We remark that the population remains discrete for the panels in the left-hand column, while expanding in powers of the system size (middle and right column) leads to continuous populations. In each panel we indicate a numerical estimate for the Jensen--Shannon divergence (JSD) of the respective stationary distribution relative to that of the full model in panel (a) \cite{lin_JSD}.

 The data in Fig.~\ref{fig:threestage_comparison} shows that the successive approximations in powers of the system size and the switching rates reduce the accuracy in reproducing the full individual-based model.  The JSD generally increases as one moves down or to the right in Fig.~\ref{fig:threestage_comparison}. For this model and parameter set, the only exception is the approximation in panel (f) which shows a smaller JSD than that in panel (c). This is due to the following effect. The full model in panel (a) can explore arbitrary numbers of mRNA and protein molecules. The stationary distribution of the PDMP in panel (c) however has bounded support, because intrinsic noise is discarded. The distribution in panel (f) does not include effects of intrinsic noise either, but the environmental stochasticity has been approximated by Gaussian noise, restoring an unbounded support. This leads to the seemingly better agreement of (f) with the full model.

We are not necessarily proposing all eight approximations in Figs. \ref{fig:table} and \ref{fig:threestage_comparison} as starting points for further analysis or simulation. For instance, it is not easy to find analytical descriptions for the stationary distribution of the piecewise diffusive description in panel (b), and the piecewise deterministic model in panel (c). This is only feasible for simple models, see also our earlier work \cite{hufton2016intrinsic}. The SDE in panel (e) on the other hand (i.e., approximating both intrinsic and extrinsic randomness as Gaussian noise) allows for the stationary distribution, among other things, to be approximated analytically; following a linearisation of the noise terms (LNA) in Eq.~$\eqref{eq:three_stage_central}$, the resulting distribution is a bivariate Gaussian. At this level of approximation the stationary distribution can be obtained analytically. In this respect, approximation (e) can be seen as a useful trade-off between accuracy and practical analytical results in our limit of interest, at least for the unimodal distribution of the current model. We will also discuss the SDE as a starting point for efficient simulations in the context of the next example.
\\

\subsection{Bimodal genetic switch}\label{sec:bimodal}
\begin{figure*}
\includegraphics[width=1\textwidth,valign=t]{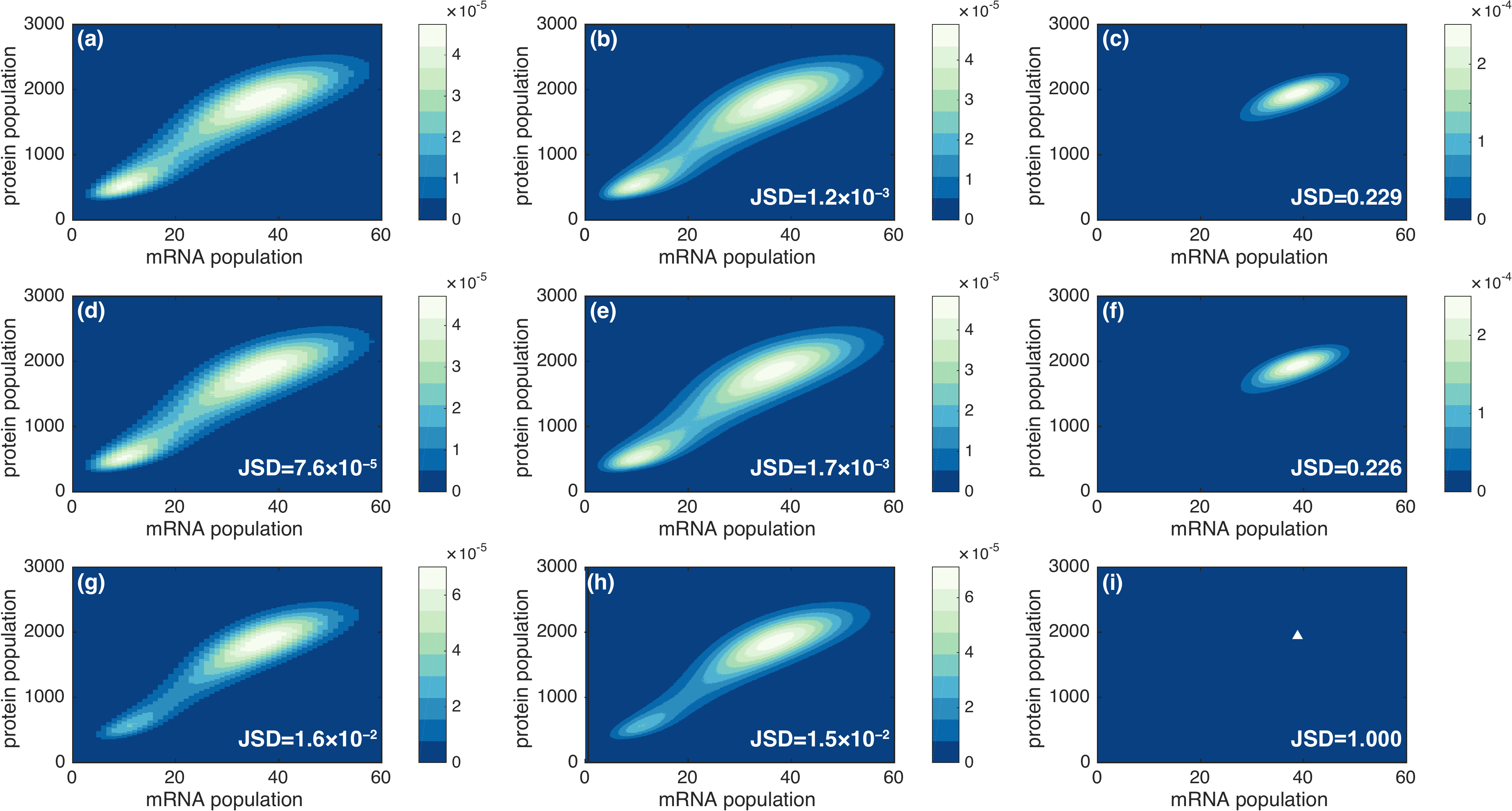}
\caption{Stationary probability distribution of the populations of mRNA and protein molecules for the full model in Sec. \ref{sec:bimodal}, and the eight levels of model reduction in Fig.~\ref{fig:table}: (a) full model; (b) piecewise-diffusive process; (c) piecewise-deterministic Markov process; (d) reduced master equation; (e) SDE with switching noise and demographic noise; (f) SDE with switching noise; (g) master equation with average rates; (h) SDE with demographic noise; and (i) rate equation ($\blacktriangle$ represents a delta peak). Parameters: $N=2, \Omega=50, b_0=b_1=1, b_2=20, d=9.2, \beta=50, \delta=1, k_-=0.025, k_+ = 1$ and $\lambda=1250$.}
\label{fig:bimodal}
\end{figure*}
\subsubsection{Model}
The simple model of protein production in the previous section shows a unimodal distribution. Pluripotent stem cells have the ability to differentiate into several possible cell types \cite{Masui2007, Kalmar2009, LinHufton2018}; the basic features of the networks of genes, transcription factors and epigenetic variables leading to these cell-fate decisions are a current focus of research \cite{Buchler2003, Cai2006, Munsky2015, Gomez2017}. Several hypotheses exist about the mechanisms leading to cell differentiation; among these it has been proposed that excursions of the genetic circuit into different areas of state space might contribute to steering cells towards distinct differentiated states \cite{Masui2007, Kalmar2009}. Bimodal distributions are observed in a variety of biological switches  \cite{gardner2000construction,roma2005optimal,assaf1,friedman,roberts2011noise}. In this Section we discuss a stylised model of processes leading to bimodal distributions; the difference to the model in the previous Section is that this extended model admits a multi-modal stationary distribution. In the context of the above hypothesis, these different peaks would lead to distinct differentiated states.

The model describes a single gene ${G}$, with a promoter site which can bind to a total of up to $N$ molecules of protein.
Each protein molecule binds with a rate $\lambda k_+/\Omega$, and unbinds with a rate $k_-$. Binding and unbinding are sequential \cite{weiss}. Depending on the current state of the gene (i.e., the number of bound proteins, $\sigma=0,1,\dots,N$), mRNA molecules are produced with rate $\Omega b_\sigma$. As in the previous section mRNA in turn decays with (per capita) rate $d$; mRNA leads to the production of protein molecules with a rate $\beta$ per mRNA molecule. Protein molecules finally decay with rate $\delta$. The model can be summarised by the following reactions 
\begin{equation}\begin{gathered}
{G}_{\rm \sigma} + P \xrightleftharpoons[\lambda k_-]{\mathmakebox[10mm]{\lambda k_+ /\Omega}}{} {G}_{\rm \sigma+1}, \quad  \text{for $\sigma<N$}\\
{G}_{ \sigma}  \xrightarrow{\mathmakebox[6mm]{\Omega b_\sigma}}{} {G}_{ \sigma} + {M},~~~~~~\\
{M} \xrightarrow{\mathmakebox[6mm]{d}}{} \emptyset, \quad\quad {M} \xrightarrow{\mathmakebox[6mm]{\beta}}{} {M} + {P}, \quad\quad
{P} \xrightarrow{\mathmakebox[6mm]{\delta}}{} \emptyset.\end{gathered}
\label{eq:bimodalmodel}
\end{equation}
where $M$ and $P$ refer to molecules of mRNA and protein, respectively. 

Mathematically, the two main differences compared to the model in the previous section are the following: (i) the environment (the gene) can take more than two states ($\sigma=0,1,\dots,N$); (ii) the overall rate with which switches from state $\sigma$ to $\sigma+1$ occur ($\sigma<N$) depends on the number of protein. Each protein molecule contributes $\lambda k_+/\Omega$ to the switching rate; the total rate of switching from state $\sigma<N$ to $\sigma+1$ is $\lambda k_+N_p/\Omega$, if the number of proteins is $N_p$. This means that the environmental switching depends on the state of the population.  

Different architectures of the genetic switching and associated mRNA-production rates are discussed in the literature, e.g. \cite{Buchler2003,Karapetyan2015,Munsky2015, Gomez2017,Lin2018}. We focus on $N=2$, i.e. there are three possible envirommental states, $\sigma=0,1,2$. We also assume that mRNA molecules are produced with a common basal rate in gene states $\sigma=0,1$, i.e. we set $b_0=b_1$. When the maximum of $N=2$ proteins are bound to the gene mRNA is produced with the activated rate $\Omega b_2$, where $b_2>b_0$ \cite{LinHufton2018}.

\subsubsection{Comparison of the different approximation schemes}
As in the previous model we test the eight different approximations in Fig.~\ref{fig:table}. In order to derive the reduced master equation, we need to go beyond the formalism of Sec.~\ref{sec:two_environments}, as the environmental switching depends on the state of the population of mRNA and proteins. The construction therefore starts from Eqs. (\ref{eq:bhelp3}) and (\ref{eq:bhelp6}), with three environmental states $\sigma\in\{0,1,2\}$. The calculation leading to the reduced master equation for this model is tedious, but straightforward. The expression for the reduced master equation is lengthy, and given in Appendix \ref{sec:redmasterbistable}.

For large but finite $\Omega$, the piecewise-diffusive process for this model is as in Eq.~(\ref{eq:three_stage_PDiffP}); the only difference is in the dynamics governing $\sigma(t)$. The approximation corresponding to the central box in Fig.~\ref{fig:table} is given by the stochastic differential equations 
\BE
\dot m&=&\left(b_{\rm avg}(p)-d m\right) +\left[g^{\rm m}_{\rm i}(m,p)^2+g^{\rm m}_{\rm e}(m,p)^2\right]^{1/2}\eta^{\rm m}(t), \nonumber \\
\dot p&=&\left(\beta m -\delta p\right) +g^{\rm p}_{\rm i}(m,p)\eta^{\rm p}(t),
\label{eq:sde_bimodal}
\EE
where
\begin{align}\begin{split}
b_\text{avg}(p)={}&\frac{b_0 k_-^2+b_0 k_- k_+ p + b_2 k_+^2 p^2}{ k_-^2 + k_-k_+ p + k_+ ^2 p^2}, \\
g^{\rm m}_{\rm e}(m,p)={}&\sqrt{\frac{2k_- k_+^2 p^2\left[k_-^2+3 k_-k_+ p + k_+^2 p^2\right]}{\lambda\left( k_-^2 + k_-k_+ p + k_+^2 p^2 \right)^3}(b_2-b_0)^2},\\
g^{\rm m}_{\rm i}(m,p)={}&\Omega^\nhalf\sqrt{b_\text{avg}(p)+dm}, \\
g^{\rm p}_{\rm i}(m,p)={}&\Omega^\nhalf\sqrt{\beta m + \delta p}.
\end{split}\end{align}
Again it is straightforward to obtain the approximations (f), (h) and (i), by either sending the amplitude of the intrinsic noise ($g^{\rm m}_{\rm i}$ and $g^{\rm p}_{\rm i}$) to zero, or of the environmental noise ($g^{\rm m}_{\rm e}$), or that of both.

Figure~\ref{fig:bimodal} shows the stationary distributions obtained for the full model, and for the different approximations.  All data is from direct simulations, except (d) which is discussed further below. As before, the arrangement corresponds to that in the schematic of Fig.~\ref{fig:table}, and for each approximation we report the JSD relative to the stationary distribution of the full model in panel (a). The JSD in panel (f) is lower than that in (d) for the same reason as in the previous section. A similar effect is seen comparing (h) and (g). The figure also demonstrates the bimodal structure of the stationary distribution is induced by the intrinsic noise; it is present in each panel in the left-hand and centre columns, but in none of the panels in the right-hand column. While the model is stylised and not intended to directly model a particular biological system Fig.~\ref{fig:bimodal} demonstrates that analyses of this type may help to establish the origin of relevant biological features---in this case bimodality linked to pluripotency and cell-fate decision making is due to intrinsic rather than extrinsic noise.

On a technical note, we add that approximation (d), the reduced master equation, does not in itself define a Markovian process for this model, due to the appearance of negative rates (see Appendix \ref{sec:redmasterbistable}). We have generated the data for the stationary distribution of the reduced master equation in two different ways. One is direct numerical integration of the reduced master equation, this leads to a JSD relative to the distribution for the full model of approximately $7.6\times 10^{-5}$. The second method consists of Gillespie simulations of an approximation to the reduced master equation (\ref{eq:redmasterbistable}), in which sub-leading terms of order $\Omega^2/\lambda$ are kept, but those of order $\Omega/\lambda$ are discarded; specifically, we have set $\varpi_1=\varpi_2=0$ in Eq.~(\ref{eq:redmasterbistable}) for the purpose of these simulations. This leads to a Markovian process, and sample paths can hence be generated using the standard Gillespie algorithm. The JSD for the stationary distribution obtained in this way from that of the full model is found to be approximately $9.1\times 10^{-5}$. Visually, the results from the two methods are indistinguishable, and their JSD from each other is approximately $1.3\times 10^{-5}$, almost an order of magnitude lower than the JSD of either of the two from the stationary distribution of the full model.
\subsubsection{Efficient simulations and required computing time}
Although in the previous two examples we have carried out all eight different approximations, we remark that some prove more useful than others in terms of providing an efficient simulation scheme for specific applications. The purpose of collating data from the different levels of model reduction in Figs. \ref{fig:threestage_comparison} and \ref{fig:bimodal} was to give an illustration of the schematic Fig.~\ref{fig:table} in the context of two concrete examples. 

The approximation as an SDE [panel (e) in Figs. \ref{fig:table}, \ref{fig:threestage_comparison} and \ref{fig:bimodal}] provides a good starting point for simulations of systems with intrinsic noise of small and moderate amplitude, and fast-switching environments. The SDE is an approximation, but it retains both intrinsic and extrinsic noise. In the context of simpler models, we have already used the SDE to carry out further mathematical analysis using the LNA (see Sec.~\ref{sec:spectraanalytic}). To further illustrate the possible advantages of the approximation as a SDE, we have investigated the amount of computing time needed to carry out simulations of the full model in Eq.~(\ref{eq:bimodalmodel}), and of the SDE (\ref{eq:sde_bimodal}). Broadly speaking, the number of environmental switching events per unit time in the full model can be expected to scale as $\lambda$, and the number of events in the population per unit time grows as $\Omega$. One would therefore expect the computing time required to generate a given number of sample paths for the full model up to a specified end time to grow when $\lambda$ or $\Omega$ are increased. This is confirmed in Table \ref{table:cpu}. As seen in Table \ref{table:cpu} the time required to generate sample paths of the SDE (\ref{eq:sde_bimodal}) is independent of $\lambda$ and $\Omega$, as these only enter in the noise strength. These results indicate that simulations of the SDE can be carried out more efficiently than those of the full model, especially when either the environmental switching is fast, or the typical population size large, or both. This is also the regime in which the SDE approximation becomes increasingly accurate.

\begin{table}
\begin{tabular}{ c | c |  C{35mm} |  C{35mm} }		
  $\lambda$ & $\Omega$ & {computation time (s) for \linebreak full model} & {computation time (s) for SDE with switching and demographic noise}\\
  \hline  
  $500$ & $50$ & 62.4 & 34.3 \\
  $1000$ & $50$ & 74.0 & 34.4 \\
  $1500$ & $50$ & 85.0 & 34.4 \\
  $2000$ & $50$ & 93.2 & 34.4 \\
 \hline
  $1250$ & $20$ & 40.4 & 34.5 \\
  $1250$ & $40$ & 67.4 & 34.7 \\
  $1250$ & $60$ & 95.7 & 34.4 \\
  $1250$ & $80$ & 123.2 & 34.3 \\
\end{tabular}
\caption{Comparison of the simulation time of the full model Eq.~(\ref{eq:bimodalmodel}) and the SDE (\ref{eq:sde_bimodal}). The Gillespie algorithm and Euler--Maruyama method ($\d t = 5\times10^{-3}$) are used, respectively, to simulate the system up to time $10^4$. While the simulation time of the full model increases with $\lambda$ and $\Omega$, the simulation time for the SDE is independent of $\lambda$ and $\Omega$. \label{table:cpu}}
\end{table}

\subsection{Genetic network with multiple genes}\label{sec:multi}
A related model, as considered in Ref.~\cite{duncan2015noise}, involves $N$ identical promoter genes, ${G}^{(i)}$ ($i=1,\dots,N$), which can each be in their `on' or `off' states, and switch between these independently. This is different from the model in the previous section where a single gene can bind up to $N$ molecules of protein. The $N$ genes operate `in parallel'; for the dynamics of the population only the total number of genes in each state matters. As a consequence, there are $N+1$ different environmental states describing the configuration of the genes. We use the number of genes in the `on' state to label these states, $\sigma\in\{0,\dots,N\}$. We leave out the mRNA dynamics, and focus only on protein production and decay. We assume that each gene in its `on' state contributes $\Omega b_1$ to the total production rate, and each gene in its `off' state contributes $\Omega b_0$. As before the parameter $\Omega$ controls the typical size of the population of protein molecules.
We then have $\Omega b_\sigma=(N-\sigma)\Omega b_0+\sigma \Omega b_1$ for the total production rate.
The model is defined by the reactions

\begin{equation}
\begin{gathered}[c]
{G}^{(i)}_{\rm off} \xrightarrow{\mathmakebox[5mm]{\Omega b_0}}{} {G}^{(i)}_{\rm off} + {\rm P},\vphantom{{G}_{\rm off} \xrightleftharpoons[\lambda k_0]{\mathmakebox[5mm]{\lambda k_1}}{} {G}_{\rm on},}\\
{G}^{(i)}_{\rm on} \xrightarrow{\mathmakebox[5mm]{\Omega b_1}}{} {G}^{(i)}_{\rm on} + {\rm P},
\end{gathered}\quad
\begin{gathered}[c]
{G}^{(i)}_{\rm off} \xrightleftharpoons[\lambda k_0]{\mathmakebox[5mm]{\lambda k_1}}{} {G}^{(i)}_{\rm on},\\
{P} \xrightarrow{\mathmakebox[5mm]{\delta}}{} \emptyset,\\
\end{gathered}
\label{eq:proteinmodel_many_genes}
\end{equation}
where the reactions for different genes $i=1,\dots,N$ run independently.
The SDE description of the model in the limit of large but finite $\Omega$ and $\lambda$ is of the form
\begin{align}
\dot p ={}&
N b_\text{avg} - \delta p + \left[g_{\rm i}(p)^2 + g_{\rm e}(p)^2\right]^{1/2} \eta(t),
\end{align}
where each gene contributes an average rate of production $b_\text{avg}=(b_0k_0+b_1k_1)/(k_0+k_1)$.
The contribution to the noise from intrinsic fluctuations has amplitude 
\begin{equation}
g_{\rm i}(p)^2=\frac{1}{\Omega}\left(N b_\text{avg}+ \delta p\right).
\end{equation}
The environmental noise comes from the switching between the $N+1$ gene configurations; each gene switches between its on and off states independently. Following the earlier examples, one expects a contribution $2k_0k_1(b_0-b_1)^2/[\lambda(k_0+k_1)^3]$ to the variance of the environmental noise from each gene, so that the total variance is
\begin{equation}\label{eq:gem}
g_{\rm e}(p)^2= \frac{2 N k_1 k_0\left(b_0-b_1\right)^2}{\lambda\left(k_0 + k_1\right)^3}.
\end{equation}
We note that the relative fluctuations of the total production rate [i.e., the ratio $g_{\rm e}(p)/(N b_{\rm avg})$] scales as $N^{-1/2}$.

Mathematically, the transition rate matrix for the ${N+1}$ environmental states may be written as the tridiagonal matrix
\begin{align}\begin{aligned}
A_{\sigma\to\sigma-1}={}&\lambda k_0 \sigma, &&\text{for}~\sigma\geq 1,\\
A_{\sigma\to\sigma+1}={}&\lambda k_1 (N-\sigma), &&\text{for}~\sigma\leq N-1, \\
A_{\sigma\to\sigma\pm j}={}&0, &&\text{for}~j\geq 2,
\end{aligned}\end{align}
together with the convention $A_{\sigma\to\sigma}=-A_{\sigma\to\sigma-1}-A_{\sigma\to\sigma+1}$.
The formalism of Sec.~\ref{sec:analysis} can then be applied, but becomes algebraically tedious. Using numerical algebra packages we have verified Eq.~\eqref{eq:gem} up to $N=100$.
\begin{figure*}[ht!!]
\includegraphics[width=0.9\textwidth,valign=t]{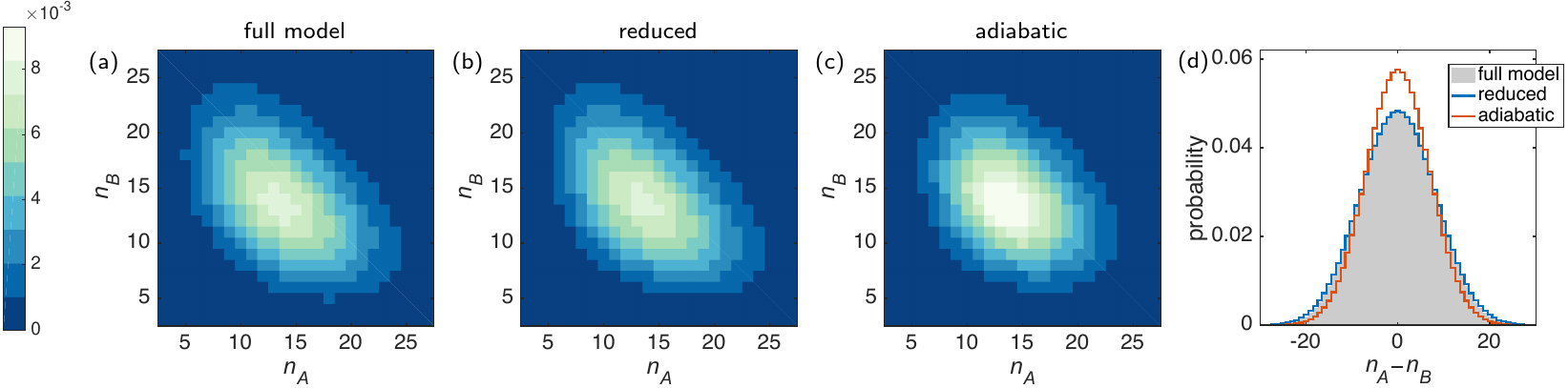}
\caption{Stationary distribution for the genetic circuit with exclusive binding (Sec.~\ref{sec:excl}) from numerical integration of (a) the full master equation with explicit environment, (b) the reduced master equation Eq.~\eqref{eq:CME_approx_exclusive}, and (c) the adiabatic approximation which considers average rates.
Panel (d) shows the marginal distribution of $n_A-n_B$ in order to compare the three distributions.
Parameters $\Omega\alpha_1=20$, $\Omega\alpha_0=0$, $\lambda \kappa_0=\Omega \lambda \kappa_1=20$.}
\label{fig:RK4_exclusive}
\end{figure*}

\subsection{Genetic circuit with exclusive binding}\label{sec:excl}

Next, we consider a circuit with exclusive promoter binding \cite{lipshtat2006genetic,warren2005chemical}.
The model describes two genes $G^A$ and $G^B$, and two corresponding proteins $P^A$ and $P^B$.
Proteins $P^A$ and $P^B$ bind to genes of the opposing type, $G^B$ and $G^A$, respectively, with (per capita) rates $\lambda\kappa_1/\Omega$ and $\lambda\mu_1/\Omega$.
They unbind from these promoters with rates $\lambda\kappa_0$ and $\lambda\mu_0$. These binding and unbinding reactions can be summarised as follows:

\begin{align}\begin{split}
G^A_{\rm unbnd.}+G^B_{\rm unbnd.} \xrightleftharpoons[\lambda \kappa_0]{\mathmakebox[12mm]{n_B\lambda \kappa_1/\Omega}}{}& G^A_{\rm bnd.}+G^B_{\rm unbnd.}, \\
G^A_{\rm unbnd.}+G^B_{\rm unbnd.} \xrightleftharpoons[\lambda \mu_0]{\mathmakebox[12mm]{n_A\lambda \mu_1/\Omega}}{}& G^A_{\rm unbnd.}+G^B_{\rm bnd.}, \\
\end{split}\end{align}
where the subscripts `bnd.' and `unbnd.' indicate whether the gene is bound to a protein or unbound, respectively, and where $n_A$ and $n_B$ are the numbers of molecules of proteins of type $A$ and $B$. In this model either gene $G^A$ or gene $G^B$ can be bound, but not both simultaneously.
This is due to spatial considerations of the binding process: owing to the proteins size and the proximity of the genes, the binding of a particular protein blocks the other protein from binding \cite{lipshtat2006genetic,warren2005chemical}.
When gene $G^A$ is bound, proteins of type $A$ are produced with rate $\Omega \alpha_0$, and when it is unbound they are produced with rate $\Omega \alpha_1$.
Similarly when gene $G^B$ is bound, proteins of type $B$ are produced with rate $\Omega \beta_0$, and when it is unbound they are produced with rate $\Omega \beta_1$. To summarise, the protein production rates in the three gene configurations are as follows:
\begin{equation*}
\begin{array}{lcc}
{} & \text{production} & \text{production} \\
{} & \text{rate $P^A$} & \text{rate $P^B$} \\
\hline
\mbox{$G^A$, $G^B$ unbound:} &\Omega \alpha_1& \Omega\beta_1 \\
\mbox{$G^A$ bound:}& \Omega \alpha_0& \Omega\beta_1\\
\mbox{$G^B$ bound:} &\Omega \alpha_1& \Omega\beta_0.
\end{array}\
\end{equation*}
In this model one protein inhibits the expression of  the other, i.e., $\alpha_0<\alpha_1$ and $\beta_0<\beta_1$.
Additionally, protein $A$ degrades with rate $\gamma$ and protein $B$ with rate $\delta$.  
 
In this model the birth rates of the two types of proteins are not independent; rather, they are connected through the state of the environment (the binding status of the two genes). Furthermore, when production of one protein is inhibited (for example protein $A$ when $G^A$ is bound), the other protein is expressed with a higher rate ($G^B$ unbound). This is an example of a model of the kind considered in Sec.~\ref{sec:twospecies} where we showed how anti-correlations lead to negative rates in the reduced master equation. The reduced master equation for this model is lengthy, we present it in Appendix~\ref{app:excl}. 

Figure~\ref{fig:RK4_exclusive} shows the results for the stationary distribution obtained from numerical integration of this reduced master equation; we also show the stationary distributions of the full model and of the adiabatic approximation.
As seen in panel (d) of the figure the reduced master equation reproduces the stationary distribution of the full model with greater accuracy than the adiabatic approximation. 
\subsection{Staged switching of the environment}
\label{sec:staged}

In many situations the switching between environmental states is not purely Markovian. Periodic switching between environmental states has been considered in experimental and theoretical studies of bacterial populations; for example the presence or absence of antibiotic treatment according to a periodic protocol. As a bet-hedging strategy, the bacteria respond to time-dependent external stresses with phenotypic heterogeneity \cite{thattai2004stochastic,acar2008stochastic,gefen2009importance,patra2015emergence,hufton2017bacteria}. In this context it is therefore important to be able to study stochastic populations coupled to environments with non-Markovian dynamics.

In this Section we consider an example in which there are two distinct environmental conditions, labelled $0$ and $1$. In contrast with the previous examples, each of these conditions consists of several identical, internal states (or stages), which are traversed in sequence. Similar setups have been used to model dynamics which fall between the purely periodic and purely Markovian limits, see e.g. Refs.~\cite{thattai2004stochastic,korobkova2006hidden,black2009stochastic}.

\begin{figure*}[ht]
{\centering
\includegraphics[width=0.9\textwidth,valign=t]{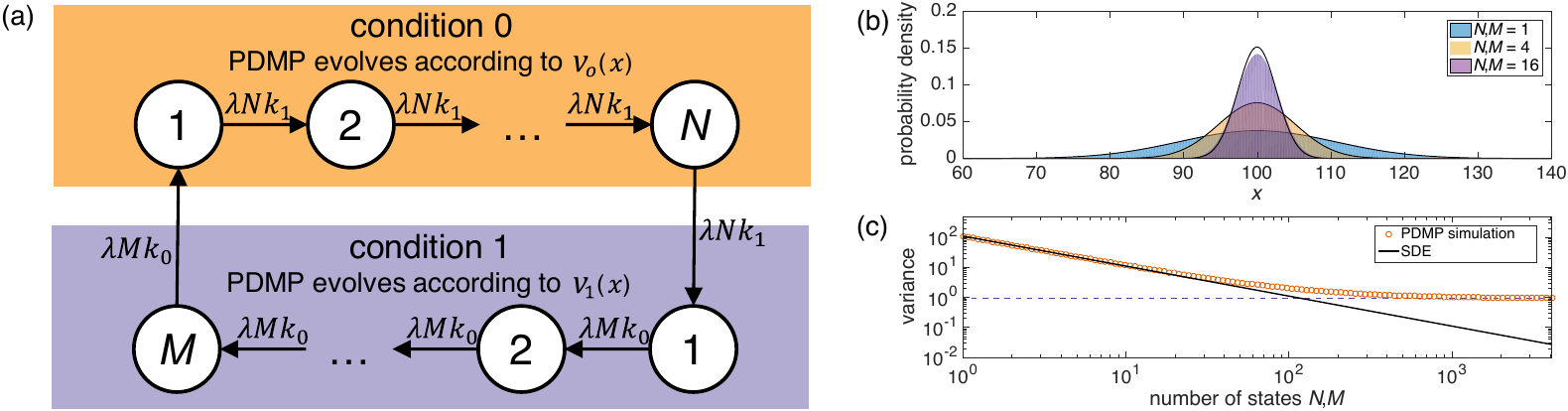}
\caption{
(a) Schematic illustrating an environment with two distinct conditions, with $N$ and $M$ identical stages, respectively.
(b) Stationary distribution for different values of $N=M$. Histograms show results of simulations, lines are from the theory described in the text.
(c) Variance of the stationary distribution as a function of $N$ (again for the case $N=M$). The black line shows the results of the theory, orange circles are from simulations of the piecewise-deterministic Markov process. Parameters $N$ and $M$ have been generalised to include non-integer values, by considering gamma-distributed waiting-time distributions in the two conditions (see text). Dashed line shows the variance of the limit cycle obtained in the limit of a periodic environment. Parameters: $b_0=100/3$ and $b_1=500/3$, $\lambda k_0=\lambda k_1=20$.
}\label{fig:gamma}}\end{figure*}

The model is illustrated in Fig.~\ref{fig:gamma}(a). There are $N$ environmental states which correspond to environmental condition $0$, and $M$ states that correspond to environmental condition $1$.
States in condition $0$ transition to the next state with rate $\lambda k_1 N$, and states corresponding to condition $1$ transition to the next state with rate $\lambda k_0 M$. The environment cycles through all states in order, as indicated in the figure.
 
In this way, the time spent in condition $0$ before switching to $1$ is $\Gamma(N,\lambda N k_1)$-distributed, and similarly the time spent in condition $1$ follows a $\Gamma(M,\lambda M k_0)$ distribution. Independent of $N$ and $M$, the environment spends an average time $(\lambda k_1)^{-1}$ in condition $0$ before it switches to $1$, and then an average time $(\lambda k_0)^{-1}$ in condition $1$ before it switches back to condition $0$.
Increasing the number of states $N$ and $M$ leads to an increased regularity of time spent in each condition.
The limit $N,M\to\infty$ in particular corresponds to periodic switching between the two conditions.

For simplicity we disregard intrinsic noise in this example and focus on a piecewise-deterministic process.
We assume that the dynamics are given by $\dot x = v_0(x)$ if the environment is in condition $0$, and by $\dot x = v_1(x)$ if it is in condition $1$.
Based on the formalism of Sec.~\ref{sec:analysis}, we use a symbolic algebra package to solve Eq.~\eqref{eq:bhelp6}, where the operator $\mathcal{M}_\sigma$ is substituted by the Liouville operator ${\cal L}_\sigma=-\partial_x v_\sigma(x)$. We use this to derive an SDE in the limit of fast but finite environmental dynamics. We find
\begin{equation}
\dot x= v_\text{avg}(x) +g_{\rm e}(x) \eta(t),
\label{eq:SDE_internal}
\end{equation}
where $\eta(t)$ is white Gaussian noise, and where the drift and diffusion terms are given by
\begin{align}\begin{split}
v_\text{avg}(x)={}&\frac{ k_0 v_0(x)+ k_1 v_1(x)}{ k_0+ k_1}, \\
g_{\rm e}(x)={}&\lambda^\nhalf \sqrt{\frac{k_1 k_0 (N+M)\left[v_0(x)-v_1(x)\right]^2}{NM\left(k_1 + k_0 \right)^3}}.
\label{eq:SDE_internal2}
\end{split}\end{align}
We have not attempted to formally prove this for general $N$ and $M$; rather, we tested this result for a range of combinations $N, M<150$ and found it to be true for all tested values.

In Fig.~\ref{fig:gamma}(b) we use a specific example, where the drift is $v_0(x)=b_0-x$ and $v_1(x)=b_1-x$; this choice corresponds to the protein production model considered in Sec.~\ref{sec:example1}.
In this figure we compare the stationary distributions obtained from simulation of the PDMP with the stationary distribution obtained analytically from solving the one-dimensional Fokker--Planck equation for Eq.~\eqref{eq:SDE_internal}. We show this data for different choices of $N$ and $M$ in Fig.~\ref{fig:gamma}(b), restricting to $N=M$ for simplicity.

Similarly, we compare the variance of the stationary distributions from the PDMP and the SDE in Fig.~\ref{fig:gamma}(c). The parameters $\lambda$, $k_1$, and $k_0$ are kept fixed; we focus again on the case $N=M$, and vary this number of internal states. Analytical results from the SDE and numerical simulation of the PDMP agree well for $N,M<100$, but there are deviations when $N=M$ becomes large. This is due to fact that the PDMP tends to a deterministic limit cycle; this limit cycles leads to a finite variance of the corresponding distribution, indicated by the dashed line of Fig.~\ref{fig:gamma}(c)]. These limit cycle dynamics are not captured by the SDE.

The model as described above is only defined for integer values of $N$ and $M$. However, the distribution of waiting times in either environmental condition can be generalised to the case of gamma distributions with non-integer shape parameters.
The interpretation as a series of internal states within conditions $\sigma=0$ and $\sigma=1$ then no longer holds, but simulations of the model can still be carried out, drawing waiting times directly from the appropriate gamma distributions. The SDE \eqref{eq:SDE_internal2} remains unaltered, and it provides an accurate description of the dynamics of the model also when $N$ and $M$ are not integers. This can be seen in Fig.~\ref{fig:gamma}(c), where many of the markers (circles) correspond to simulations for non-integer values of $N$ and $M$.
 
\subsection{Reliability analysis and crack propagation}\label{sec:crack}
\begin{figure*}
\includegraphics[width=0.85\textwidth,valign=t]{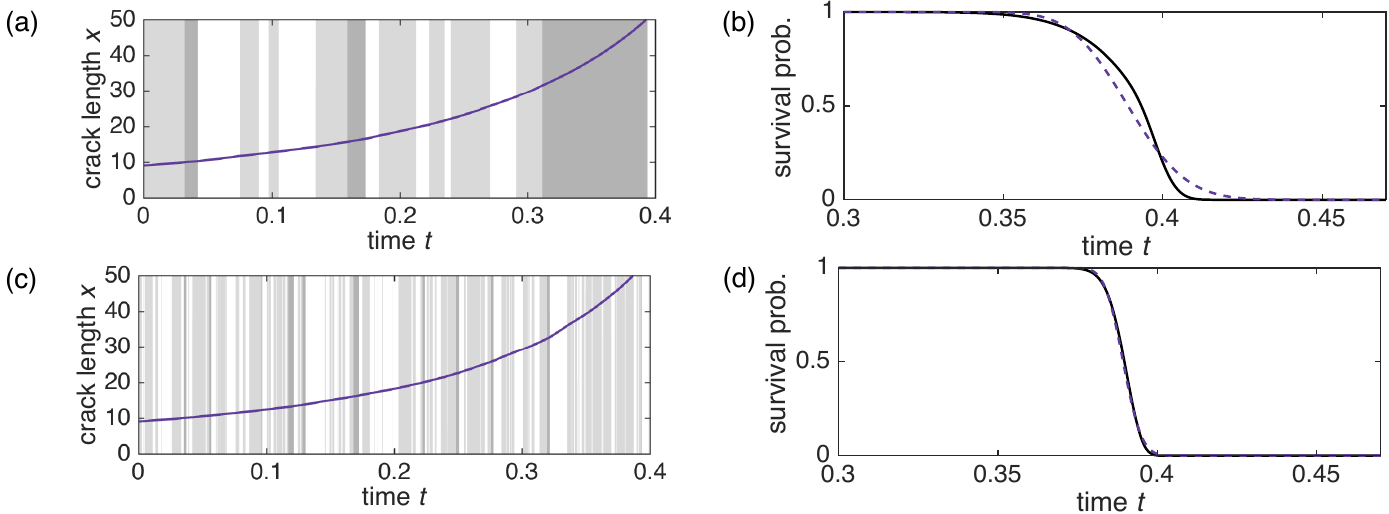}
\caption{
Panel (a): Sample path of the model of crack growth (Sec.~\ref{sec:crack}). Background shading indicates the state of the environment, with states $0$, $1$ and $2$ shown progressively darker. Panel (b): survival probability as a function of time. The black line shows results from Monte Carlo simulations, the dashed line is the prediction of Eq.~\eqref{eq:growth_passage_time}. Panels (c) and (d) show the same quantities for tenfold increased switching rates. Model parameters are given in the text.}
\label{fig:crack_modeling}
\end{figure*}
The formalism we have developed can also be applied to the calculation of time-to-failure in models of industrial systems.
One of the challenges in this field is to capture features of real-world systems in tractable mathematical models.
In this context, many authors have used piecewise-deterministic processes with Markovian external environments.
These models incorporate discrete environmental effects such as different modes of operation, external stresses or loads \cite{zhang2008piecewise,lorton2013methodology,lorton2013computation}.
In these applications there is often a clear separation of time scales, the environmental switching is a much faster process than the degradation of the system.
For example a piece of material may be subject to mechanical load which changes several times a day or hour, and the degradation of the material occurs over months or years.

Specifically, we consider the example of fatigue crack growth; this is an engineering problem describing the growth in the length of a crack in a mechanical component.  One such model uses a piecewise-deterministic Markov process to describe the growth of the length of a crack~\cite{chiquet2006estimating,chiquet2008modelling,chiquet2008method,chiquet2009piecewise} as follows,
\begin{equation}
\dot x =x^b \times v_{\sigma(t)},
\label{eq:crack_PDMP}
\end{equation}
where $x$ is the crack length, the exponent $b>0$ is a constant, and where as before $\sigma(t)$ represents the state of the environment at time $t$.
The factor $v_\sigma$ takes into account that the crack grows faster in some environments than in others. Transitions from state $\sigma$ to $\sigma'$ occur with rate $\lambda A_{\sigma\to\sigma'}$.

Given an initial length $x_0$, we are interested in the time it takes to reach the threshold length $x=L$; this is when the component is deemed unreliable. We use the formalism of the earlier sections to approximate the PDMP as an SDE in the limit of fast (but not infinitely fast) environmental switching ($\lambda \gg 1$). We then find the first-passage time of this SDE through the threshold value. While diffusive processes have been used as starting points in models of reliability \cite{lin1985stochastic,spencer1988markov}, we systematically reduce the PDMP to an effective stochastic differential equation.

In the simplest case of two environmental states (and writing $A_{0\to 1}=k_1, A_{1\to 0}=k_0$ as before), Eq.~\eqref{eq:crack_PDMP} can be approximated as by the SDE
\begin{align}
\dot x= x^b v_\text{avg} + g\, x^b \eta(t),
\label{eq:crack_SDE}
\end{align}
where
\begin{align}
v_\text{avg}=\frac{k_{0}v_0+k_{1}v_1}{k_{0}+k_{1}}, &&
g^2=\frac{2k_{0} k_{1} (v_0-v_1)^2 }{\lambda(k_{0}+k_{1})^3}.
\end{align}
Higher-order terms in $\lambda^{-1}$ have been discarded. For the special case of exponential growth, $b=1$, the SDE approximation turns into geometric Brownian motion.  In a different context this has been implemented in Ref.~\cite{Bressloff2014}. 
We proceed to find the first-passage time of the process in Eq.~\eqref{eq:crack_SDE} through the threshold $L$. This can be done following Ref. \cite{capocelli1974diffusion}, but with a modification allowing for $b\neq 1$. As a first step we apply the transformation
\begin{equation}\label{eq:trafo}
y=\left\{ \begin{array}{cc}
\ln x & b=1\\
~\\
\frac{x^{1-b}-1}{1-b} & b\neq 1.
 \end{array} \right.
\end{equation}
The SDE \eqref{eq:crack_SDE} can then be written
\begin{equation}\label{eq:y}
\dot y = v_{\rm avg} + g\, \eta(t).
\end{equation}
For such a process, the distribution of first passage times through a given threshold is known \cite{capocelli1974diffusion}. Returning to the original variables, we obtain the probability density $Q(x_0,t)$ of first-passage times of the process Eq.~\eqref{eq:crack_SDE} through 
$L$, if started at position $x_0$ at time $t=0$. For $b=1$ one finds
\begin{align}
Q(x_0,t)=
 \dfrac{\left| \ln(L/x_0)\right|}{g t(2\pi t)^{1/2}}
\exp\left( -\frac{ \left[ \ln(L/x_0) - (v_\text{avg}-\frac{g^2}{2}) t \right]^2}{2g^2 t}\right),
\end{align}
and for $b\neq 1$ one has
\begin{multline}
Q(x_0,t)=\dfrac{1}{g t(2\pi t)^{1/2}}
\left| \dfrac{L^{1-b}-x_0^{1-b}}{1-b}\right| \\
\hspace{-5em}\times \frac{ -\left[ (L^{1-b}-x_0^{1-b})/(1-b) - (v_\text{avg}-\frac{g^2}{2}) t \right]^2}{2g^2 t}.
\label{eq:growth_passage_time}
\end{multline}

This approach can be extended to models with more than two environmental states, leading to modifications in the noise strength $g$. We demonstrate this with a numerical example. We use the parameters suggested in Ref.~\cite{chiquet2006estimating}, in particular $b=1.5$, and 
\begin{align*}
 \lambda {\mathbf A}=\left[ \begin{array}{rrr}
\!-40 & \!40	& \!0 \\
\!54	& \!-60 & \!6 \\
\!20	& \!0 	& \!-20 \end{array} \right]\!\!,
 ~v=\left[ \begin{array}{l}
1.0\\
0.9\\
1.2\\
\end{array} \right]\!\!.
\end{align*}
The initial crack length is $\xx_0=9$, and we use $L=50$ as the threshold for the onset of failure.
Compared to Ref.~\cite{chiquet2006estimating} we have rescaled time. Implementing our theory shows that the process can be approximated by the SDE \eqref{eq:crack_SDE} 
where $v_\text{avg}={69}/{70}$ and $g^2={141}/{274400}$.
This is obtained from solving Eq.~\eqref{eq:bhelp6} with a numerical algebra package, again substituting the operator $\M_\sigma$ with the appropriate Liouville operator.
Figure~\ref{fig:crack_modeling}(a) shows a sample path of the PDMP generated by Monte Carlo simulation, while the background indicates the state of the environment. Figure~\ref{fig:crack_modeling}(b) shows the probability that a given component is still reliable at time $t$.
The black line is obtained through Monte Carlo simulations, whereas the dashed line is the prediction of Eq.~\eqref{eq:growth_passage_time}.
For the specified parameters, the two lines show agreement. Increasing the switching rate [Fig.~\ref{fig:crack_modeling}(d)] strengthens this agreement.

\section{Summary and conclusions}\label{sec:concl}
We have developed methods for the reduction and approximation of the dynamics of discrete stochastic systems coupled external environments with a finite number of discrete states. Our analysis focuses on the limit in which the environmental dynamics are fast relative to that of the system, but where the time scale separation is not necessarily infinite. In particular, we have derived reduced dynamics for the open system, capturing next-order corrections to the adiabatic limit.

The model reduction leads to master equations with bursting, and---in some cases---with negative transition `rates'. Our analysis shows that negative (pseudo-) probabilities can arise from such non-Markovian reduced dynamics, and it suggests that these negative transients only occur on time scales shorter than than that of the environment. The reduced dynamics---obtained by coarse graining the environmental process---does not resolve the physics of the problem on such fine time scales. The occurrence of bursting reactions can be understood further by looking at the time evolution of individual sample paths of system and environment over a finite time interval. This leads to a discrete-time approximation for the dynamics of the open system. The path of the environment in one time step can be approximated by Gaussian random variables; bursting in the system results from fluctuations of this discrete-time Gaussian process.

We find that trajectories obtained using a simulation algorithms adapted from open quantum systems to the classical case do not reproduce statistical features of sample paths of the full model of system and environment. This suggests further work on the relation of full and reduced dynamics in the quantum context. We note one potentially important difference between the classical and the quantum cases; the origin of non-Markovianity in open quantum systems is often attributed to a two-way exchange of information between the system and the environment \cite{Breuer2016,Vega2017}. This mechanism is not available in some of the examples we have looked at, even though these models still lead to negative rates in the reduced dynamics (Sec.~\ref{sec:twospecies}).

We placed our reduction schemes in the context of existing work on piecewise-deterministic Markov processes, and piecewise-diffusive processes. In particular we study combined expansions in the relative time-scale of the environment and/or the strength of the intrinsic noise. This provides a more complete picture of different approximations for systems with intrinsic noise and environmental fluctuations. We also introduced a scheme approximating such a process as a SDE, capturing both switching and demographic noise. We expect this tool to facilitate efficient simulation of open systems.

We have demonstrated how these results can be used to study a number of problems in different areas. In particular, the reduction schemes we have proposed allow for an approximation of the open system in terms of stochastic differential equations. The approximation is valid when populations are large and the environmental process fast. In this situation simulations of the full dynamics of the population and the environment are particularly costly. The stochastic differential equation approximates both the intrinsic and the extrinsic randomness as Gaussian noise, and it can be used to carry out simulations more efficiently. As we have shown, it also allows for analytical progress in some cases. We have used the different approximation schemes for a varied set of applications, including models of genetic circuits, cases in which the switching between external conditions is non-Markovian, and a model of crack propagation. These applications are only a selected set of examples of situations in which switching environments play a role. We expect that the model-reduction schemes we have developed will be of use for further open classical systems in biology and the physical sciences, and in other disciplines. In Sec.~\ref{sec:km} we have related the different approximations to each other, and to existing work. This may help to select the most appropriate approximation method for specific applications.

Our work raises a number of questions for future work. For example, it would be interesting to study in more detail the analogies and differences between non-Markovian reduced dynamics for open quantum systems and for classical systems coupled to fast environments. As a first step, one might focus on classical systems in which the dynamics of the environment depends on the state of the system itself (such as in the examples in Secs.~\ref{sec:bimodal} and \ref{sec:multi}), and try to characterise the mathematical structure of the resulting reduced dynamics, and the information flow between system and environment. We also note that we have found non-Markovian features in the reduced dynamics at sub-leading order only when the system itself has discrete states, but not when the population is described by a piecewise diffusive or piecewise deterministic process. Based on the Pawula theorem \cite{risken1984fokker, gardiner1985handbook} we expect unphysical terms in the latter cases when the expansion is truncated at higher-orders. Further work is required to understand in more detail how these features non-Markovian features emerge in the combined coarse graining process for the population and the environment. A separate further line of research might focus on systems in which the environment takes continuous states (see e.g. Ref.~\cite{assaf1,assaf2,assaf3}), and the comparison with the discrete case.

{\em Acknowledgements.} P.G.H., Y.T.L., and T.G. thank the Engineering and Physical Sciences Research Council (EPSRC) for funding
(PhD studentship, and Grant No. EP/K037145/1). Y.T.L. was supported by the Center for Nonlinear Studies. 

{\em Author contributions.} All authors contributed to this work equally.

\onecolumngrid
\begin{appendix}
\section{State-dependent environmental process}\label{sec:gen}
In this Section of the Appendix we briefly consider the case in which the transition matrix for the environmental process depends on the state of the system proper, i.e., $A_{\sigma\to\sigma'}=A_{\sigma\to\sigma'}(\ell)$. From Eq.~\eqref{eq:bhelp3} we have
\begin{equation}\label{eq:bhelp3prime}
 \frac{\d}{\d t}\Pi(\ell,t) =\sum_\sigma\cM_\sigma \left[\rho^\star(\sigma|\ell)\Pi(\ell,t) \right]+ \frac{1}{\lambda}\sum_\sigma \cM_\sigma w_\sigma(\ell,t), 
\end{equation}
and from Eq.~\eqref{eq:bhelp6}
\begin{equation}\label{eq:bhelp6prime}
\sum_{\sigma'}A_{\sigma'\to\sigma}(\ell) w_{\sigma'}(\ell,t) = \rho^\star(\sigma|\ell)\sum_{\sigma'}\cM_{\sigma'} \left[\rho^\star(\sigma'|\ell)\Pi(\ell,t) \right]-\cM_\sigma \left[\rho^\star(\sigma|\ell)\Pi(\ell,t) \right].
\end{equation}
This serves as a starting point for the further analysis.
\subsection{Adiabatic limit}
It is useful to define the following operators, acting on functions $f(\ell)$,
\begin{equation}
\widehat \cM_\sigma f(\ell) = \cM_\sigma[ \rho^\star(\sigma|\ell) f(\ell)],
\end{equation}
where the right-hand side indicates that the operator $\cM_\sigma$ acts on the object inside the square bracket. In the adiabatic limit one finds [by sending $\lambda\to\infty$ in Eq.~\eqref{eq:bhelp3prime}]
\begin{equation}
\frac{\d}{\d t}\Pi(\ell,t)=\widehat\cM_{\rm avg} \Pi(\ell,t),
\end{equation}
where we now have
\begin{equation}
 \cM_{\rm avg} = \sum_\sigma \widehat\cM_\sigma.
\end{equation}

\subsubsection*{Example}
To illustrate the principle we use a population with $n$ individuals of a single species, and a birth reaction with rate $b_\sigma(n)$. We then have $\cM_\sigma = [\E^{-1}-1]b_\sigma(n)$. We find
\begin{equation}
\cM_{\rm avg} \Pi(n,t)=[\E^{-1}-1]b_{\rm avg}(n)\Pi(n,t),
\end{equation}
where
\begin{equation}
b_{\rm avg}(n)=\sum_\sigma \rho^\star(\sigma|n)b_\sigma(n).
\end{equation}
We note that $b_{\rm avg}(n)$ carries a dependence on $n$, even if $b_\sigma(n)$ is itself independent of $n$.

\subsection{Next-order contribution}
In order to address the sub-leading term in $1/\lambda$, we focus on the case of two environmental states, with switching rates $A_{1\to 0}(\ell)=k_0(\ell)$ and $A_{0\to 1}(\ell)=k_1(\ell)$. In this case we have $\rho^\star(0|\ell)=k_0(\ell)/[k_0(\ell)+k_1(\ell)]$, and $\rho^\star(1|\ell)=k_1(\ell)/[k_0(\ell)+k_1(\ell)]$. From Eq.~\eqref{eq:bhelp6prime} one then finds
\begin{equation}
w_0(\ell,t)=-w_1(\ell,t)=\frac{1}{k_0(\ell)+k_1(\ell)}\left[\rho^\star(1|\ell) \widehat\cM_0 - \rho^\star(0|\ell)\widehat\cM_1\right]\Pi(\ell,t).
\end{equation}
Inserting into Eq.~\eqref{eq:bhelp3prime} we have
\begin{equation}
\frac{\d}{\d t}\Pi(\ell,t)=\cM_{\rm avg} \Pi(\ell)+\frac{1}{\lambda}(\cM_0-\cM_1)\frac{1}{k_0(\ell)+k_1(\ell)}\left[\rho^\star(1|\ell) \widehat\cM_0 - \rho^\star(0|\ell)\widehat\cM_1\right]\Pi(\ell,t),
\end{equation}
which can be written as
\begin{equation}
\frac{\d}{\d t}\Pi(\ell,t)=\cM_{\rm avg} \Pi(\ell)+\frac{1}{\lambda}\left[\widehat \cM_0\rho^*(0|\ell)^{-1}-\widehat\cM_1\rho^*(1|\ell)^{-1}\right]\frac{1}{k_0(\ell)+k_1(\ell)}\left[\rho^\star(1|\ell) \widehat\cM_0 - \rho^\star(0|\ell)\widehat\cM_1\right]\Pi(\ell,t).
\end{equation}
While this object is quite lengthy, it formally describes the reduced dynamics to sub-leading order in $1/\lambda$, and can be used for further analysis. The next steps would then depend on the nature of the specific example at hand.
 \section{Further remarks relating to power spectra in Sec.~\ref{sec:sim_paths}}
 \subsection{Cross spectra in adiabatic limit}\label{app:xspec}
We find in simulations of the model in Sec.~\ref{sec:twospecies:model} that the cross spectrum $S_{AB}(\omega)$ vanishes in the adiabatic limit (see Fig.~\ref{fig:corspec}). This can be understood by inspecting the master equation in the adiabatic limit [obtained from Eq.~(\ref{eq:cme_ab_approx}) by sending $\lambda\to\infty$],
\begin{align}
\frac{\d}{\d t} \Pi={}&\gamma(\E_{A}-1) n_A \Pi+\delta(\E_{B} -1) n_B \Pi \nonumber \\
&+\Omega\alpha_{\rm avg} (\E_{A}^{-1}-1) \Pi+\Omega\beta_{\rm avg}(\E_{B}^{-1}-1) \Pi.
\label{eq:ab_adiab}
\end{align}
No reaction in this equation involves both types of particles; hence there are no correlations between $n_A$ and $n_B$, leading to $S_{AB}(\omega)=0$.

\subsection{Indepence of $S_{AA}(\omega)$ from $\Delta\beta$}\label{app:saa}

 We focus on the power spectral density $S_{AA}(\omega)$ of the dynamics defined by Eq.~\eqref{eq:cme_ab_approx}. Since $S_{AA}(\omega)$ is a feature only of the dynamics of species $A$, we can integrate out the variable $n_B$ in Eq.~\eqref{eq:cme_ab_approx}. We obtain the following equation for the marginal distribution $\Pi_A(n_A)$:
\begin{align}
\frac{\d}{\d t} \Pi_A(n_A)={}&\gamma(\E_{a}-1) n_A \Pi_A(n_A) \nonumber \\
&+\Omega \left[\alpha_\text{avg} - \frac{\Omega\theta^2}{\lambda}(\Delta\alpha)^2 \right] (\E_{a}^{-1}-1)\Pi_A(n_A) \nonumber\\
&+ \frac{\Omega^2 \theta^2 }{2 \lambda}(\Delta\alpha)^2 (\E_{a}^{-2}-1) \Pi_A(n_A) .
\label{eq:marg_a}
\end{align}
In particular, all terms proportional to $\Delta\alpha\Delta\beta$ have cancelled out, so that Eq.~\eqref{eq:marg_a} should apply to both cases, $\Delta\alpha\Delta\beta>0$ and $\Delta\alpha\Delta\beta<0$. Thus, one would expect the power spectral density $S_{AA}(\omega)$ to be independent of the choice of $\Delta\beta$ (all other parameters kept fixed).
This in turn indicates that the spectra $S_{AA}(\omega)$ in panels (a) and (c) of Fig.~\ref{fig:corspec} should come out as identical, if the modified Gillespie algorithm is a valid method of generating sample paths of the reduced master equation~\eqref{eq:cme_ab_approx}. At sufficiently low frequencies one would also expect these spectra to agree with those obtained from simulating the full model.
The observations in Fig.~\ref{fig:corspec} indicate that (i) the spectra $S_{AA}(\omega)$ for the reduced model for $\Delta\alpha\Delta\beta>0$ and $\Delta\alpha\Delta\beta<0$ are markedly different from each other [compare panels (a) and (c)]; (ii) for $\Delta\alpha\Delta\beta>0$, the spectrum $S_{AA}(\omega)$ from the reduced model agrees to a good approximation with that from the full model in the low-frequency range [panel (a)]. For $\Delta\alpha\Delta\beta<0$, these findings suggest a problem in approximating sample paths of the full model from the reduced master equation, using the renormalisation technique.
\\

\section{Kramers--Moyal expansion}
\subsection{Kramers--Moyal expansion of reduced master equation}\label{sec:appkm}
In this Appendix we carry out a direct Kramers-Moyal expansion of the reduced master equation (\ref{eq:masterexample2}). This master equation can be written as
\begin{align}\label{eq:masterexample3}
\frac{\d}{\d t}\Pi(n,t) ={}& \Omega\beta ({\cal E}^{-1}-1)\Pi(n)+ ({\cal E}-1) \delta_{\rm eff}n\Pi(n,t)+ \frac{1}{2}\frac{\theta^2}{\lambda}\Delta^2 \left[{\cal E}^{2}-1\right]n(n-1)\Pi(n,t),
\end{align}
where $\Delta=\delta_0-\delta_1$, and
\begin{align}
\delta_{\rm eff} ={}& \delta_{\rm avg} - \frac{1}{2}\frac{\theta^2}{\lambda} \Delta^2(2n-1).
\end{align}
To carry out the expansion we write ${\cal E}^2=1+\frac{2}{\Omega}\partial_x + \frac{2}{\Omega^2}\partial_x^2 +\dots$, and obtain (writing $x=n/\Omega$)
\begin{align}\label{eq:help2}
\frac{\partial}{\partial t}\Pi(x) ={}& \beta \left(-\partial_x+\frac{1}{2\Omega}\partial_x^2\right)\Pi(x) + \left(\partial_x+\frac{1}{2\Omega}\partial_x^2\right) \delta_{\rm eff}x\Pi(x)+\frac{1}{2}\frac{\theta^2}{\lambda}\Omega\Delta^2 \left(2\partial_x+\frac{2}{\Omega}\partial_x^2\right)x\left(x-\frac{1}{\Omega}\right)\Pi(x),
\end{align}
where neglected terms are of order $1/\Omega^2$ or of order $\frac{\theta^2}{\lambda}/\Omega\propto 1/(\lambda\Omega)$. There will be further terms in Eq (\ref{eq:help2}) which can be neglected at the order we are working at. Next we collect terms
\begin{align}\label{eq:help3}
\frac{\partial}{\partial t}\Pi(x) ={}& -\partial_x \left\{\left[\beta-\delta_{\rm eff}x-\frac{\theta^2}{\lambda}\Omega\Delta^2 x\left(x-\frac{1}{\Omega}\right)\right]\Pi(x) \right\}+\frac{1}{2\Omega}\partial_x^2\left\{\left[\beta+\delta_{\rm eff}x +2\frac{\theta^2}{\lambda}\Omega\Delta^2 x^2\right]\Pi(x)\right\},
\end{align}
where another term of order $1/(\lambda\Omega)$ has been dropped. Now we use $\delta_{\rm eff} = \delta_{\rm avg} - \frac{1}{2}\frac{\theta^2}{\lambda} \Omega\Delta^2\left(2x-\frac{1}{\Omega}\right)$, and find
 \begin{align}\label{eq:help4}
\frac{\partial}{\partial t}\Pi(x) ={}& -\partial_x \left\{\left[\beta-\delta_{\rm avg}x+\frac{1}{2}\frac{\theta^2}{\lambda}\Omega \Delta^2x\left(2x-\frac{1}{\Omega}\right)-\frac{\theta^2}{\lambda}\Omega\Delta^2 x\left(x-\frac{1}{\Omega}\right)\right]\Pi(x) \right\} \nonumber\\
&+\frac{1}{2\Omega}\partial_x^2\left\{\left[\beta+\delta_{\rm avg}x - \frac{\theta^2}{\lambda}\Omega \Delta^2x^2 +2\frac{\theta^2}{\lambda}\Omega\Delta^2 x^2\right]\Pi(x)\right\},
\end{align}
where yet another term of order $1/(\lambda\Omega)$ has been dropped.
This is the same as
\begin{align}\label{eq:help5}
\frac{\partial}{\partial t}\Pi(x) ={}& -\partial_x \left\{\left[\beta-\delta_{\rm avg}x+\frac{1}{2}\frac{\theta^2}{\lambda} \Delta^2x\right]\Pi(x) \right\}+\frac{1}{2}\partial_x^2\left\{\left[\frac{1}{\Omega}(\beta+\delta_{\rm avg}x) +\frac{\theta^2}{\lambda}\Delta^2 x^2\right]\Pi(x)\right\},
\end{align}
i.e., we recover Eq.~\eqref{eq:fpecentre}.

\subsection{Reduced Liouville equation}\label{sec:appliouville}
Using $\cL_\sigma\Pi(x)=-\partial_x (\beta-\delta_\sigma x)\Pi(x)$ in Eq.~\eqref{eq:2statesliouville} gives
\begin{equation}
\frac{\partial}{\partial t}\Pi=-\partial_x(\beta-\delta_{\rm avg}x)\Pi(x)+\frac{1}{2}\frac{\theta^2}{\lambda}\Delta^2\partial_x x \partial_x \Pi(x).
\end{equation}
Next we use $\partial_x \left(x \partial_xx\Pi(x)\right)=\partial_x^2 \left(x^2\Pi\right)-\partial_x\left(x\Pi\right)$ to write this as
\begin{equation}
\frac{\partial}{\partial t}\Pi=-\partial_x\left(\beta-\delta_{\rm avg}x+\frac{1}{2}\frac{\theta^2}{\lambda}\Delta^2x\right)\Pi(x)+\frac{1}{2}\frac{\theta^2}{\lambda}\Delta^2\partial_x^2 [x^2 \Pi(x)].
\end{equation}
This is Eq.~\eqref{eq:centreright}.
\subsection{Kramers--Moyal expansion for two-species model}\label{sec:kmab}
Carrying out the Kramers--Moyal expansion on Eq.~\eqref{eq:cme_ab_approx} 
we find
\begin{align}
\frac{\partial}{\partial t} \Pi(x)={}&\gamma\left(\partial_x+\frac{1}{2\Omega}\partial_x^2\right) x \Pi+\delta\left(\partial_y+\frac{1}{2\Omega}\partial_y^2\right) y \Pi +\alpha_{\rm eff} \left(-\partial_x+\frac{1}{2\Omega}\partial_x^2\right) \Pi+\beta_{\rm eff}\left(-\partial_y+\frac{1}{2\Omega}\partial_y^2\right) \Pi\ \nonumber\\
&+ \frac{\Omega\theta^2}{2\lambda}(\Delta\alpha)^2 \left(-2\partial_x+\frac{2}{\Omega}\partial_x^2\right) \Pi (t)+ \tfrac{\Omega\frac{\theta^2}{\lambda}}{2}(\Delta\beta)^2 \left(-2\partial_y+\frac{2}{\Omega}\partial_y^2\right) \Pi \nonumber\\
&+\frac{\Omega\theta^2}{\lambda}\Delta\alpha\Delta\beta\left(-\partial_x-\partial_y+\frac{1}{2\Omega}\partial_x^2+\frac{1}{2\Omega}\partial_y^2+\frac{1}{\Omega}\partial_x\partial_y\right) \Pi.
\label{eq:cme_ab_approx3}
\end{align}
Using Eq.~\eqref{eq:effrates}, this simplifies to
\begin{align}
\frac{\partial}{\partial t}\Pi(x)={}& - \partial_x\left(\alpha_{\rm avg}-\gamma x\right)\Pi - \partial_y\left(\beta_{\rm avg}-\delta y\right)\Pi \nonumber \\
&+\frac{1}{2}\partial_x^2 \left(\frac{\alpha_{\rm avg}+\gamma x}{\Omega}+\frac{\theta^2}{\lambda}\Delta\alpha^2\right)\Pi+\frac{1}{2}\partial_y^2 \left(\frac{\beta_{\rm avg}+\delta y}{\Omega}+\frac{\theta^2}{\lambda}\Delta\beta^2\right)\Pi+\partial_x\partial_y\left(\frac{\theta^2}{\lambda}\Delta\alpha\Delta\beta\right)\Pi,
\end{align}
which describes the dynamics of the stochastic differential equations in Eqs. (\ref{eq:kmtoy1},\ref{eq:kmtoy2}).
 \section{Applications---Further details}
 \subsection{Reduced master equation for bi-stable genetic circuit}\label{sec:redmasterbistable}
 In this Appendix we report the reduced master equation for the model described in Sec.~\ref{sec:bimodal}. The reduced master equation is obtained starting from Eq.~(\ref{eq:bhelp3}), where the $w_\sigma(\ell)$ are determined from (\ref{eq:bhelp6}). We do not report the full calculation; it is laborious, but ultimately straightforward. The final result for the reduced master equation reads:
\BE
&&\frac{\d}{\d t}\Pi(N_p,N_m,t) \nonumber \\
&=&(E_m^{-1}-1)\left\{\Omega\beta_{\rm avg}(N_p)-\frac{1}{\lambda}\Omega^2 (\beta_2-\beta_0)^2\frac{1}{k_-}\frac{2\psi^2}{(1+\psi+\psi^2)^3}[\psi^2+3\psi+1 ]\right\}\Pi  \nonumber  \\
&&+(E_m-1)[\delta_m N_m\Pi]    \nonumber \\
&&+ (E_p^{-1}-1)[\alpha N_m+\varpi_1] \Pi   \nonumber  \\
&&+(E_p-1)[\delta_p N_p+\varpi_2]\Pi     \nonumber \\
&& +(E_m^{-2}-1) \left[ \frac{1}{\lambda}\Omega^2 (\beta_2-\beta_0)^2\frac{1}{k_-}\frac{\psi^2}{(1+\psi+\psi^2)^3}\left[\psi^2+3\psi+1 \right]
\Pi\right]   \nonumber   \\
 &&+(E_m^{-1}E_p^{-1}-1) (-\varpi_1 \Pi)   \nonumber \\
 &&+(E_m^{-1}E_p-1)(-\varpi_2 \Pi),\label{eq:redmasterbistable}
\EE
where we have introduced the following short-hands ($\sigma=0,1,2$),
\BE
\psi(N_p)&=&\frac{k_+ N_p}{\Omega k_-}, \nonumber \\
\rho_\sigma^\star(N_p)&=&\frac{\psi(N_p)^{\sigma-1}}{1+\psi(N_p)+\psi(N_p)^2}, \nonumber \\
\Delta_\sigma(N_p)&=&\rho^\star_\sigma(N_p+1)-\rho^\star_\sigma(N_p),\nonumber \\
\beta_{\rm avg}(N_p)&=&\sum_\sigma \rho_\sigma^*(N_p)\beta_\sigma,\nonumber \\
\varpi_1&=& \frac{1}{\lambda}\Omega (\beta_2-\beta_0) \frac{1}{k_-}\left\{[\rho_0^\star(N_p+1)+\rho_1^\star(N_p+1)]  \Delta_2 -\rho_1^\star(N_p+1)  \Delta_0 \right\}\alpha N_m.  \nonumber \\
\varpi_2&=& \frac{1}{\lambda}\Omega (\beta_2-\beta_0) \frac{1}{k_-}  \left\{\rho_1^\star(N_p-1)  \Delta_0(N_p-1)  -[\rho_0^\star(N_p-1)+\rho_1^\star(N_p-1)]   \Delta_2(N_p-1)  \right\}\delta_p N_p.
\EE
We note that $\varpi_1>0$, irrespective of the choice of $\lambda$, so that the rate of the penultimate reaction in Eq.~(\ref{eq:redmasterbistable}) is negative. The rates of all other reactions are non-negative, provided $\lambda$ is large enough (all other parameters fixed).
\subsection{Gene circuit with exclusive binding}\label{app:excl}
In this Appendix we report the reduced master equation for the gene circuit with exclusive binding, discussed in Sec.~\ref{sec:excl}. Labelling the states $G^A$ and $G^B$ not occupied, only $G^A$ occupied, and only $G^B$ occupied as $\sigma=0,1,$ and $2$, respectively, we have the transition matrix elements
\begin{align}
A_{0\to 1}= n_A \mu_1/\Omega, \quad A_{0\to 2}= n_B \kappa_1/\Omega, \quad A_{1\to 0}=\kappa_0, \quad A_{2\to 0}=\mu_0,
\end{align}
where all other off-diagonal entries are zero, and the diagonal elements follow from the convention $\sum_{\sigma'} A_{\sigma\to\sigma'}=0$. For the purposes of the numerical analysis we make the simplification $\alpha_0=\beta_0, \alpha_1=\beta_1$, and $\kappa_0=\mu_0, \kappa_1=\mu_1$, as well as $\gamma=\delta$. The reduced master equation in the limit of large but finite $\lambda$ is then obtained as
\begin{align}\begin{split}
\frac{\d}{\d t}P_{n_A,n_B}(t)={}&\left(\E_{ A }^{-1}-1\right)
\left\{ \Omega\frac{ n_B \alpha_1 \tilde{\kappa}_1 + \alpha_0 (\kappa_0 + n_A \tilde{\kappa}_1)}{\kappa_0 + ( n_A + n_B ) \tilde{\kappa}_1}
 -\frac{\Omega^2}{\lambda}\frac{2 n_B \kappa_0 \tilde{\kappa}_1 (\alpha_0 - \alpha_1)^2}{\left[\kappa_0 + ( n_A + n_B ) \tilde{\kappa}_1\right]^3 } \right\}P_{n_A,n_B}(t) \\ 
& + \left(\E_{ B }^{-1}-1\right)\left\{ \Omega\frac{ n_A \alpha_1 \tilde{\kappa}_1 + \alpha_0 (\kappa_0 + n_B \tilde{\kappa}_1)}{\kappa_0 + ( n_A + n_B ) \tilde{\kappa}_1} -
 \frac{\Omega^2}{\lambda} \frac{ 2 n_A \kappa_0 \tilde{\kappa}_1 (\alpha_0 - \alpha_1)^2 }{\left[\kappa_0 + ( n_A + n_B ) \tilde{\kappa}_1\right]^3 } \right\}P_{n_A,n_B}(t) \\
&+ \left(\E_{ A }^{-2}-1\right)\frac{\Omega^2}{\lambda} \frac{ n_B \tilde{\kappa}_1 \left[\kappa_0^2 + 2 n_A \kappa_0 \tilde{\kappa}_1 + n_A ( n_A + n_B ) \tilde{\kappa}_1^2\right](\alpha_0 - \alpha_1)^2}{\kappa_0 \left[\kappa_0 + ( n_A + n_B ) \tilde{\kappa}_1\right]^3 }P_{n_A,n_B}(t) \\
&-\left(\E_{ A }^{-1}\E_{ B }^{-1}-1\right) \frac{\Omega^2}{\lambda}\frac{2 n_A n_B \tilde{\kappa}_1^2 \left[2 \kappa_0 + ( n_A + n_B ) \tilde{\kappa}_1\right](\alpha_0 - \alpha_1)^2 }{\kappa_0 \left[\kappa_0 + ( n_A + n_B ) \tilde{\kappa}_1\right]^3 }P_{n_A,n_B}(t) \\
&+\left(\E_{ B }^{-2}-1\right) \frac{\Omega^2}{\lambda} \frac{n_A \tilde{\kappa}_1 \left[\kappa_0^2 + 2 n_B \kappa_0 \tilde{\kappa}_1 + n_B ( n_A + n_B ) \tilde{\kappa}_1^2\right] (\alpha_0 - \alpha_1)^2}{\kappa_0 \left[\kappa_0 + ( n_A + n_B ) \tilde{\kappa}_1\right]^3 }P_{n_A,n_B}(t) \\
&+\gamma (\E_{ A }^{+1}-1) n_A P_{n_A,n_B} (t)+\gamma (\E_{ B }^{+1}-1) n_B P_{n_A,n_B} (t),
\label{eq:CME_approx_exclusive}
\end{split}\end{align}
where $\tilde{\kappa}_1$ has been introduced as shorthand for $\kappa_1/\Omega$. We have discarded terms of order $\Omega/\lambda$.

\end{appendix}

\end{document}